\documentclass[useAMS,usegraphicx,usenatbib,usedcolumn]{mn2e}

\usepackage{times}

\newcommand{\oi}{[O \textsc{i}]6300}
\newcommand{\oii}{[O \textsc{ii}]3727}
\newcommand{\oiii}{[O \textsc{iii}]5007}

\newcommand{\sii}{[S \textsc{ii}]6716}
\newcommand{\nii}{[N \textsc{ii}]6584}

\newcommand{\ha}{H\ensuremath{\alpha}}
\newcommand{\hb}{H\ensuremath{\beta}}
\newcommand{\ew}[1]{EW\ensuremath{_{abs}}(#1)}

\newcommand{\Mstar}{\ensuremath{M_{*}}}
\newcommand{\hii}{H\textsc{ii}}

\newcommand{\Msun}{\ensuremath{M_\odot}}
\newcommand{\pasp}{PASP}
\newcommand{\aj}{AJ}
\newcommand{\mnras}{MNRAS}
\newcommand{\apj}{ApJ}
\newcommand{\apjl}{ApJL}
\newcommand{\apjs}{ApJS}
\newcommand{\aap}{A\&A}
\newcommand{\araa}{ARA\&A}
\newcommand{\sdssu}{\ensuremath{u}}
\newcommand{\sdssg}{\ensuremath{g}}
\newcommand{\sdssr}{\ensuremath{r}}
\newcommand{\sdssi}{\ensuremath{i}}
\newcommand{\sdssz}{\ensuremath{z}}

\newcommand{\gone}{\ensuremath{^{0.1}{g}}}
\newcommand{\rone}{\ensuremath{^{0.1}{r}}}
\newcommand{\ione}{\ensuremath{^{0.1}{i}}}

\newcommand{\grone}{\ensuremath{^{0.1}(g-r)}}
\newcommand{\rione}{\ensuremath{^{0.1}(r-i)}}

\newcommand{\sSFR}{\ensuremath{r_{\mathrm{SFR}}}}
\newcommand{\rhoSFR}{\ensuremath{\rho_{\mathrm{SFR}}}}
\newcommand{\sfre}{\ensuremath{\mathrm{SFR}_\mathrm{e}}}
\newcommand{\sfrd}{\ensuremath{\mathrm{SFR}_\mathrm{d}}}

\begin{document}

\graphicspath{{./SFRPaper/}}

\title[The physical properties of star forming galaxies]{The physical
  properties of star forming galaxies in the low redshift universe}
\author[J. Brinchmann et
al.]{J.~Brinchmann,$^{1,2}$\thanks{\texttt{jarle@astro.up.pt}. Present
    address: Centro de Astrof{\'\i}sica da Universidade do Porto}
S.~Charlot,$^{1,3}$ S.~D.~M.~White,$^1$, C.~Tremonti,$^{4,5}$
\newauthor
G.~Kauffmann,$^1$, T.~Heckman,$^4$ J.~Brinkmann,$^6$
\\
$^1$ Max-Planck-Institut f{\"u}r Astrophysik,
Karl-Schwarzschild-Str. 1, 85740 Garching bei M{\"u}nchen \\
$^2$ Centro de Astrof{\'\i}sica da Universidade do Porto, Rua das
Estrelas - 4150-762 Porto, Portugal\\
$^3$ Institut d'Astrophysique du CNRS, 98 bis Boulevard Arago, F-75014
Paris, France\\ 
$^4$ Department of Physics and Astronomy, Johns Hopkins University,
Baltimore, MD 21218\\
$^5$ Steward Observatory, University of Arizona, Tucson, AZ 85721\\
$^6$ Apache Point Observatory PO Box 59, Sunspot, NM 88349-0059, USA}
\maketitle

\begin{abstract}
  We present a comprehensive study of the physical properties of $\sim
  10^5$ galaxies with measurable star formation in the SDSS. By
  comparing physical information extracted from the emission lines
  with continuum properties, we build up a picture of the nature of
  star-forming galaxies at $z<0.2$. We develop a method for aperture
  correction using resolved imaging and show that our method takes out
  essentially all aperture bias in the star formation rate (SFR)
  estimates, allowing an accurate estimate of the total SFRs in
  galaxies.  We determine the SFR density to be $1.915^{+0.02}_{-0.01} \mathrm{(rand.)}^{+0.14}_{-0.42} \mathrm{(sys.)}$ h$_{70} 10^{-2}$
  \Msun/yr/Mpc$^{3}$ at $z=0.1$ (for a Kroupa IMF) and we study the
  distribution of star formation as a function of various physical
  parameters.  The majority of the star formation in the low redshift
  universe takes place in moderately massive galaxies
  ($10^{10}$--$10^{11}\Msun$), typically in high surface brightness
  disk galaxies.  Roughly 15\% of all star formation takes place in
  galaxies that show some sign of an active nucleus. About 20\% occurs
  in starburst galaxies.  By focusing on the star formation rate per
  unit mass we show that the present to past-average star formation
  rate, the Scalo $b$-parameter; is almost constant over almost three
  orders of magnitude in mass, declining only at
  $\Mstar>10^{10}\Msun$. The volume averaged $b$ parameter is
  $0.408^{+0.005}_{-0.002} \mathrm{(rand.)}^{+0.029}_{-0.090} \mathrm{(sys.)} \mathrm{h}_{70}^{-1}$. We use this value constrain
  the star formation history of the universe.  For the concordance
  cosmology the present day universe is forming stars at least 1/3
  of its past average rate. For an exponentially declining cosmic star
  formation history this corresponds to a time-scale of
  $7_{-1.5}^{+0.7}$Gyr. In agreement with other work we find a
  correlation between $b$ and morphological type, as well as a tight
  correlation between the 4000\AA\ break (D4000) and $b$. We discuss
  how D4000 can be used to estimate $b$ parameters for high redshift
  galaxies.
\end{abstract}

\section{Introduction}
\label{sec:intro}

One of the most active areas of research during the last decade has
been the study of the present and past star
formation histories of galaxies. This includes detailed studies of
the stellar population of our own Galaxy
\citep[e.g.][]{2000A&A...358..869R} and 
of Local Group dwarfs, both individually and as relic evidence of the
global star formation history of the Universe
\citep{2000MNRAS.317..831H,2001ApJ...558L..31H}. While potentially
very powerful, the techniques used in these studies are difficult to
extend much beyond the Local Group.

At somewhat larger distances, efforts have gone towards understanding
the global properties of more massive galaxies and the dependence of
star formation on mass and/or morphological classification
\citep[e.g.][]{1983ApJ...272...54K,1994ApJ...435...22K,1996A&A...312L..29G,2001AJ....121..753B}.
This has been accompanied by careful mapping of both the star formation
and the gas distribution in galaxies 
\citep[e.g.][]{2001ApJ...555..301M,2002ApJ...569..157W,1995ApJS...98..219Y}. 
Based on these studies, a picture has
emerged in which more massive galaxies undergo a larger fraction of
their star formation at early times than less massive ones. This has often been 
said to present a challenge for existing
models of galaxy formation.

These studies have traditionally been carried out with relatively
small samples of local galaxies. For example, the comparatively large
sample of \citet{2002A&A...396..449G} contains 369 spiral galaxies.
With the advent of large fiber-fed surveys such as the 2dF
\citep{2001MNRAS.328.1039C} and the SDSS \citep{2000AJ....120.1579Y},
it is now possible to extend such studies dramatically in size.  Large galaxy surveys
have most recently been exploited by \citet{Hopkins-et-al-2003}, who 
carried out a comprehensive study of different              
SFR indicators, and by
\citet{Nakamura-Fukugita-2003} who studied  the local \ha\
luminosity density.

Alongside these careful studies of individual local galaxies, a global
picture of the star formation history of the universe has been
assembled through careful studies of star forming galaxies as a
function of look-back time
\citep{1988prun.proc....1C,1990ApJ...348..371S,1996ApJ...460L...1L,1996MNRAS.283.1388M,Hogg-SFR,1998ApJ...498..106M}.
The main emphasis of this work has been the star formation history of
the universe as a whole. Some workers have also tried to study the
evolution of different morphological types
\citep{1998ApJ...499..112B,1997ApJ...489..559G}.  These studies have
adopted a wide variety of star formation indicators
\citep{1999MNRAS.306..843G,2001ApJ...558...72S,2001AJ....122..288H,Glazebrook-cosmspec,2000ApJ...536...68A,1999ApJ...517..148F}.
Despite differences between different indicators it is clear that
there has been a decline in the rate of star formation since $z\sim
1$, although there is still no consensus on how rapid this decline
was. There is also no clear consensus on the star formation history
before $z=2$. Studies in the literature give answers ranging from a
clear increase back to $z=8$ \citep{2002ApJ...570..492L}, through a
constant rate of star formation back to $z=5$
\citep[e.g.][]{2002ApJ...569..582B,2003astro.ph..9065G}, to a decline
at high $z$ \citep[e.g.][]{1998ApJ...498..106M}.

One of the key unanswered questions is what physical
parameter(s) drive changes in the star formation rate in individual
galaxies. Although an increasing body of evidence points to turbulence
being a key influence on star formation on small scales \citep[e.g.][
and references therein]{2003astro.ph..1093L}, the origin of this
turbulence and its relevance for star formation on galaxy-sized scales
is unclear. It is therefore of major interest to investigate
empirically what global quantities correlate strongly with star
formation activity. It is logical, however, to expect that the star
formation rate is a function of many variables. To investigate this
dependence it is necessary to study the star formation rate density
function in various projections \citep[see also the approach
by][]{2002ApJ...570..492L}. This is also becoming a favoured
approach to analyse theoretical models
\citep[e.g.][]{2003MNRAS.339..312S,2003MNRAS.339.1057Y}.

We will pursue this goal here by trying to tie together detailed
information on individual galaxies with their contribution to the overall
cosmic star formation activity. This is possible both through studying a
snapshot in time and by comparing current and past average star
formation distributions. On a 
global scale, this provides a nearly model-independent complement to
the studies of the cosmic spectrum by \citet{2002ApJ...569..582B} and
\citet{Glazebrook-cosmspec}. 

This kind of study requires a survey that spans a large range in
galaxy properties and with accurate photometric and spectroscopic
information.  The Sloan Digital Sky Survey \citep[SDSS
][]{2000AJ....120.1579Y} is an ideal candidate for this kind of work.
Its accurate photometry allows good characterisation of the selection
function and the structural parameters of galaxies. The large
database of accurately calibrated spectra allows a much more careful
study of the spectral properties of galaxies than has hitherto been
possible. 

With such unprecedented quality of data it is pertinent to improve
modelling over what has generally been done before. One aspect in
particular that is generally missing in earlier work is a rigorous
treatment of line formation and dust. This is especially important for
the SDSS as it spans a large range in luminosity so the physical
parameters of the galaxy can be reasonably expected to vary. In this
paper we have therefore built on the work of
\citet{2002MNRAS.330..876C} and fit all strong emission lines
simultaneously to simultaneously constrain these physical properties.

Our paper is structured as follows: We start in section~\ref{sec:data}
by briefly describing the data used.  Section~\ref{sec:models} gives
an overview of the models and the modelling needed for our analysis.
This includes a discussion of the relationship between \ha-luminosity
and SFR.  Section~\ref{sec:agn_contrib} discusses in some detail the
influence of AGN activity on our estimates of star formation rates
(SFRs). Section~\ref{sec:aperture_effects} discusses the very
important question of aperture effects in a fibre survey such as the
SDSS and sets out our method for removing these.  With these
preliminaries out of the way, we discuss the local SFR density in
Section~\ref{sec:sfr_density} and give an inventory of star formation
in Section~\ref{sec:overall_prop}. The specific star formation rate
and its variation with mass is discussed in section~\ref{sec:spec_sfr}
and the variation with other physical parameters is covered in
Section~\ref{sec:b_vs_phys_param}.  The paper is concluded in
Section~\ref{sec:discussion}.  We use a cosmology with
$H_0=70$km/s/Mpc, $\Omega_M=0.3$, $\Omega_\Lambda=0.7$.  Where
appropriate we define $h=H_0/100$km/s/Mpc.  We use the
\citet{2001MNRAS.322..231K} universal IMF. To convert to the popular
choice of a \citet{1955ApJ...121..161S} IMF between 0.1 and 100\Msun,
one should multiply our SFR estimates by 1.5, the ratio of mass in the
two IMFs for the same amount of ionising radiation. We will also make
use of stellar mass estimates below and to convert these to a Salpeter
IMF it is again necessary to multiply the masses by 1.5.

\section{The data used}
\label{sec:data}

The Sloan Digital Sky Survey (SDSS) (\citet{2000AJ....120.1579Y};
\citet{2002AJ....123..485S}) is an ambitious survey to obtain
spectroscopic and photometric data across $\pi$ steradians. The survey
is conducted using a dedicated 2.5m telescope at Apache Point
Observatory.  The photometry is obtained using drift-scanning with a
unique CCD camera \citep{1998AJ....116.3040G}, allowing
near-simultaneous photometry in five bands (\sdssu, \sdssg, \sdssr,
\sdssi, \sdssz,
\cite{1996AJ....111.1748F,2001AJ....122.2129H,2002AJ....123.2121S}).
The resulting data are reduced in a dedicated photometric pipeline,
\texttt{photo} \citep{Photo-Ref} and astrometrically calibrated
\citep{2003AJ....125.1559P}. We use the results of \texttt{photo
  v5.4}. To estimate total magnitudes we use the \texttt{cmodel}
magnitudes recommended for use by the DR2 Paper \citep{DR2Paper}. This
is a weighted combination of the exponential and de Vaucouleurs fits
to the light profile provided by \texttt{photo} and has the advantage
over Petrosian magnitudes that they are higher S/N \citep[cf. the
discussion in][]{2003astro.ph..9710B}. We have also performed all
calculations with the Petrosian total magnitudes and found that the
results in the paper change if these are used although we get a few
more outliers.  Fiber magnitudes in $g,r,i$ are measured directly from
the observed spectrum and the $\sdssr$-band magnitude is normalised to
the fiber magnitude of \texttt{photo}\footnote{For those galaxies
  where we cannot measure fibre magnitudes from the spectra we adopt
  the fibre magnitudes from \texttt{photo}. This has no effect on the
  results in this paper.}.  When estimating fixed-frame magnitudes we
follow \citet{2003ApJ...592..819B} and k-correct our magnitudes to
$z=0.1$ to minimise the errors on the k-correction. We will also adopt
their notation and refer to these redshifted magnitudes as \gone,
\rone and \ione.

The spectroscopic observations are obtained using two fiber-fed
spectrographs on the same telescope with the fiber placement done
using an efficient tiling algorithm \citep{2003AJ....125.2276B}.  The
spectroscopic data are reduced in two independent pipelines
\texttt{spectro1d} (SubbaRao et al, in preparation) and
\texttt{specBS} (Schlegel et al, in preparation) --- we use the
\texttt{specBS} redshifts, but the differences between the two
pipelines are entirely negligible.

The dataset used in the present study is based on the main SDSS galaxy
sample, described by \citet{2002AJ....124.1810S}. We use a subset of
\citet{2003ApJ...592..819B}'s \emph{Sample 10} consisting of 149660
galaxies with spectroscopic observations, $14.5<\sdssr<17.77$ and
$0.005<z<0.22$. This encompasses slightly more galaxies than the SDSS
Data Release 1 \citep{2003AJ....126.2081A}. The overall continuum
properties of this sample are discussed in some detail by
\citet{2003MNRAS.341...33K,2003MNRAS.341...54K} and the sample
selection and spectrum pipeline analysis are discussed by
\citet{Tremonti-etal-2002}. The spectrophotometric calibration we use
will also be used in Data Release 2 \citep{DR2Paper}. We make
extensive use of stellar masses for the galaxies taken from
\citet{2003MNRAS.341...33K}. To be consistent with the latter we will
throughout use the narrow definition of the 4000\AA-break from
\citet{1998ApJ...504L..75B} and denote this as D4000.

As detailed in \citet{Tremonti-etal-2002} we have chosen to re-analyse
the 1D spectra using our own optimised pipeline. This allows us to take
more care in the extraction of emission line fluxes than is done in the
general purpose SDSS pipeline. The key differences for the work
presented here are as follows: we remove the smear exposures so that the
spectra are uniformly exposed within a 3'' fibre\footnote{This is the
  default procedure in DR2.} and 
we use the latest high-resolution population synthesis models by
\citet{2003MNRAS.344.1000B} to fit the continuum using a non-negative linear-least-
squares routine \citep{Lawson-Hanson-73}.  This provides an excellent
fit to the continuum and lets us perform continuum subtraction with
unprecedented accuracy. The fitting procedure automatically accounts for weak metal
absorption under the forbidden lines and for Balmer absorption 
(see e.g.\ \citet{2003MNRAS.344.1000B} for examples of the
quality of fit).

The ability to extract very weak line emission turns out to be very
important when studying trends with mass (and hence increasing
continuum contribution) and when estimating dust attenuation. We will
see below that it is also important to identify possible AGN
contamination of the emission lines.

A crucial aspect of the SDSS for studies of the nearby galaxy
population is that it has a well-defined and well-studied selection
function \citep{2002AJ....124.1810S} and covers a large range in
absolute magnitude. This means that we can use the galaxy sample not
only to study individual galaxies, but to construct distribution
functions for volume-limited samples with high accuracy. 
Note that one can usually
extract \emph{trends} from a judiciously chosen, but statistically
poorly characterised sample, but the \emph{distribution} of a
parameter requires a statistically well-defined sample. We will
therefore focus our attention on distribution functions
throughout our analysis.
 

We will limit our study to galaxies with $0.005<z<0.22$. The lower
limit reflects our wish to include galaxies of the lowest possible
luminosity. At the same time, we wish to avoid redshifts where deviations
from Hubble flow are substantial and the accessible volume is very
small.  At the low redshift limit, a galaxy with the 
mass-to-light ratio appropriate  for a 13.6 Gyr old stellar population and
$\sdssr=17.77$ would have a stellar mass just below $10^8$ solar
masses. We should therefore be complete to this mass
limit\footnote{Dust attenuation should be of minor importance in this
  estimate given the observed low attenuation in old stellar
  systems.}. 
Our sample can thus be used to reconstruct the properties of a
volume limited sample of galaxies with $\Mstar > 10^8\Msun$. As we
will see below, we also include the vast majority of all star
formation in the nearby Universe.

\subsection{Sample definitions}
\label{sec:sample_def}

\begin{figure}
  \centering
  \includegraphics[width=84mm]{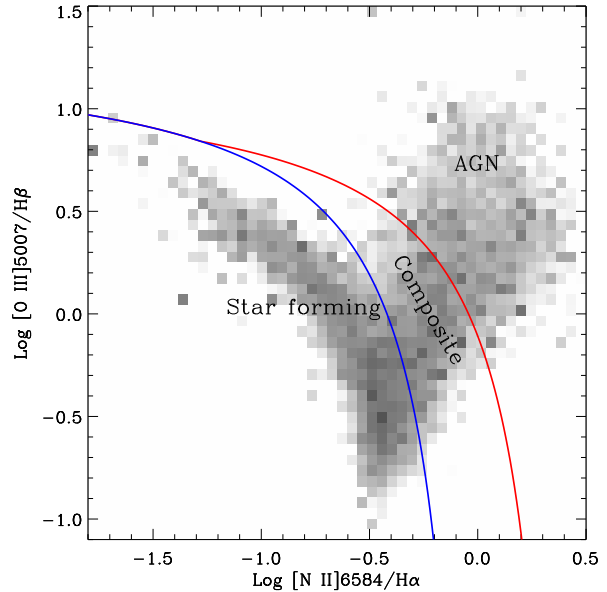}
  \caption{The distribution of the galaxies in our
    sample in the BPT line-ratio diagram. The two lines shows the
    division of our sample into the three subsamples discussed in the
    text. An unweighted version of this diagram can be seen in
    Figure~\ref{fig:agn_illustration}. The galaxies plotted here have
    S/N$>3$ in all four lines.} 
  \label{fig:subsample_illustration}
\end{figure}

\begin{table}
  \centering
  \begin{tabular}{|lrrr|}\hline
    \multicolumn{1}{|c}{Subsample} &  \multicolumn{1}{c}{Number} &
     \multicolumn{1}{c}{Percent} & 
     \multicolumn{1}{c|}{Percent of mass density} \\ \hline
     All           &  146994  &  100.0\%  &  100.0\%\\ 
     SF            &   39141  &   26.6\%  &   20.6\%\\ 
     C             &   14372  &    9.8\%  &   11.3\%\\ 
     AGN           &    8836  &    6.0\%  &    7.1\%\\ 
     Low S/N LINER &   11752  &    8.0\%  &    9.5\%\\ 
     Low S/N SF    &   29115  &   19.8\%  &   20.0\%\\ 
     UnClass       &   43778  &   29.8\%  &   31.6\% \\
 \hline
  \end{tabular}
  \caption{Basic data about the subsamples discussed in the text.}
  \label{tab:subclasses}
\end{table}

We will attempt to
treat the entire sample in a uniform way in our SFR calculations. It
will, however, prove necessary in the following discussion to define several
subsamples of objects based on their emission line properties. These
are defined on the basis of the \citet[][ BPT]{1981PASP...93....5B}
diagram shown in Figure~\ref{fig:subsample_illustration}. The diagram
has been divided into three regions which we discuss further
below. 

Although a classification can be made on the basis of
Figure~\ref{fig:subsample_illustration} regardless of the
signal-to-noise (S/N) in the lines, we find that requiring S/N$>3$ for
all lines is useful.  Below this limit, a rapidly increasing fraction
of galaxies have negative measured line fluxes and the non-symmetric
distribution of galaxies along the y-axis of the BPT diagram leads to
classification biases.  We have adopted the following subsamples (cf.\ 
Table~\ref{tab:subclasses}):
\begin{description}
\item[\textbf{All}] The set of all galaxies in the sample regardless
  of the S/N of their emission lines.
\item[\textbf{SF}] The star forming galaxies. These are the galaxies
  with S/N$>3$ in all four BPT lines that lie below our most
  conservative AGN rejection criterion.  
  As discussed in
  section~\ref{sec:agn_contrib}, they are expected to have very
  low ($<1$\%) contribution to \ha\ from AGN. 
\item[\textbf{C}] The objects with S/N$>3$ in all four BPT lines that
  are between the upper and lower lines in
  Figure~\ref{fig:subsample_illustration}. We refer to these as
  composite galaxies. Up to 40\% of their \ha-luminosity might come
  from an AGN. The lower line is taken from
  \citet{2003astro.ph..4239K} and is a shifted version of the upper
  line. It does lie very close to an empricial determination of the SF
  class (see section~\ref{sec:agn_contrib} below) but we have kept the
  Kauffmann et al line to be consistent with that work. No results
  below are affected by this choice.
\item[\textbf{AGN}] The AGN population consists of the galaxies above
  the upper line in Figure~\ref{fig:subsample_illustration}. This line
  corresponds to the theoreical upper limit for pure starburst models
  so that a substantial AGN contribution to the line fluxes is
  required to move a galaxy above this line. The line has been taken
  from Equation~5 in \citet[][]{2001ApJS..132...37K}, but our models
  have an identical upper limit.
\item[\textbf{Low S/N AGNs}] A minimum classification for AGN galaxies
  is that they have \nii/\ha $>0.6$ (and S/N$>3$ in both lines)
  \citep[e.g.][]{2003astro.ph..4239K}. It is therefore possible to
  classify these even if \oiii\ and/or \hb\ have too low S/N to be
  useful (cf.\ Figure~\ref{fig:sn_vs_mass}). A similar approach is
  taken by~\citet{2003astro.ph..7124M}. In general, we will include
  these galaxies together with the AGN class.
\item[\textbf{Low S/N SF}] After we have separated out the AGN, the composites and the
  low S/N AGNs, we have thrown out most galaxies with a possible AGN
  contribution to their spectra. The remaining galaxies with S/N$>2$
  in \ha\ are considered  low S/N
  star formers. We can still estimate the SFR of these galaxies from
  their line strengths, even though we cannot use the full modelling
  apparatus described in the following section.
\item[\textbf{Unclassifiable}] Those remaining galaxies that are
  impossible to classify using the BPT diagram. This class is mostly
  made up of galaxies with no or very weak emission lines.
\end{description}

\begin{figure*}
  \centering
  \includegraphics[height=0.9\textheight]{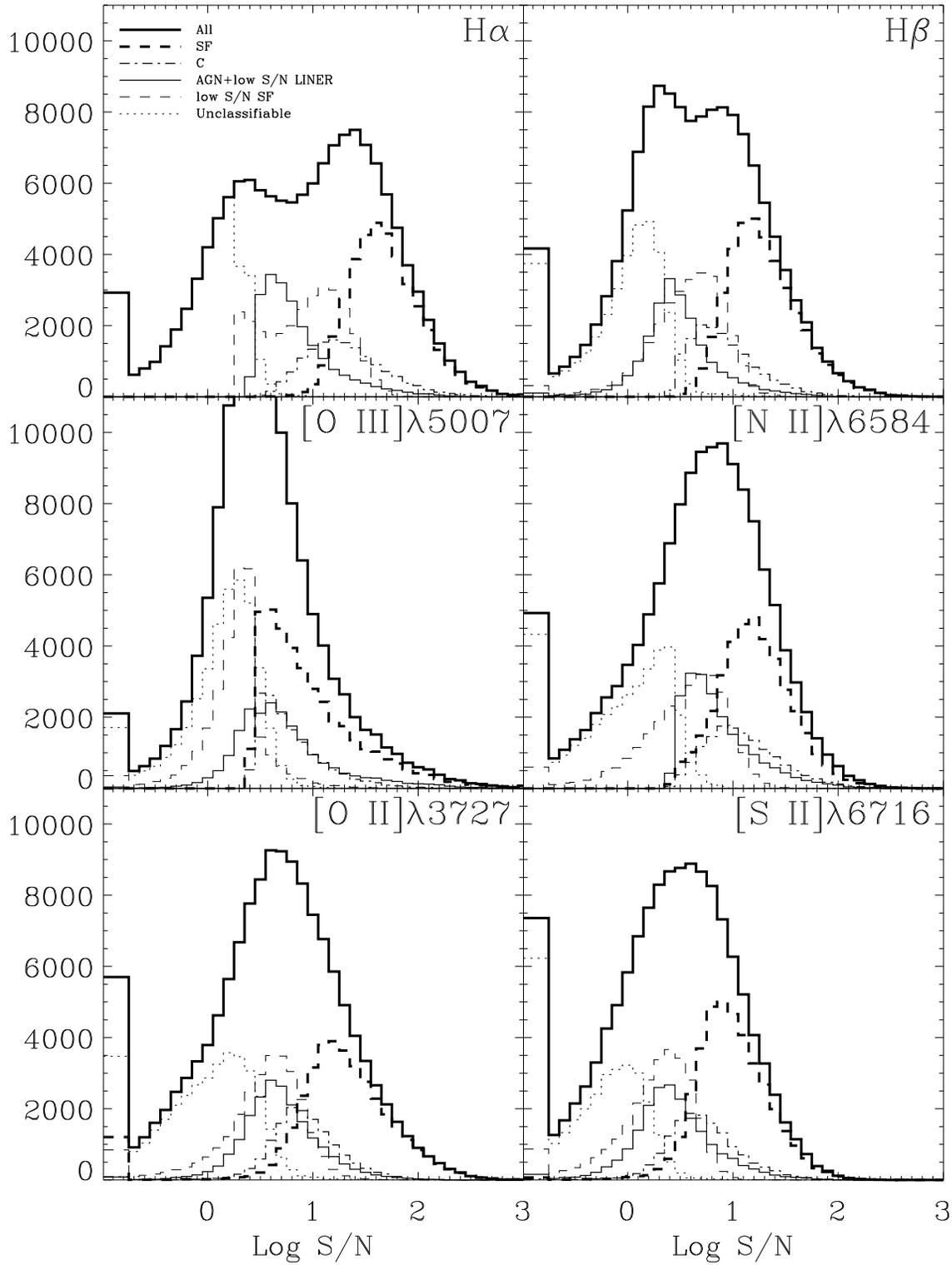}
  \caption{The distribution of signal-to-noise for the six
    lines that we include in our fits. The thick solid line shows the
    overall distribution and the distributions for the different sub
    categories are shown with different line styles as indicated in
    the legend. The objects with S/N in the emission line outside the
    displayed range are included in the extreme three bins of the
    distributions.}
  \label{fig:sn_histograms}
\end{figure*}

Note that this is merely a `'nuclear'' classification --- it does not
make any statement about the properties of the parts of the galaxy
outside the region sampled by the fibre. In particular, we expect 
the Unclassifiable category to include a substantial number of galaxies
where we only sample the central bulge --- for such galaxies there
may be considerable amounts of star formation outside the fibre. We
will return to this point when discussing aperture corrections below.

Figure~\ref{fig:sn_histograms} shows the distribution of log S/N in
each line considered for the galaxies in our sample. All galaxies
falling outside the plotted range have been included in the three bins
at the extremes of the plot. The thick solid line shows the overall
S/N distribution and the distributions for each of the different
classes defined above are shown with different line styles as
indicated by the legend in the first panel. For clarity, the low S/N
AGNs have been grouped with the AGN.

These distributions are of considerable interest to understand
possible biases in our classification scheme. The most notable results
from these panels can be summarised as follows
\begin{itemize}
\item  The SF \& C classes are limited by the strength of the
  \oiii-line (see the sharp edge for the thick dashed line at S/N=3
  in the \oiii-panel). This is because the low S/N SF\&C galaxies tend to
  be in the bottom right of the BPT diagram where \oiii$<$\hb.
\item Since the AGN class has \oiii$>$\hb\ the main limitation for the
  AGN class turns out to be the S/N in \hb. This is why using a AGN
  definition that only depends on \ha\ and \nii\ can help us identify
  a larger number of AGN.
\item There is clear evidence that there is a considerable number of
  SF galaxies that are thrown out because they have low S/N in \oiii.
  The inclusion of the low S/N SF class is therefore very beneficial.
  As Table~\ref{tab:subclasses} shows this almost doubles the number
  of galaxies for which we can use \ha\ to determine the SFR.
\end{itemize}

\begin{figure}
  \centering
  \includegraphics[angle=90,width=84mm]{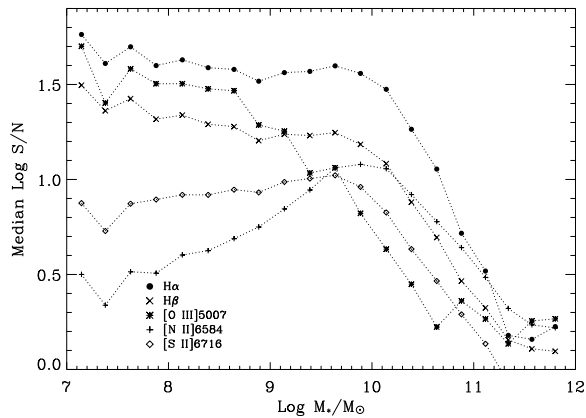}
  \caption{The median signal-to-noise ratio for  five of our lines
   that can be observed over the entire redshift range spanned by our sample. 
   Notice in particular how \oiii\ goes from being very strong
   at low masses to being the weakest line at high masses.}
  \label{fig:sn_vs_mass}
\end{figure}

Our sample covers a large range in galaxy luminosity and the
distributions shown in Figure~\ref{fig:sn_histograms} suppress this
dimension. A complementary view of these data is shown in
Figure~\ref{fig:sn_vs_mass}. This shows the trend in the median S/N of
the different lines as a function of the stellar mass of the galaxies.
We have suppressed the results for the \oii\ line as it falls out of the
sample at low redshift and also lies in a region of the spectrum that
has rather different noise characteristics to that where the other
lines reside. It shows a similar trend to to \oiii.

The most important point in this figure is that the fraction of
galaxies for which we can assign a nuclear classification is a declining
function of mass. With any fixed S/N cut, the unclassifiable
category will be biased towards more massive galaxies.

The decline in S/N in the lines with mass is seen at $\Mstar
>10^{10}\Msun$. For lower masses, we note that the S/N in \ha\ is
constant over a large range in mass. We will return to this point
later.  \oiii\ changes from being the strongest line at low mass to
being the weakest at high mass. This is caused by the decreasing
electron temperature. At high mass, most of the cooling takes place in
mid-IR fine-structure lines, depressing the \oiii-flux strongly.  At
the very highest masses the trend reverses because of the higher
fraction of AGN in the most massive galaxies
\citep[c.f.][]{2003astro.ph..4239K}. The \nii\ and \sii\ lines
increase in strength as a function of mass because of the increasing
metallicity (see the discussion in CL01). At high masses, the
\nii-line is similar in strength to the \ha-line, consistent with an
increasing fraction of LINERs at high masses
\citep[c.f.][]{2003astro.ph..4239K}.

\section{Modelling}
\label{sec:models}

Our aim is to use the data presented in the previous section to
constrain the physical properties of the galaxies in our sample. The
main emphasis in this paper is on the SFR. In the following we will
sometimes distinguish between SFRs derived directly from the emission
lines as discussed in this section and those derived indirectly from
the 4000\AA-break as discussed in
section~\ref{sec:alt_sfr_indicators}. The former we will refer to as
\sfre\ and the latter as \sfrd. In the majority of the paper we will
use \sfre\ for some galaxies and \sfrd\ for others and in that case we
will use SFR for simplicity, the details will be given towards the end
of section~\ref{sec:aperture_effects}.

We will build on the
methodology of \citet[][ hereafter C02]{2002MNRAS.330..876C} and model
the emission lines in our galaxies with the \citet[][
CL01]{2001MNRAS.323..887C} models. These combine galaxy evolution
models from \citet[][ version BC02]{1993ApJ...405..538B} with emission
line modelling from Cloudy (Ferland 2001; see CL01 for details). These
are similar in spirit to other models recently presented in the
literature
\citep{2001ApJ...556..121K,2001A&A...375..814Z,2002A&A...392..377B,2001A&A...365..347M,2003A&A...409...99P},
but they contain a physically motivated dust model and are optimised
for integrated spectra as opposed to individual HII regions.  The
modelling is discussed in more detail in \citet[][ hereafter Paper I,
see also C02]{Paper-I-SDSS}; here we will only discuss those aspects
important for our current undertaking.

\begin{table*}
  \centering
  \begin{tabular}{l|p{5cm}l}
\textbf{Parameter} & \textbf{Description} & \textbf{Range}\\
\hline
$Z$ & The metallicity & $-1 < \log Z/Z_\odot < 0.6$ in 24 steps \\
$U$ & The ionisation parameter & $-4.0 < \log U < -2.0$ in 33 steps \\
$\tau_V$ & The total dust attenuation & $0.01 < \tau_V < 4.0$ in 24
steps \\
$\xi$ & The dust-to-metal ratio & $0.1 <\xi < 0.5$ in 9 steps \\
  \end{tabular}
  \caption{The model grid calculated for the present work. This is
    calculated for a constant star formation history at $t=10^8$yrs
    --- see text for details. }  
  \label{tab:model_parameters}
\end{table*}

We have generated a model grid using the following \emph{effective}
(ie.\ galaxy-wide) parameters: metallicity, $Z$, the ionization
parameter at the edge of the Str{\"o}mgren sphere, $U$, the total dust
attenuation in the $V$-band, $\tau_V$, and the dust-to-metal ratio of
the ionized gas, $\xi$. The parameter ranges and grid sizes adopted
are given in Table~\ref{tab:model_parameters}.  Since we will only
study relative line fluxes in this paper, we are not sensitive to
stellar age, star formation history and the relative attenuation by
dust in the stellar birthclouds (i.e.\ giant molecular clouds)
and the ambient ISM.  In total our grid contains $\sim 2\times 10^5$
models.

Given the major uncertainty in traditional star formation estimates
due to dust attenuation
\citep{2001ApJ...558...72S,2001AJ....122..288H,2001ApJ...548..681B,1997AJ....113..162C},
we should briefly summarise our dust treatment which follows that of
\citet[][ hereafter CF00]{2000ApJ...539..718C}. The CF00 model
provides a consistent model for UV to far-IR emission. It does this by
incorporating a multi-component dust model where the birth clouds of young
stars have a finite lifetime. This ensures                           
a consistent picture for the attenuation of continuum and line
emission photons. In addition the model deals with angle-averaged,
effective quantities, which depend on the combined optical properties
and spatial distribution of the dust.

Each model in our grid has been calculated with a particular dust
attenuation and these attenuated line ratios are compared directly to
the observed spectrum. This means that all lines contribute to the
constraint on the dust attenuation. To first approximation, however,
our dust corrections are based on the \ha/\hb\ ratio.  Note that we do
\emph{not} interpret the derived dust attenuations as coming from a
simple foreground screen, a procedure which has been rightfully
criticised as simplistic
\citep[e.g.][]{2000ApJ...528..799W,2001ApJ...548..681B,2002MNRAS.330..621S}.

This model grid is somewhat larger than that used by C02, but the main
difference from C02 is that we adopt a Bayesian approach to calculate
the likelihood of each model given the data. This is similar to
the approach used by \citet{2003MNRAS.341...33K} in their calculation
of stellar masses. If, for notational simplicity, we enumerate the models
with a single index, $j$, we can write the likelihood of a model given
the data, $P(M_j| \left\{L_i\right\})$, as
\begin{equation}
  \label{eq:p_model}
  \ln P(M_j | \left\{L_i\right\}) = -\frac{1}{2} \sum_{i=0}^{N_{\mathrm{lines}}}
  \left(\frac{L_i - A M_{i,j}}{\sigma_i}\right)^2,
\end{equation}
where the sum is over all lines observed for that galaxy, $L_i$ is the
luminosity of line $i$, $A$ is an overall scaling factor and $M$ is
the model grid --- in practice this is a 4D array.  The prior adopted
is  given in Table~\ref{tab:model_parameters} and is
essentially a maximum ignorance, flat prior for all parameters. The
choice of prior has very little effect for the high S/N galaxies as
long as we span the range of physical parameters in the sample. However,
as we will explain below, it can make a difference for low S/N galaxies.

We fit in linear flux units rather than to flux ratios, because the
approximation of Gaussianity is more appropriate there. (By comparing
the observed ratio of the [O\textsc{iii}] lines at 4959\AA\ and
5007\AA\ with the theoretical value\footnote{The likelihood
  distribution of the ratio of two Gaussians is, of course, not Gaussian
  and we compare with the appropriate distribution for the ratio of
  two Gaussians.},  we find that the errors on our flux measurements
become non-Gaussian below S/N $\approx 2$. Since we limit our SF
sample to S/N$>3$, this is not a major concern)\footnote{Few galaxies
  have S/N $<3$ in \oii\ or \sii\ when the S/N $>3$ in the other lines
  (1--2\% for \oii\ and $\sim 4$\% for \sii). These do not affect our
  fits as the other lines dominate the fits for these objects.}.

\begin{figure*}
  \centering
  \includegraphics[angle=90,width=184mm]{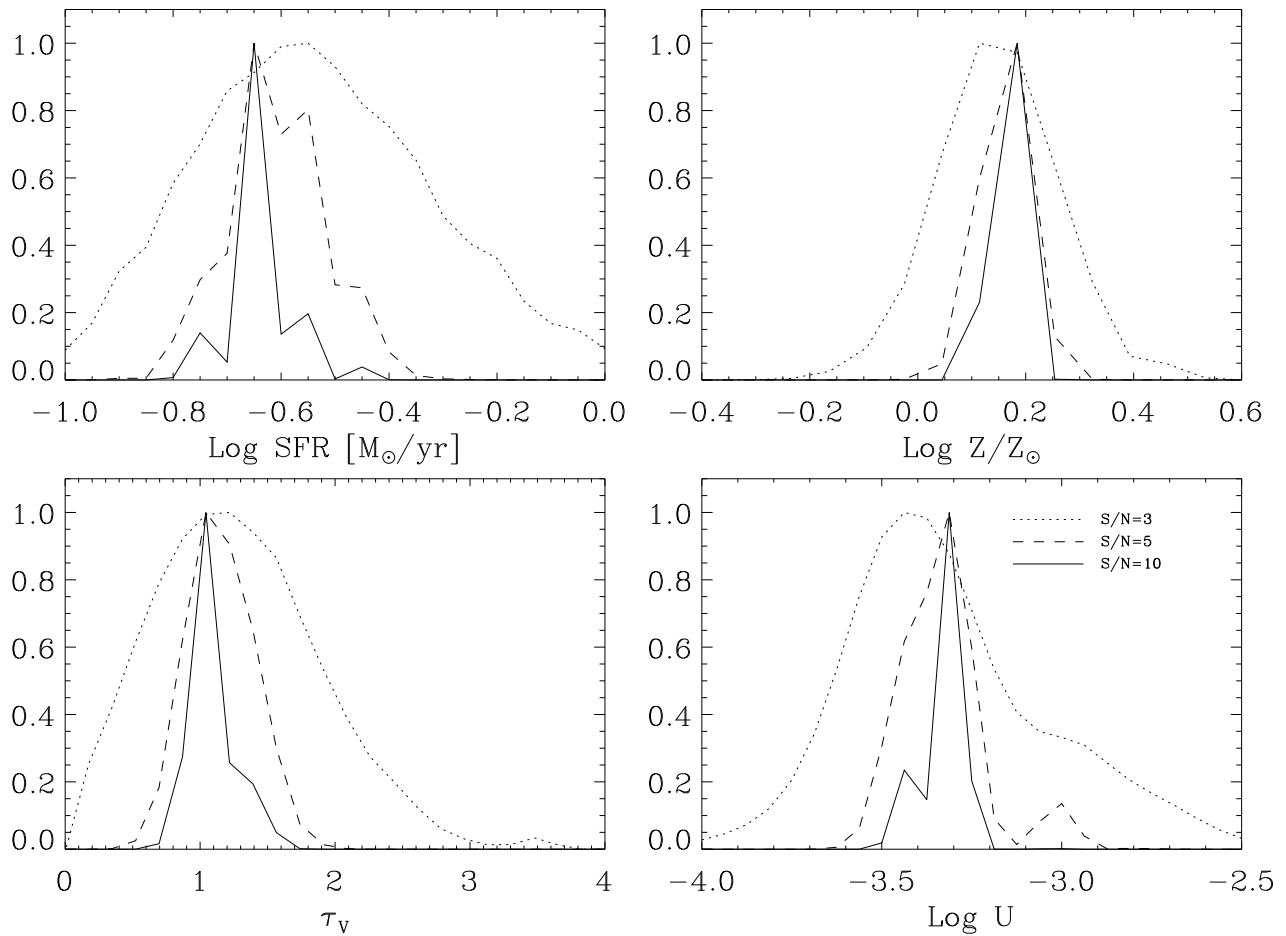}
  \caption{The likelihood distributions derived for model parameters
    and the \sfre\ as a function of the S/N of the weakest line (here
    \hb) for one high S/N galaxy. Each distribution is the average of
    100 realisations with that S/N. The top left shows the likelihood
    distribution of log \sfre\, the top right the logarithm of the gas
    metallicity in units of solar. The bottom left panel shows the
    dust attenuation in the emission lines at $V$ and finally the
    bottom right shows the logarithm of the ionization parameter at
    the edge of the Str{\"o}mgren sphere. The galaxy is SDSS
    J234342.75-001255.9 at $z=0.026$.}
  \label{fig:fit_vs_sn}
\end{figure*}

The important advantage of our approach is that it generates the full
likelihood distribution of a given parameter and from this we can rigorously
define confidence intervals for our estimates.  For high
S/N objects, these likelihood distributions are symmetric around a
best-fit value. In this situation it is possible to summarise the
distribution using the best-fit value and the spread around this
value. However, for lower S/N galaxies, the distributions may be
double-peaked or non-symmetric and it is necessary to keep the entire
likelihood distribution in further calculations. The main disadvantage
of working with full distributions is that they are somewhat
cumbersome to manipulate. The basic theory is covered in introductory
statistics books \citep[e.g.][]{Rice-1995}, but for the benefit of the
reader we have collected the results most important for the present
paper in Appendix~\ref{sec:manipulate_likelihoods}.

Figure~\ref{fig:fit_vs_sn} shows the result of taking a high S/N
spectrum and decreasing the S/N of the emission lines. Each likelihood
distribution is the result of 100 realisations (assuming Gaussian
noise on the line flux).  The figure shows S/N$=2,5$ and 15, where the
S/N is that of the weakest line of those contributing to the BPT
diagram, for this galaxy this is \hb.

\begin{figure*}
  \centering
  \includegraphics[angle=90,width=0.9\textwidth]{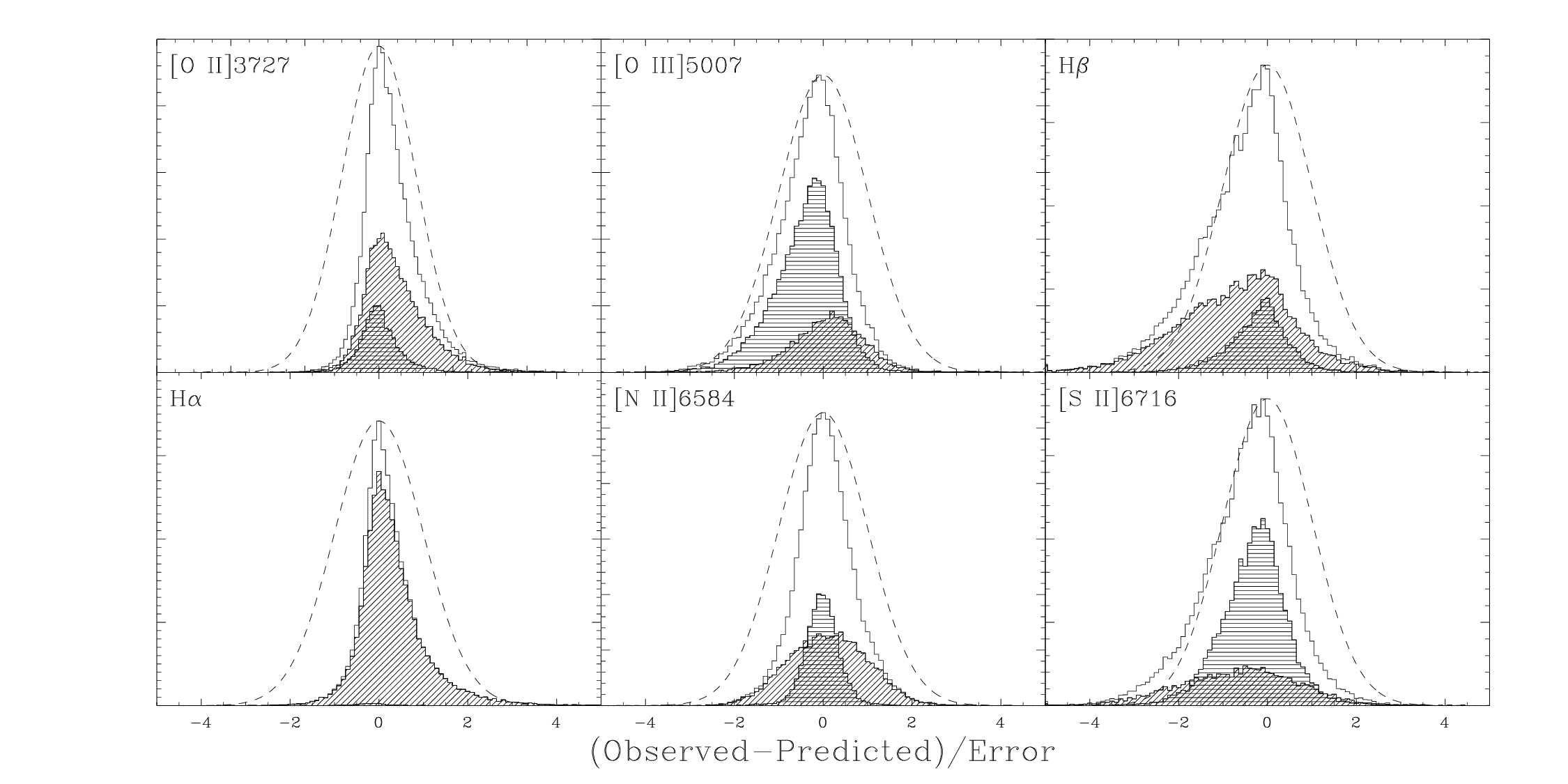}
  \caption{The distribution of the difference between the observed
    value of the line fluxes and that of the the best-fit model,
    divided by the error for all galaxies in the SF class. The smooth
    dashed curve shows a unit Gaussian for reference. The horizontally
    shaded histogram shows the results for the galaxies that have S/N
    in the line between 3 and 10. The obliquely shaded histogram is
    for galaxies with S/N$>20$ in the line.}
  \label{fig:pred_vs_obs}
\end{figure*}

The distributions broaden as expected when the S/N decreases. There is
also a slight shift in the estimates with decreasing S/N --- in
particular the estimate of the \sfre\ goes up by 0.1 dex in the mode,
somewhat more in average. Although the magnitude and sign of this
shift depends on the particulars of the line fluxes of the galaxy, it is an
indication that our apparatus does not function well at low S/N. For
this reason, and because we cannot reliably classify galaxies at low
S/N, we only use the results of the model fits for the SF class.

Figure~\ref{fig:pred_vs_obs} shows the distribution of the difference
between the observed value of the line fluxes and that of the the
best-fit model, divided by the error for all galaxies in the SF class.
The agreement is excellent except perhaps the models slightly
overpredict the \hb-flux for the highest S/N objects. This could be
caused by uncertainties in the continuum subtraction, but it has
negligible effect on our \sfre\ estimates.

This modelling approach is necessary to get accurate \sfre\ estimates
because the SDSS spans a range in galaxy properties such as Hubble
type, mass and emission-line characteristics. Previous studies of star
formation activity \citep[see][ for a review]{1998ARA&A..36..189K}
using optical spectra have usually assumed a fixed conversion factor
between \ha\ luminosity and \sfre. Dust attenuation is estimated from
\ha/\hb\ assuming a fixed unattenuated Case B ratio.  The fixed
conversion factor appears to be a reasonable approximation when
stellar absorption in the Balmer lines is taken into account
\citep[e.g.][]{2002MNRAS.330..876C}.  However, as is well known, and
shown quantitatively by CL01, it is not correct in detail.  The
approximation neglects the effect of metallicity and ionization state.
It also neglects the diffuse emission in galaxies.  More careful
modelling is required in order to avoid creating biases in the \sfre\ 
estimates as function of metallicity or stellar mass.

However, in many situations, in particular at high redshift, detailed
modelling in the way we have done is very hard to carry out. It is
therefore of considerable interest to look at the relationship between
the \ha-luminosity and \sfre\ found in our study.

\subsection{The relationship between \ha-luminosity and \sfre}
\label{sec:ha_conversion_factor}

\begin{figure}
  \centering
  \includegraphics[width=84mm]{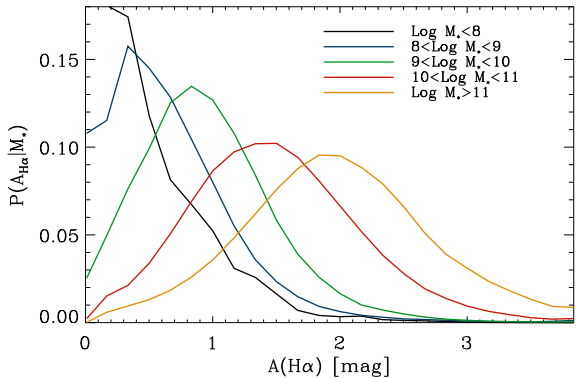}
  \caption{The likelihood distributions of the dust attenuation as a
    function of stellar mass. These are derived from the model fits to
    the SF class.}
  \label{fig:dust_likelihoods}
\end{figure}

\begin{figure}
  \centering
  \includegraphics[width=84mm]{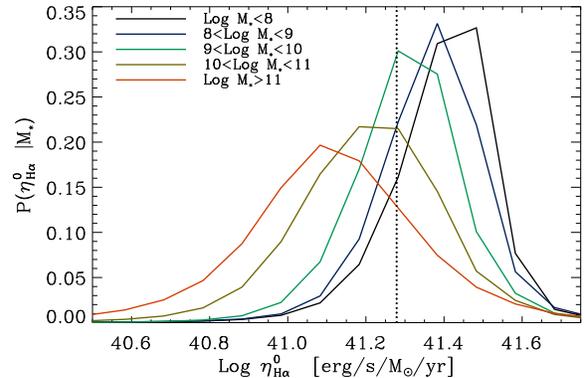}
  \caption{The likelihood distribution of the conversion factor
    between \ha-luminosity and \sfre, $\eta^0_{H\alpha}$. These were again
    derived from the model fits to the SF class.}
  \label{fig:eta_distrib}
\end{figure}

We will follow CL01 and denote the ratio of observed \ha-luminosity to
\sfre\ as, $\eta_{H\alpha}$:
\begin{equation}
  \label{eq:eta_ha_def}
  \eta_{H\alpha} = L_{H\alpha}/\mathrm{\sfre}.
\end{equation}
In this form the \ha-luminosity is assumed to be uncorrected for dust
attenuation. We now wish to study how this parameter varies with the
physical properties of the galaxy. It will prove convenient to split
this into the conversion factor between \emph{unattenuated}
\ha-luminosity and \sfre\ which we will refer to as $\eta^0_{H\alpha}$
and the dust attenuation at \ha\ in magnitudes, $A_{H\alpha}$, which
we will take as 
\begin{equation}
A_{H\alpha} \approx 0.96 \tau_V,
\label{eq:Ahalpha_def}
\end{equation}
This assumes an attenuation law of the form $\tau(\lambda) \propto
\lambda^{-0.7}$, which is a good approximation to the CF00 model over
the wavelength range we consider.

Figure~\ref{fig:dust_likelihoods} shows the likelihood distributions
for $A_{H\alpha}$ in five bins in stellar mass. These have been
constructed by co-adding the individual likelihood distributions for
$A_{H\alpha}$ for all SF galaxies in the relevant mass range. We note
that ignoring the metallicity-dependence of the Case B \ha/\hb\ ratio
would lead to an overestimate of the dust attenuation by up to $\sim
0.5$ magnitudes for the most metal rich galaxies.  The most striking
result from Figure~\ref{fig:dust_likelihoods} is the clear increase in
dust content at high stellar masses.  \citet{Tremonti-etal-2002} has
shown that there is a strong correlation between metallicity and mass
for star-forming galaxies.  One would therefore expect that the main
physical reason for the correlation between dust attenuation and
stellar mass is an increase in the metallicity of the emitting gas. It
is also interesting that the width of the A$_{H\alpha}$ distribution
increases strongly as one goes to more massive galaxies. Some of this
is due to the more massive galaxies having lower S/N emission lines
and hence more extended likelihood distributions. However, note that
we require S/N$>3$ for all four lines in the BPT diagram, so it is
unlikely that this is the main cause.

Figure~\ref{fig:eta_distrib} shows the corresponding likelihood
distributions for $\eta^0_{H\alpha}$. We have compared this to the
conversion factor advocated by \citet{1998ARA&A..36..189K}
($\eta_{\ha}^0(\mathrm{Kennicutt}) = 10^{41.28}$erg/s/\Msun/yr for our adopted
IMF). This value is indicated as the vertical dotted line in
Figure~\ref{fig:eta_distrib}. Clearly the \citet{1998ARA&A..36..189K}
conversion factor is a very good \emph{typical} value and compares
well with the median value for our sample $\log \eta^0_{\ha}=41.27$.
However, the peak values of $\log \eta^0_{H\alpha}$ do vary by nearly 0.4
dex, with the most massive/most metal rich galaxies producing less
\ha-luminosity for the same \sfre\ than low mass/metal poor galaxies.
This correlation is also likely to be driven in main by the changes in
gas-phase metallicity with mass, but as discussed by CL01  and
\citet{2002MNRAS.330..876C} a changing escape fraction for ionizing
photons is also likely to contribute to these changes in
$\eta^0_{H\alpha}$. 

\subsection{A comparison to other estimators}
\label{sec:a_comparison}

These trends with mass suggest that any simple recipes for conversion
of \ha-luminosity into \sfre\ are likely to show some biases with stellar
mass.  To quantify these possible biases, it is useful to compare our
\sfre\ estimates with those obtained from previous methods to convert
observed \ha-flux to SFRs \citep[see also][]{2002MNRAS.330..876C}. We
noted above that the \citet{1998ARA&A..36..189K} conversion factor is
a good typical value so we will use this in what follows, ie.\ 
\begin{equation}
  \label{eq:typica_conv}
  \sfre = L_{\ha}^0/10^{41.28},
\end{equation}
where $L_{\ha}^0$ is the estimated emission luminosity of \ha.
Following the discussion above we will also fix the Case B \ha/\hb\ 
ratio to be 2.86 (appropriate for an electron density of
$n=100$cm$^{-3}$ and electron temperature $T_e=10^4$K; Osterbrock
1989), where relevant below.  We have also adopted the
\citet{1979MNRAS.187P..73S} attenuation law in the following. These
choices have been made to ensure that we will use a method to estimate
SFRs which is very similar to those commonly used in the literature
\citep[e.g.][]{2000MNRAS.312..442S}.  The residual difference between
methods therefore lie in the estimates of dust attenuation and stellar
absorption under the Balmer lines. This leads to the following list of
methods of \sfre\ estimation:
\begin{enumerate}
  \renewcommand{\theenumi}{(\arabic{enumi})}
\item Assume that the observed flux can be converted to a luminosity
  without any correction for dust or continuum absorption. This
  underestimates the \sfre, but might be reasonable for strongly
  star-forming low metallicity galaxies.
\item Correct for dust attenuation by assuming a default attenuation
  of $A_V=1$mag in the emission line and correct for absorption at
  \ha\ by assuming a typical stellar absorption, \ew{\ha}, of 2\AA. If
  the assumptions are correct in the mean for the sample this is
  likely to give a good estimate of the \sfre's, but the scatter around
  the mean value might be large depending on the sample.
\item Improve on the dust corrections by using \ha/\hb\ (or some other
  Balmer line combination) to determine the dust attenuation using a
  fixed dust-free Case B reference value. We can make two assumptions
  about the absorption underlying \hb:
  \begin{enumerate}
  \item We can assume that  \ew{\ha}$=$\ew{\hb}, this is the most
    common assumption in the literature.
  \item \ew{\ha}$=0.6$\ew{\hb}. This is a better assumption for the
    BC03 models and presumably for our galaxies as well.
  \end{enumerate}
\item Finally, with good spectra such as the SDSS it is possible to
  perform a careful estimate of the absorption at \ha\ and \hb, which 
  is what we have done for the present paper. These fluxes can be used
  to determine the dust attenuation as above and thence the \sfre. 
\end{enumerate}

\begin{figure*}
  \centering
  \includegraphics[angle=90,width=184mm]{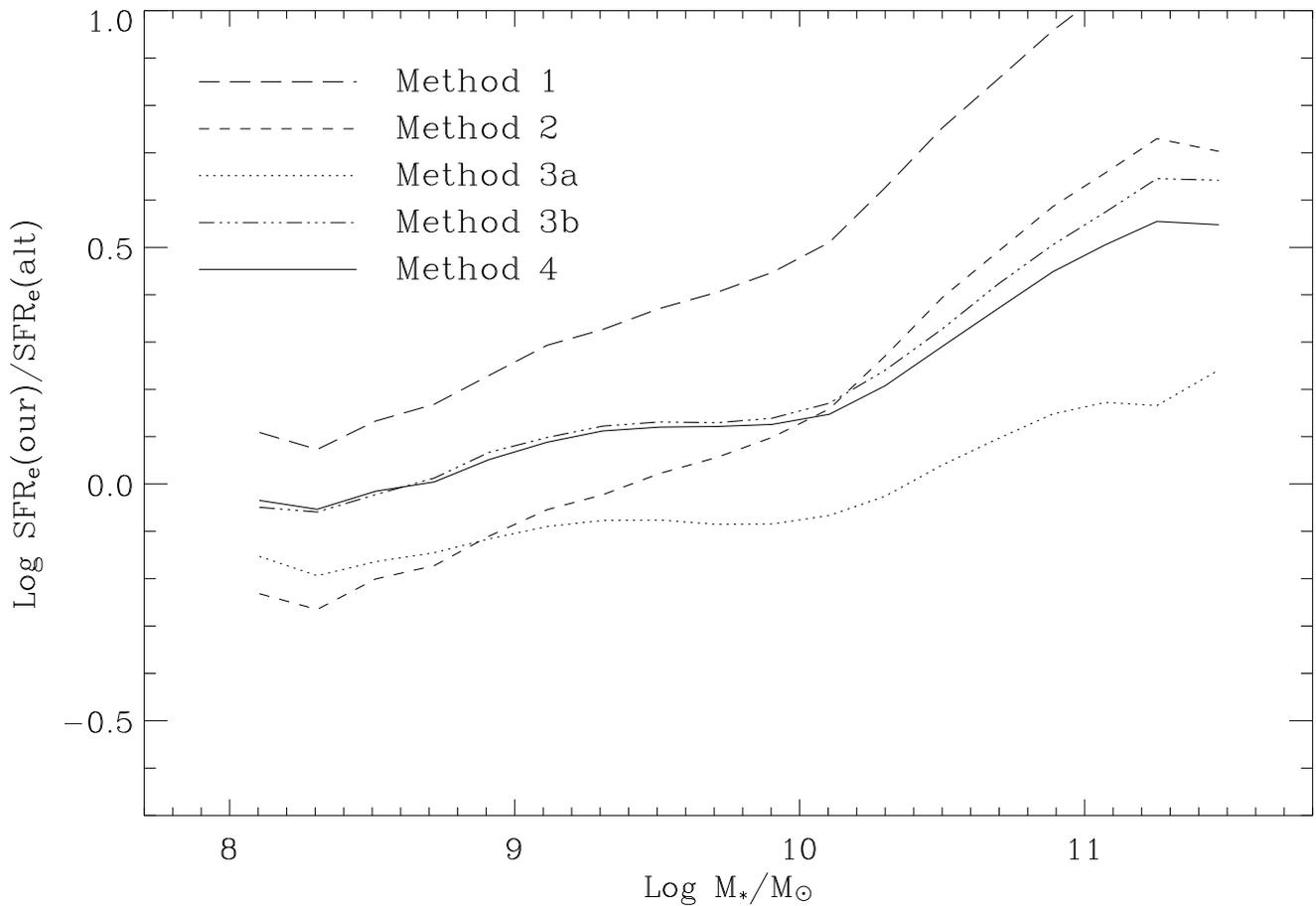}
  \caption{Our estimates compared with the simpler \sfre\
    estimates discussed in the text as a function of stellar mass. We
    have used all SF galaxies with S/N \hb\ $>10$. The figure shows
    the logarithm of the ratio of our estimate of the \sfre\ to that
    obtained by the alternative method, so a value of
    zero corresponds to equality. See text for details.}
  \label{fig:comp_sfr_estimators}
\end{figure*}

We compare these four estimates of \sfre's with our best estimates in
Figure~\ref{fig:comp_sfr_estimators}. To produce this figure we took
all SF galaxies with S/N in \hb$>10$ (to ensure that dust attenuations
from \ha/\hb\ are reliable) and compared each of the
methods 1-4 above to the mode of our likelihood distribution for the
\sfre's. The figure shows the median of the logarithm
of the ratio of the two \sfre\ estimates in bins of 0.2 dex in stellar
mass. 

It is immediately clear that methods 1, 3b and 4 all agree well with
our estimates for low mass galaxies. This is expected because the
equivalent width of stellar absorption is very small compared to that
of the emission line (the median ratio is 2--6\% at $\log \Mstar<10$)
and the dust attenuation is very low in these low metallicity
galaxies.  But it is clear that the assumption (method 3a) of equal
absorption under \ha\ and \hb\ leads to an overestimate of the dust
attenuation and hence an overestimate of the \sfre. Method 2 does
\emph{not} perform well at low masses either, but this is again
expected since an average dust attenuation of 1 magnitude at $V$ is
too high for these galaxies so again we overestimate the \sfre.

As expected method 1 starts to diverge from the other estimators when
going to more massive and dusty systems (cf.
Figure~\ref{fig:dust_likelihoods}). Likewise method 3a gives
systematically discrepant results as explain above. Method 2 is
similar to method 1 but agrees with the more sophisticated estimators
in a different mass range since the underlying assumptions are more
appropriate for more massive systems.

Finally the remaining methods, 3b and 4, agree fairly well with the
results of the full fits except at the highest masses. This is because
the assumption of a fixed $\eta^0_{\mathrm{\ha}}$ and Case B ratio for
all masses is flawed as we noted in the discussion of
Figure~\ref{fig:eta_distrib}. However it is interesting to note that
the changes in the Case B ratio and in $\eta^0_{H\alpha}$ with mass
nearly cancel when one estimates the \sfre\ over a large range in mass.
This happy coincidence comes about because the standard Case B ratio
will tend to overestimate the unattenuated \ha-flux, whereas ignoring
the variation on $\eta_{H\alpha}^0$ will require a larger
\ha-flux for the same \sfre. Thus it is a fairly good approximation to
use a fixed Case B ratio and $\eta_{H\alpha}^0$ value, but one should
never use only one and then apply the relations derived here. Also, it
is necessary to include the uncertainty in the Case B ratio when
quoting uncertainties, this uncertainty can only be accurately
assessed using the full model fits.

At the very highest masses it is however worth noting that ignoring
these factors might lead to up to a factor of 3 systematic
underestimate of the \sfre's. It is therefore important to adjust the
conversion factor and Case B ratio to that expected for the sample
galaxies in which case the simple methods will work well.

We should also note that similar trends are seen with the emission
equivalent width of \ha, with the result that method 2, 3b and 4 all
give \sfre\ estimates in good agreement with the full model fits when the
\ha\ emision line has an equivalent width $\ga 20${\AA}.

As remarked above, there are a large number of galaxies with
significant \ha\ which cannot be classified in the BPT diagram, the
low S/N SF class. We would nevertheless still like to use the \ha\ 
line to constrain the \sfre\ of these galaxies. We do this by adopting
the average likelihood distribution for the set of high S/N SF
galaxies with masses that lie within a factor of 3 of the mass of the
object in question.  The range is increased if it contains less than
50 galaxies.

\subsection{Additional uncertainties}
\label{sec:add_uncert}

There are several residual uncertainties in our determination of \sfre\
that warrant discussion.  First, we have considered only one IMF, the
\citet{2001MNRAS.322..231K} universal IMF.  The emission lines we
consider here depend on the massive end of the IMF.  As long as this
does not vary dramatically, the effect of changing the IMF is mostly
to change the overall normalisation of our \sfre's.  To correct from our
choice of IMF to the \citet{1955ApJ...121..161S} IMF between 0.1 and
100\Msun, one should multiply our \sfre\ estimates by 1.5. This is the
ratio of mass in the two IMFs for the same amount of ionising
radiation.

Second, our model assumes that essentially no ionising radiation
escapes the galaxies (see CL01). This assumption is necessary in order
to relate \ha-luminosity to \sfre\ and seems reasonable for our sample of
relatively massive galaxies.  Even local starburst galaxies have
escape fractions $<10$\% \citep[][ see also
CL01]{1995ApJ...454L..19L,2001ApJ...558...56H}.

Finally, our model uses angle-averaged, ``effective'' parameter values, whereas
any particular galaxy is only seen from one angle. For a given
galaxy, there may be a mismatch between the modelling and the
observations. We have verified that no significant systematic is
present with respect to the inclination of the galaxies, but a
residual scatter is likely to remain. This can not be quantified
without rigorous modelling of the spatially resolved
properties of galaxies.

As explained in Paper I, additional uncertainties come from the mix
of stellar populations within the fibre, from uncertainties
related to interpolation of our model grid (1--2 \%), and
from uncertainties in stellar tracks and theoretical population
modelling ($\sim 2$ \%). We include these uncertainties below by
adding a 4 \% uncertainty in quadrature.

The largest source of uncertainty in the estimates of \sfre's within in
the fibre, however, is that due to the
possible contamination of our spectra by other sources of ionising
radiation.  We now turn to a discussion of these.

\section{Other sources of emission line flux}
\label{sec:agn_contrib}

As with any other non-resolved spectroscopic survey of galaxies, the
emission line fluxes of the galaxies in our sample will contain a 
component due to sources not directly connected to star formation.
These include planetary nebulae (PNe), supernova remnants
(SNRs), the diffuse ionized gas (DIG) and active galactic nuclei
(AGN). 
The contributions of SNR, PNe and DIG are discussed in
Appendix~\ref{sec:PNe_contribution}, where we show that they are of
minor importance for our work. Here we will quantify the possible
level of AGN contribution in each of the classes defined in
section~\ref{sec:sample_def}. We will show                             
that for the Composite and AGN classes, blind
application of the modelling machinery described in the previous section is
likely to lead to incorrect results.  
Note that a comprehensive study of the AGN activity in our sample
is presented in ~\citet{2003astro.ph..4239K}.

There are at least three methods commonly employed to deal with the
presence of AGN in a galaxy survey:
\begin{enumerate}
\item Remove galaxies with AGNs by cross-correlating the sample with
published AGN catalogues 
 \citep[e.g.][]{2002AJ....124..675C,2002MNRAS.330..621S}.
\item Identify galaxies with non-stellar ionising spectra using a
  diagnostic diagram, typically the so-called BPT diagram introduced
 by \citet{1981PASP...93....5B}. This is possible so long as one can
 detect the lines required for classification (see
 section~\ref{sec:sample_def}). 
  At $z\ga 0.5$, where \ha\ is
 redshifted out of the optical this method is less accurate
  \citep[see however ][]{1997MNRAS.289..419R,1996MNRAS.281..847T}. 
\item Finally AGN can be subtracted in a statistical manner. This is
  necessary when no other methods are applicable
  \citep[e.g.][]{1998ApJ...495..691T}. The method is, of course, sensitive to
  evolution in  AGN activity with redshift and it is therefore 
  necessary to include the related uncertainties in the error budget
  for the derived SFRs
  \citep[e.g.][]{2002MNRAS.337..369T,1999ApJ...517..148F}.
\end{enumerate}
All of these methods tend to classify galaxies as either AGN or
starbursts, but as shown  by \citet{2003astro.ph..4239K}, many AGN
have young stellar populations and ongoing star formation.

\begin{figure}
  \centering
  \includegraphics[width=84mm]{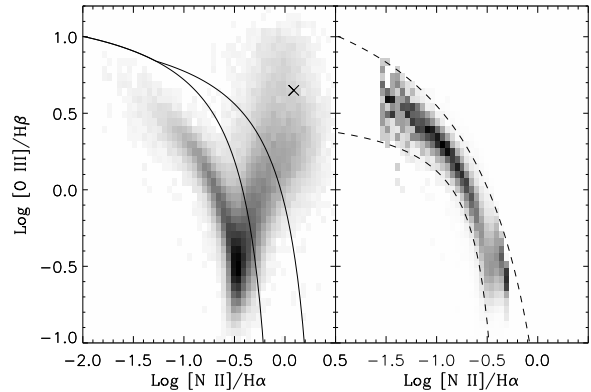}
  \caption{Left: The standard BPT excitation diagram for those
    galaxies in our sample with S/N$>3$ in all lines used here (70189
    in total).  The grayscale shows the square root of the number of
    galaxies in each bin of 0.05 in the x and y-axes. The cross shows
    the location of the average AGN used to assess the AGN
    contribution to the \ha\ flux. The lines are the same as those in
    Figure~\ref{fig:subsample_illustration}.  Right: The same plot
    after removal of the composites and AGN (leaving 42262 galaxies;
    cf. Table~\ref{tab:subclasses}) and after normalising the
    distribution in each bin along the x-axis to 1.  The thick dashed
    lines show fits to intervals containing 95\% of the galaxies as
    described in the text. }
  \label{fig:agn_illustration}
\end{figure}

Our analysis will be based on the BPT diagram.  The left panel of
Figure~\ref{fig:agn_illustration} shows the BPT diagram as a 2D
histogram for galaxies in our sample with S/N $>3$ in all lines.  The
shading is based on the square root of the number of galaxies in each
bin and the grid spacing is 0.05 along the x and y-axes. The lines
delineate our AGN, C and SF classes. The cross shows the location of
an ``average'' AGN which is constructed by taking the mean of the
luminosities of all AGN galaxies with sigma-clipping of outliers using
$\sigma=3.5$. We use this AGN below, the results are robust to the
exact average AGN used.

\citet{2001ApJS..132...37K} advocate the use of several
diagnostic diagrams to rigorously isolate AGNs. However, diagnostic
diagrams involving
\sii/\ha\ or \oi/\ha\ are much less effective at assessing the degree
to which AGN may contaminate our estimated \sfre. This is because a change
in AGN fraction
moves a galaxy almost parallel to the SF locus in these diagrams.
In the BPT diagram of Figure~\ref{fig:agn_illustration}, a change in AGN fraction 
moves the galaxy  perpendicular to
this locus.  

The upper line in Figure~\ref{fig:agn_illustration}
shows the uppermost region that can {\it in principle} be reached by 
systems ionized by normal stellar populations. However, real star
forming galaxies (and ionization-bounded \hii-regions) populate a
narrow sequence in the BPT diagram, because there are strong
correlations between ionization parameter, dust-depletion and
metallicity. A galaxy with both an AGN and ongoing star formation   
will be located to the right of this line
\citep[see also][]{2001ApJS..132...37K}.

The right-hand panel in
Figure~\ref{fig:agn_illustration} shows the BPT diagram for the
SF class, but this time as a conditional density diagram (i.e. the distribution
in each bin along the x-axis has been normalised to unity).
This distribution is very similar to that found for \hii-regions in nearby galaxies
\citep[e.g.][]{2002ApJ...572..838B,2000ApJ...537..589K}.  The dashed
lines show fits to the intervals enclosing 95\% of the
distribution. The fits use only the data with $\log \nii/\ha < -0.5$.
This ensures that the fits are the same if the AGN and C classes are
included in the distribution.

We will take the upper dashed line in the right hand panel as the
upper limit for our pure star formation sequence. For each galaxy
above this limit, we ask what AGN contribution needs to be subtracted
to move the galaxy below the limit. 
The average AGN we use  is
indicated with a cross in the left-hand panel in
Figure~\ref{fig:agn_illustration}. 
The results are fairly robust to the exact location of this average AGN
until one gets close to the AGN region.

\begin{figure}
  \centering
  \includegraphics[width=84mm]{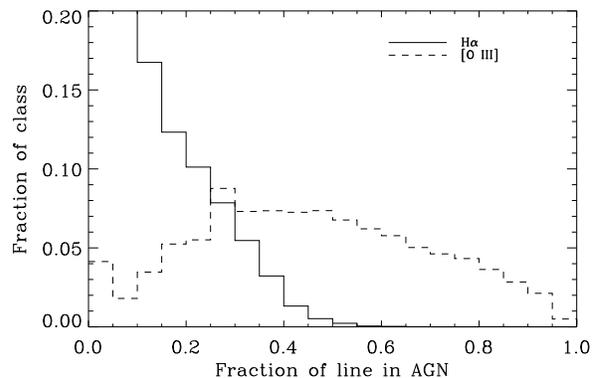}
  \caption{The fraction of \oiii\ and \ha\ luminosity estimated to come from
    the AGN in the composite galaxies (see text for details). The
    thick lines refer to \ha\ and the thin to \oiii.}
  \label{fig:ha_o3_subtraction_agn}
\end{figure}

In Figure~\ref{fig:ha_o3_subtraction_agn} we show the results of this
exercise for the composite class. The solid histogram shows the
fraction of galaxies that have a given fraction of their \ha-flux
coming from an AGN.  The dashed histogram shows the same for \oiii. At
the top of the AGN plume in Figure~\ref{fig:agn_illustration}, all the
\oiii-flux can be expected to come from an AGN, but we do not expect
more than $\sim 40$\% of the \ha-flux to have a non-stellar origin for
any galaxy\footnote{This number depends somewhat on the average AGN
  spectrum adopted.}.

In summary, we find that 11\% of the (observed) \ha-luminosity density
in the composites is likely to come from an AGN. For \oiii\ AGNs
contribute 41\% of the luminosity density in the composites.  This
means that using the \ha-luminosity for the Composite class without
any correction for AGN activity will only mildly overestimate the \sfre.

Because different lines are affected by AGN in different ways, model
fitting may give unreliable results. We therefore estimate the
in-fibre star formation in the Composite and AGN classes using their
measured D4000 values, as discussed below.

We should also comment that the pure star forming sequence differs
slightly from the definition of the SF class, but we have checked that
no more than 0.2\% of the \ha-luminosity density in the SF galaxies is
expected to be of AGN origin, which is well below our systematic
uncertainties and we will ignore this henceforth. The AGN contribution
to the \oiii-luminosity of the SF galaxies can also be shown to be
well below the level necessary to bias our model fits.

\subsection{Estimating the SFR/\Mstar\ from D4000}
\label{sec:alt_sfr_indicators}

\begin{figure}
  \centering
  \includegraphics[width=84mm]{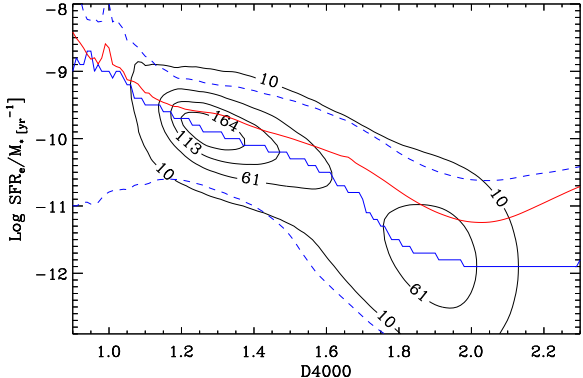}
  \caption{The observed relationship between D4000 and specific \sfre\ (both
    inside the fibre). The contour lines are
    equally spaced from 10 galaxies per bin to the maximum. The bin
    size is $0.01\times 0.1$ in the units given in the plot. The red
    line shows the average at a given D4000 and the blue
    line shows the mode of the distribution. The dashed lines show the
    limits containing 95\% of the galaxies at a given D4000.
  } 
  \label{fig:d4000_vs_specsfr}
\end{figure}

For the Composite galaxies and AGN in the sample, we cannot use the
model fits to determine the \sfre's, because the line fluxes are likely
to be affected by the AGN component. In principle it is possible to
subtract off an AGN component, but in practice this is a very
uncertain procedure. We have therefore decided to use the measured
D4000 value to estimate the SFRs and will denote this \sfrd.

This method uses the relationship between \sfre/$\Mstar$ and D4000
shown in Figure~\ref{fig:d4000_vs_specsfr} to estimate the specific
\sfrd\ of a galaxy,  and from there its total \sfrd. 
This relation shown in the figure  has been constructed using the derived
\sfre\ likelihood distributions and the  measured D4000 values
(we assume Gaussian errors on the latter). The contours
show the sum of the PDF in each bin.  The red line shows the average of
the distribution at a given D4000 and the blue line the mode. The
dashed lines show the 95\% limits at a given D4000. To derive the
likelihood distribution of the \sfrd\ of a given galaxy, we convolve the
likelihood distribution in Figure~\ref{fig:d4000_vs_specsfr} with the
likelihood distribution of D4000 for that galaxy.

To minimise the possibility of biases we did not make a S/N cut when
constructing Figure~\ref{fig:d4000_vs_specsfr}, but we did throw out
all AGN, C and low S/N AGN.  To test the reliability of this
procedure, we have checked that if we co-add the spectra of the low
S/N galaxies in the SF region of the BPT diagram and run these spectra
through our pipeline, we retrieve the same \sfre's (within a couple of
percent) that are obtained if we co-add the \sfrd's derived for individual
galaxies using the D4000 calibration outlined here.

In what follows we will use \sfrd\ for the AGN, C and Unclassifiable
classes and \sfre\ for the SF and low S/N SF classes as our best
estimate of the fibre SFR, and we will refer to this as the SFR.

\section{Aperture effects}
\label{sec:aperture_effects}

The SDSS is a fiber based survey.  At the median redshift of the
survey, the spectra only sample $\approx 1/3$ of the total galaxy
light. The presence of radial gradients in galaxy properties can
therefore lead to substantial uncertainties when one wants to correct
to total quantities. The implications of these biases have been
discussed in some detail by several authors
\citep[e.g.][]{Kochanek-Pahre-Falco-01,2003ApJ...584..210G,2002ApJ...569..582B,2003ApJ...591..827P,Nakamura-Fukugita-2003}.

We now discuss our method for aperture correction which uses the
resolved colour information available for each galaxy in the SDSS. We
will show that by using this empirically based aperture correction we
can remove the aperture bias.

Our aperture correction scheme focuses on the likelihood distribution
$P(\mbox{SFR}/L_i|\mbox{colour})$, i.e.  the likelihood
of the specific star formation rate for a given set of colours. It
is essential to keep the entire likelihood distribution in order to avoid biases
in our estimates, as the measured distributions of SFR/$L_i$ for certain colours turn
out to be multi-peaked.
We choose to normalise to the \ione-band luminosity. This is the
reddest band with uniformly small photometric uncertainties. We
have verified that the results do not depend significantly on the choice of photometric
band. 

In theory we could construct likelihoods for
SFR/$L_i$ using all the colour information available. This has the
disadvantage that even with our large sample, many bins have very few
galaxies. The method is then sensitive to outliers.
It turns out, however, that \grone\ and \rione\ contain all the useful
information necessary for our purposes. Adding other colours does not
significantly improve our constraints, this is also true if we use the
model magnitudes which give somewhat higher S/N in the \sdssu\ and
\sdssz\ magnitudes \citep{2003astro.ph..9710B}.

\begin{figure*}
  \begin{center}
    \includegraphics[angle=90,width=184mm]{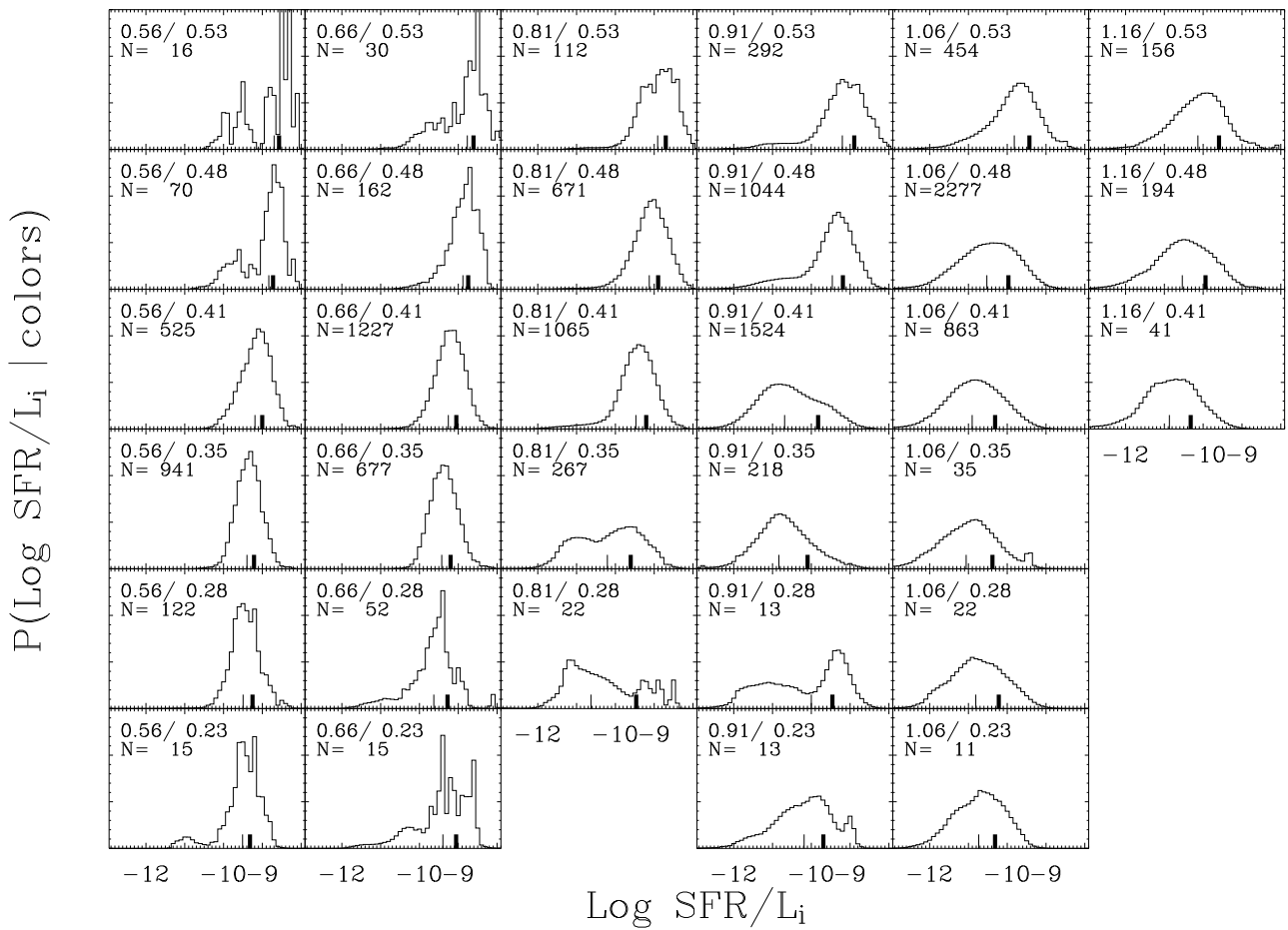}
    \caption{The likelihood distribution of log SFR$/L_i$ in bins of
      \grone\ and \rione.. The \grone\ colour increases to the right
      and is indicated as the first number in the top left corner of
      each panel. The second number gives the \rione\ colour and this
      increases upwards in the figure. The thick black mark on the
      x-axis shows the average SFR/$L_i$ whereas the thin mark shows
      the median. Note that as one moves towards redder colours the
      distributions become highly non-symmetrical and there is a
      substantial difference between the average and the median. The
      number of objects contributing to the distribution shown in a
      given panel is indicated in each panel as $\mathrm{N}=X$. Note
      also that this is a subset of the full $41\times 45$ grid.}
    \label{fig:likelihoods_sfrli_grid}
  \end{center}
\end{figure*}

We therefore construct $P(\mbox{SFR}/L_i|\mbox{colour})$ on a grid
with bins of size 0.05 in \grone\ and 0.025 in \rione. We add together
the likelihood distributions for SFR$/L_i$ (inside the fibre) for all
galaxies in each bin (we exclude AGN and low S/N AGN, including them
with their SFRs estimated from D4000 does not change the results).
The result of this procedure is illustrated in
Figure~\ref{fig:likelihoods_sfrli_grid}.  This shows
$P(\mbox{SFR}/L_i|\mbox{colour})$ for a subset of the full \grone,
\rione\ grid. The colour is indicated as \grone/\rione\ in the top
left hand corner of each panel and increases to the right for \grone\ 
and upwards for \rione. The number of galaxies in each bin is also
indicated in each panel as well as the number of galaxies that have
colours in this bin \emph{outside} the fibre (see below). The thick
and thin bars on the x-axes show the average and median SFR/$L_i$ for
that bin respectively.  Because of the non-symmetric nature of these
distributions, the average differs noticeably from the median.

The shape of the distributions varies strongly with colour and it is
often not well represented by a Gaussian. In addition, higher values
of SFR$/L_i$ often occur at \emph{redder} \rione\ colours. This is
because the emission lines around \ha\ contribute significantly to the
\ione-flux during an episode of star formation.  The likelihood
distribution also becomes very wide for galaxies with $\grone \ga
0.7$, $\rione \la 0.5$ and their star formation rates are constrained
to no better than a factor of 10.  This poor constraint on SFR/$L_i$
is simply due to the degeneracy between age, metallicity and dust.
This region of colour space is populated both by dusty star-forming
galaxies and by galaxies with old stellar populations.

We might consider improving our estimates by including the estimated
gas-phase metallicity as a constraint. We cannot implement this
procedure for the full sample, however, because gas-phase
metallicities can only be determined for the SF class.  Including the
stellar mass as an additional constraint gives little apparent
improvement.

We have also verified that k-correction uncertainties do not
appreciably affect our aperture correction estimates. We did this by
calculating $P(\mbox{SFR}/L_i|\mbox{colour})$ from galaxies with
appropriate colours in a bin of width $\Delta z=0.01$ in redshift
around the galaxy in question. This method uses only observed colours
but agrees very well with the procedure using k-corrected colours. The
disadvantage of the model is that for a subset of the data there are
not enough galaxies with the appropriate colours to construct reliable
likelihood distributions. We have therefore settled for the simplest
solution; we use only \grone, \rione\ to constrain SFR$/L_i$.

We calculate the colour of the galaxy outside the fibre by subtracting
off the fiber magnitudes from the \texttt{cmodel} (ie.\ total)
magnitudes and we convolve this with $P(\mbox{SFR}/L_i|\mbox{colour})$
to get an estimate of SFR$/L_i$ outside the fibre.  We require that at
least 5 galaxies contribute to the estimate of
$P(\mbox{SFR}/L_i|\mbox{colour})$ in a given bin.  For the bluest
colours, there are bins where no empirical estimate of
$P(\mbox{SFR}/L_i|\mbox{colour})$ exists, as few galaxies in the
survey have very blue fibre colours. For these bins we use the closest
bin. We have tested that this procedure does not introduce any
significant bias into our estimates. For a few galaxies (176) this
procedure failed because of photometric problems, these have been
excluded from the further analysis. No noticeable differences would be
found if they were included with an average aperture correction for
their mass.

\begin{figure*}
  \centering
  \includegraphics[angle=90,width=184mm]{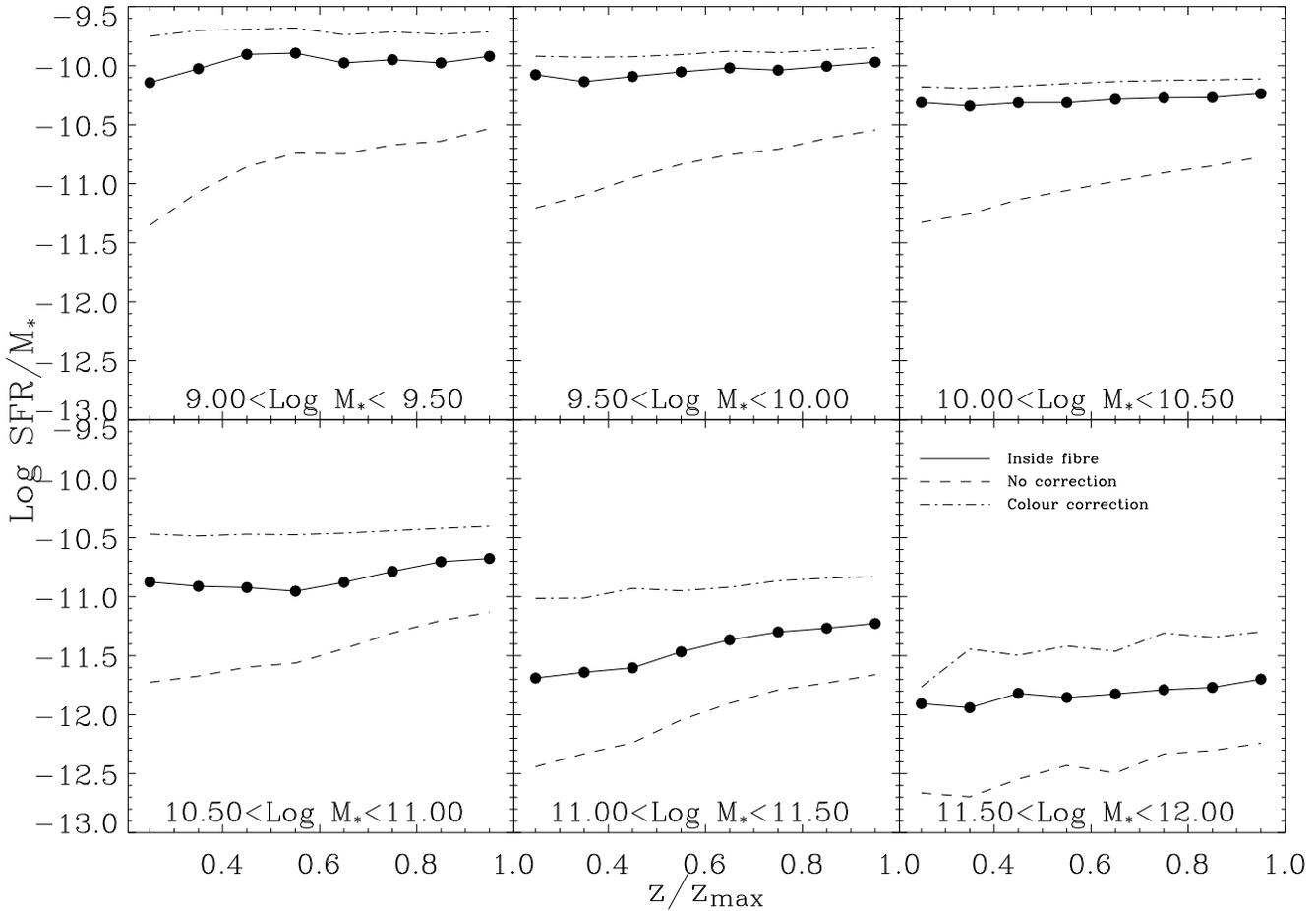}
  \caption{The value of the specific SFR, SFR/$\Mstar$ as a function
    of $z/z_{\mathrm{max}}$. The solid line with dots show the trend
    within the fibre, ie.\ both \Mstar\ and SFR calculated within the
    fibre and a clear aperture effect is seen. The dashed line shows
    the result when assuming that \emph{all} the SFR takes place
    inside the fibre, ie.\ \Mstar\ is for the galaxy as a whole and
    the SFR for the fibre only, and the dash-dotted line shows the
    result from our aperture corrections. Notice that we very nearly
    take out all aperture biases.}
  \label{fig:z_zmax_aperture_corr}
\end{figure*}

We would like to emphasise that our method of aperture correction is purely empirical.
Our main assumption is that the \emph{distribution}
of SFR/$L_i$ for given \grone, \rione\ colours is similar inside and
outside the fibre.  This assumption and the reliability of this
procedure should clearly be tested using resolved spectroscopy. We
are carrying out such a program and will report on it in an
upcoming paper.

For now, the easiest way to test our assumption is to calculate
$P(\mathrm{SFR}/L_i|\mathrm{colours})$ in bins of
$(z/z_{\mathrm{max}})^2$, where $z_{\mathrm{max}}$ is the highest
redshift at which the galaxy in question would pass the sample
selection criteria. This way we can compare $P({\rm SFR}/L_i|{\rm
  colours})$ derived from galaxies where the fibre only samples the
innermost regions with that derived from galaxies where the fibre
sample a larger fraction of the galaxy.  We have done this in 8 bins
in $(z/z_{\mathrm{max}})^2$ and compared the resulting predicted log
SFR$/L_i$ for the sample galaxies in each of these bins with that
derived from the total sample.  There is a spread, reflecting the
intrinsic scatter in the method, but for $(z/z_{\mathrm{max}})^2 \ga
0.4$ individual bins gave results in agreement with the overall
distributions. This indicates that the distribution of SFR$/L_i$ at a
given \grone, \rione\ is fairly similar at different radii as long as
we sample $\ga 20$\% of the total \sdssr-band light (the median
fraction of light within the fibre for galaxies with
$(z/z_{\mathrm{max}})^2 \ga 0.4$). This is the case for majority of
the sample.

Figure~\ref{fig:z_zmax_aperture_corr} shows that our method removes
essentially all the aperture bias from the sample. This figure shows the median
SFR$/\Mstar$ as a function of $z/z_{\mathrm{max}}$ in six different
mass bins. All galaxies have been included. 
The specific SFR is estimated from the measured value of
D4000  for the AGN and C classes (cf.\ Section~\ref{sec:alt_sfr_indicators}). The
likelihood distributions of SFR/$\Mstar$ have been co-added at each bin
in $z/z_{\mathrm{max}}$.

In the absence of aperture effects (and evolution), this plot should
show no trend with $z/z_{\mathrm{max}}$. Note that evolution is
unlikely to provide more than 0.1 dex change over the range plotted in
these diagrams.  The dashed lines in
Figure~\ref{fig:z_zmax_aperture_corr} show the trend if one assumes
that {\em all} star formation activity takes place within the fibre
but we use the stellar mass for the whole galaxy. This will clearly
underestimate the true SFR/$M_*$.
Figure~\ref{fig:z_zmax_aperture_corr} shows that this quantity
decreases towards lower values of $z/z_{\mathrm{max}}$, as expected.

The solid line with dots in the diagram shows SFR/$\Mstar$ when both
quantities are calculated inside the fibre and it is clear that there
are still strong aperture effects for galaxies with $\log \Mstar\ga
10.5$. This is expected since these galaxies often have prominent
bulges, in which the specific SFR is expected to be low. For lower
mass galaxies, the aperture problems are considerably smaller and it
is clear that a simple scaling of the fibre SFR by the \sdssr-band
flux, as done for example by \citet{Hopkins-et-al-2003} is an
acceptable aperture correction. This is also the case for galaxies
that show no sign of star formation activity when more than a third of
the total \sdssr-band light is sampled. It is not correct for other
galaxies.

Finally, the dot-dashed lines show the result after applying our aperture
corrections. Our correction has removed the aperture bias, at
least in a statistical way.
We note that  the spread in
SFR/$\Mstar$ at a given $z/z_{\mathrm{max}}$ is substantial, typically
2 orders of magnitude. We have also tested whether
the \emph{distribution} of points is the same at all $z/z_{\mathrm{max}}$
after application of our aperture corrections. We find that 
that it is, and that Figure~\ref{fig:z_zmax_aperture_corr} is an
appropriate summary of the process even though it only shows the median.

\begin{figure}
  \centering
  \includegraphics[width=84mm]{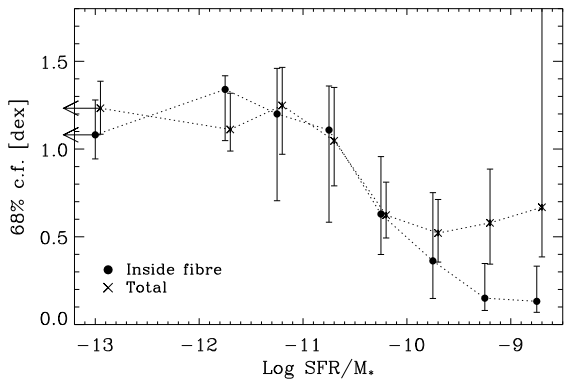}
  \caption{The distribution of the errors on the SFR estimates. The
    line with solid points show the uncertainties derived from the likelihood
    distributions inside the fibre. The line with crosses points shows the same
    for the likelihood distributions for the total SFR derived as
    described in the text. We have grouped together all galaxies with
    log SFR$/\Mstar<-12$ into one point and placed this at log
    SFR$/\Mstar=-13$. The x-axis shows log SFR$/\Mstar$,
     measured inside the fibre for the solid points and total for the
     crosses. The points for the total SFR$/\Mstar$ are offset slightly
     for clarity.}
  \label{fig:error_distributions}
\end{figure}

At this point, it is  useful
to summarise the SFR indicators that are used for each subclass. 
Inside the fibre we have the following
situation: 
\begin{enumerate}
\item For the \textbf{SF} class we use the results of the full model
  fits. 
\item For the \textbf{Low S/N SF} class we use the observed
  \ha-luminosity and convert this into a SFR using the
  $\eta_{H\alpha}$ distribution functions derived for the SF class 
  (section~\ref{sec:ha_conversion_factor}) .
\item For the \textbf{AGN}, \textbf{Composite} and
  \textbf{Unclassifiable} classes we use the D4000 value to
  estimate SFR$/\Mstar$ and SFR using  the method outlined in
  Section~\ref{sec:alt_sfr_indicators}.
\end{enumerate}
Outside the fibre, we always use the colour method  to
estimate the SFR. 

The uncertainties on our individual galaxy SFR estimates are
summarised in Figure~\ref{fig:error_distributions}. To construct this
figure we have calculated the 68\% confidence interval for the log SFR
estimate for each galaxy and calculated the spread in this value in
bins of log SFR$/\Mstar$ (where the latter is taken to be the median
of the distribution, but the results are the same with the mode or the
average).  All galaxies with log SFR$/\Mstar<-12$ have been placed in
one bin at log SFR$/\Mstar=-13$.

The solid points show the median 68\% confidence interval in dex for
the estimate of log SFR \emph{inside the fibre} for each bin in log
SFR$/\Mstar$. In the case of Gaussian errors this would correspond to
$2\sigma$. The error bars show the 68\% spread in error estimates
within that bin.  The crosses show the same for the log SFR estimated
after aperture corrections (i.e.\ total) and are slightly offset for
clarity.

Most of the trends are easy to understand. Firstly the SFR estimates
have lowest uncertainty at high SFR$/\Mstar$ because the S/N is high
in most lines and tight constraints can be obtained.  For low values
of SFR$/\Mstar$, the SFR inside the fibre is estimated from the D4000
value. This causes the error estimates to rise because D4000 provides
a poorer constraint on SFR$/\Mstar$. Finally, the error decreases when
the D4000 is so large that there is no longer any room for significant
star formation.

The uncertainties on the total log SFRs are larger than for log SFR
measured inside the fibre at high SFR$/\Mstar$ because the aperture
corrections are significantly more uncertain than the SFR estimates
from the spectra.  The spread in the uncertainty estimates is also
increasing for increasing SFR$/\Mstar$ because the outer parts of
these galaxies are very blue and few galaxies are available for the
calibration of the aperture corrections at these very blue colours
(cf.  Figure~\ref{fig:likelihoods_sfrli_grid}).  The median
uncertainties on the total SFR$/\Mstar$ are moderate, however, since
the aperture corrections are in average well constrained as
Figure~\ref{fig:likelihoods_sfrli_grid} shows.

At low values of SFR$/\Mstar$ the total and fibre estimates have
comparable uncertainties\footnote{If we had plotted the uncertainty
  on the SFR rather than log SFR this would of course not be the
  case.}. This is because the aperture corrections for these galaxies
are calibrated by the fibre estimates with similar and broad likelihood
distributions. The total log SFR therefore gets a very similar
likelihood distribution. Some galaxies with poorly constrained log SFR
in the fibre might nevertheless get better constrained log SFR total
because they have bluer outsides than insides and hence better
constrained aperture corrections.

The best test of our error estimates for the fibre SFRs is repeat
observations.  Approximately $5000$ of our targets have been observed
at least twice during the SDSS survey. We have used these duplicate
observations to check that the errors we derive for the fiber SFRs are
consistent with the dispersion between repeat observations.

\section{The local SFR density}
\label{sec:sfr_density}

\begin{table*}
  \caption{The best estimate of the star formation density at $z=0.1$
    from our survey. We show the mode and the 68\% confidence
    intervals as well as the percentage contribution to 
   the total for the different classes. The total for $\beta=0$ (no
   evolution) is shown on the first row, the others use
   $\beta=3$.}
  \begin{tabular}{|r|rrr|rrr|r|}\hline
    \multicolumn{1}{|c|}{Sample} & \multicolumn{3}{c}{Fibre} & 
    \multicolumn{3}{c|}{Total} & \multicolumn{1}{c|}{Percent}\\ 
    & \multicolumn{6}{c}{$\rhoSFR$ [$10^{-2}$ \Msun/yr/Mpc$^3$]} &
    \\ 
    & 16\% & Mode & 84\%  & 16\% & Mode & 84\%
    &  \multicolumn{1}{c|}{of total} \\ 
\hline
   Total ($\beta=0$) & 0.490 & 0.496 & 0.509 & 1.840 & 1.855 & 1.870 &   \ldots \\
   Total ($\beta=3$) & 0.507 & 0.509 & 0.528 & 1.962 & 1.974 & 1.998 & 100.00\% \\
                  SF & 0.317 & 0.320 & 0.324 & 1.001 & 1.018 & 1.024 &  51.58\% \\
          Low S/N SF & 0.063 & 0.064 & 0.065 & 0.445 & 0.448 & 0.464 &  22.69\% \\
          Composites & 0.078 & 0.080 & 0.092 & 0.203 & 0.207 & 0.210 &  10.48\% \\
                 AGN & 0.026 & 0.026 & 0.028 & 0.082 & 0.084 & 0.086 &   4.26\% \\
      Unclassifiable & 0.019 & 0.019 & 0.020 & 0.213 & 0.216 & 0.228 &  10.96\% \\
    \hline
  \end{tabular}
  \label{tab:SFR_density}
\end{table*}

We now turn to estimating the total SFR density of the local universe.
As discussed above, we have a likelihood distribution for the SFR for each
galaxy. This includes both uncertainties in the model fit and the
aperture correction. We also need to make a correction for
evolution in the SFR over the 2.5 Gyr spanned by our redshift limits.
We do this by assuming that the SFRs of all galaxies evolve as:
 \begin{equation}
  \label{eq:rho_sfr_evol}
  \mathrm{SFR}(z) \propto (1+z)^\beta.
\end{equation}
For the \textit{integrated} SFR density, the favoured evolution
exponent up to $z\sim 1$ is currently $\beta \sim 3$
\citep{Glazebrook-cosmspec,Hogg-SFR}, although substantial uncertainty
exists. We will therefore adopt this form for individual galaxies as
well. We will calculate our results at $z=0.1$.  The differences
between $\beta=0$ and $\beta=3$ are then fairly small ($\sim$ 6\% ,
see Table 3 below).  Note that a \emph{single} evolutionary correction
for \emph{individual} galaxies is likely to be wrong. As we will see
below, there are clear differences in the star formation histories of
galaxies of different mass, so although the \emph{integrated} star
formation density evolves as a power-law, the individual SFRs cannot
show the same evolution.  Fortunately, most of the results in the
following sections are not dependent on the evolution within the
sample.

Our results for the total SFR density are summarised in
Table~\ref{tab:SFR_density}. This table gives the mode of the
likelihood distribution of $\rhoSFR$ as well as the 68\% confidence
intervals. We show the result for $\beta=0$ on the top line and the
following lines all assume $\beta=3$ --- the relative contributions of
the different classes do not depend on $\beta$. To estimate the star
formation density of the survey, we add up the likelihood
distributions, using $1/V_{\rm max}$ as a weight
\citep{1976ApJ...207..700F}.  As discussed in
Appendix~\ref{sec:manipulate_likelihoods}, the summation is carried
out with a Monte Carlo method. To calculate errors on the final SFR
density, we carry out 100 bootstrap repetitions. For each repetition
we draw 149660 objects from our sample with replacement which makes up
the bootstrap sample. For this sample we carry out 30 Monte Carlo
summations as discussed in Appendix~\ref{sec:manipulate_likelihoods}
--- the results do not change when these numbers are increased. We
then calculate the mean, mode, median and confidence intervals from
the distribution of results obtained for each bootstrap repetition and
Monte Carlo summation. The confidence intervals reflect only random
errors.  The systematic uncertainties are substantially larger.

It is, of course, difficult to estimate accurately the systematic
uncertainties. The first is the uncertainty in the estimator used to
calculate the SFR inside the fibre for the Composite, AGN and
Unclassifiable class. We have experimented with various estimators
before settling on D4000. They all agree fairly well with a
spread of 2\% which has to be taken as a systematic uncertainty.

Our SFR density estimate is sensitive to our aperture
correction scheme. We remarked above that our main assumption is that
the distribution of log SFR$/L_i$ is the same at a given colour inside
the fibre as outside. This is not necessarily so, something which is
hinted at in the clear difference between the mode, median and average
for the likelihood distributions. These indicate that the likelihood
distribution for a given colour might be composed of at least two
components and the balance between these could be different in the
outer regions of the galaxy, cf.\ the double-peaked nature of some
distributions in Figure~\ref{fig:likelihoods_sfrli_grid}.  It is
therefore possible to take the difference between the average (the
formally correct estimate to use) and the mode as indicating the
importance of this assumption. We have therefore carried out the aperture
corrections also using the mode as an estimator of log SFR$/L_i$. In
this case we find a SFR density 15\% lower than that in
Table~\ref{tab:SFR_density}.

In addition there is a 6\% difference between $\beta=0$ and
$\beta=3$ which we take as an additional systematic uncertainty of
$\pm 3$\%. Finally differences between the linear interpolation scheme
we use to draw numbers from the likelihood distribution and higher
order schemes also produce a scatter of 1--2\% percent in the
calculation of the average. We will therefore adopt $^{+6}_{-21}$\% as
indicative of the expected range of the systematic uncertainties.

This gives our best estimate for the total SFR density at $z=0.1$ as 
\begin{equation}
  \label{eq:total_SFR_density}
  \rhoSFR = 1.915^{+0.02}_{-0.01} \mathrm{(rand.)}^{+0.14}_{-0.42} \mathrm{(sys.)},
\end{equation}
in units of $10^{-2}$ h$_{70}$ \Msun/yr/Mpc$^{3}$. The random errors
correspond to the 68\% confidence interval, while the larger systematic
error is the expected total range. We have here taken the average of
the $\beta=3$ and $\beta=0$ estimates since this difference is
included in the systematic uncertainties.

\begin{figure}
  \centering
  \includegraphics[width=84mm]{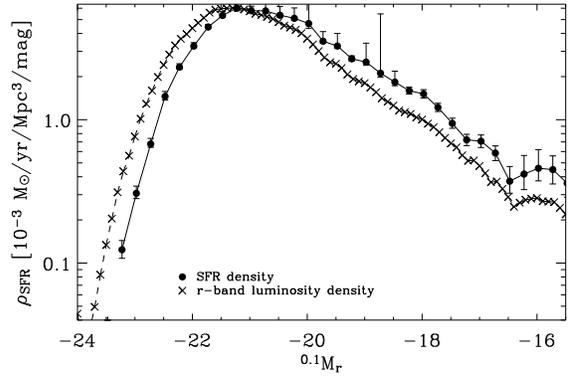}
  \caption{The star formation density as a function of \sdssr-band
    absolute magnitude compared with the \sdssr-band luminosity
    density scaled to have the same peak value. The error bars on the
    SFR density are derived from bootstrap resampling. The error-bars
    on the luminosity density estimate are suppressed for clarity. }
  \label{fig:sfr_density}
\end{figure}

We have calculated the SFR density in bins of absolute \rone-band
magnitude and this is plotted as solid circles in
Figure~\ref{fig:sfr_density}. For comparison, we plot the $r$-band
luminosity density for our sample \citep[see ][ for a more in-depth
discussion]{2003ApJ...592..819B} scaled to the same peak value as the SFR
density (crosses).  If we fit both distributions with a
\citep{1976ApJ...203..297S} function of the form
\begin{equation}
  \label{eq:schechter_LF}
  \phi(L) dL = \phi_0 \left(\frac{L}{L_*}\right)^\alpha_{\mathrm{FS}} e^{-L/L_*} \frac{dL}{L_*}.
\end{equation}
we find that the faint-end slopes, $\alpha_{\mathrm{FS}}$, are the same for the two
functions, but that $\L_*$ is 0.27 magnitudes fainter for the
star formation density. The error bars on the SFR density are derived
from bootstrap resampling of the sample and do not include systematic
errors. These will to good approximation only affect the overall
normalisation, and not the shape.

The curves are very similar in shape. At faint magnitudes where the
majority of galaxies are strongly star-forming the \sdssr-band light
is a good tracer of the SFR activity, but low luminosity galaxies are
somewhat more important for the overall SFR density than for the
\sdssr-band luminosity density. At the bright end of the distribution,
there is less star formation per unit \sdssr-band luminosity.  We will
see later that this is even more prominent when plotted against the
stellar mass.

Very little of the total SFR density occurs outside our sample
\citep[see also][ their Figure 10]{Glazebrook-cosmspec}.  When
integrating up the Schechter function, we find that only 1--2\% of the
total SFR density originates outside our selection limits. We have
therefore decided \emph{not} to correct our values for incompleteness.

\begin{figure}
  \centering
  \includegraphics[angle=90,width=84mm]{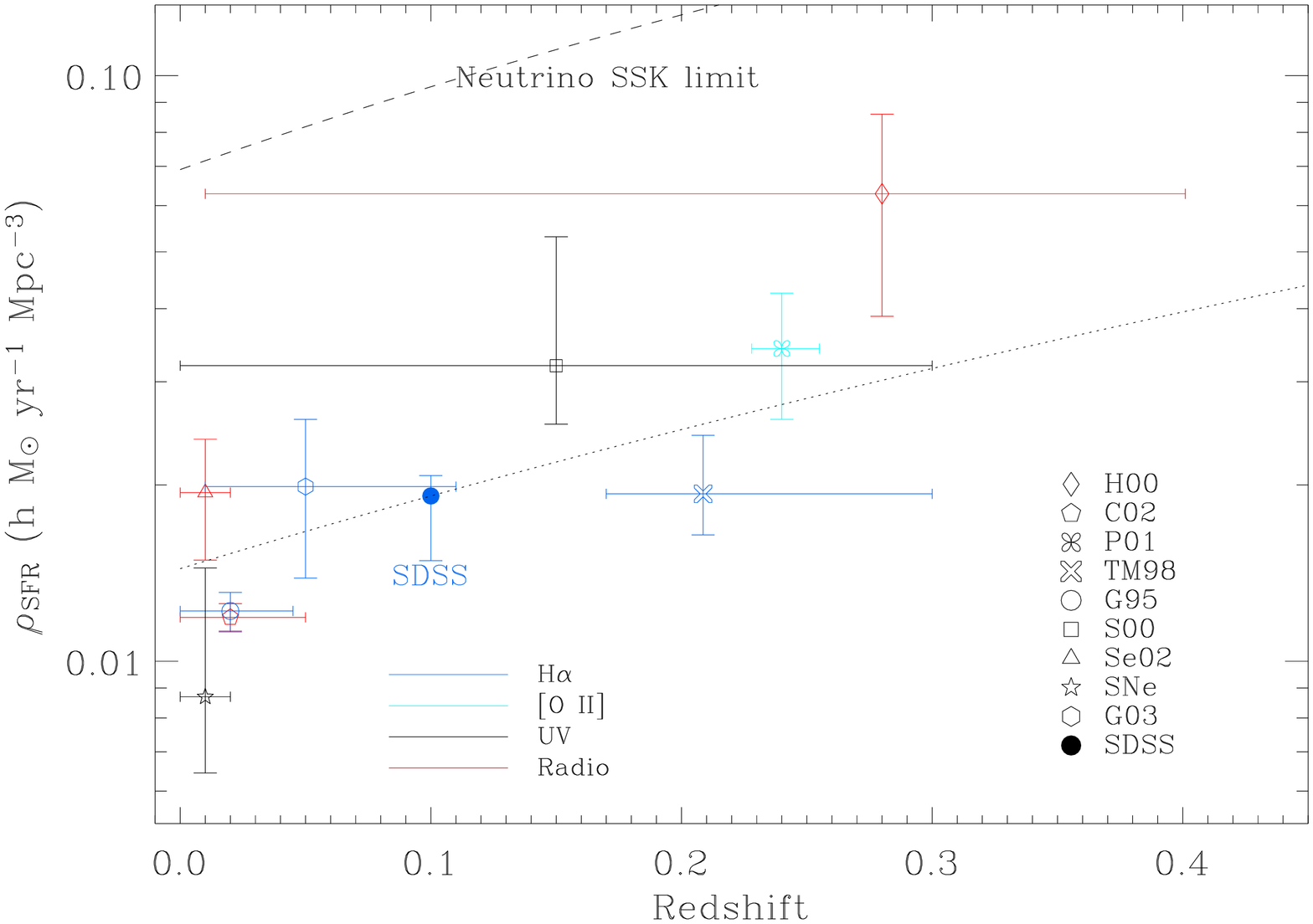}
  \caption{The star formation density in the present study. The filled
    circle is our preferred estimate with the systematic and random
    errors included in the error bar. This is compared with several
    other recent determinations of the local SFR. The papers indicated
    in the legend are: H00: \citet{2000ApJ...544..641H}, C02:
    \citet{2002AJ....124..675C}, P01: \citet{2001A&A...379..798P},
    TM98: \citet{1998ApJ...495..691T}, G95:
    \citet{1995ApJ...455L...1G}, S00: \citet{2000MNRAS.312..442S},
    Se02: \citet{2002MNRAS.330..621S}, SNe: SFR derived from SNe are
    as reported by \citet{2003MNRAS.340L...7F} who are also
    responsible for the upper limits from neutrino observations at
    SuperKamiokande, G03: \citet{Glazebrook-cosmspec} for whom we show
    the full range of the \ha-derived local SFR. The dotted line shows
    a $(1+z)^3$ evolution of $\rho_{\mathrm{SFR}}$ for
    comparison. All values have been corrected to the
    \citet{2001MNRAS.322..231K} universal IMF and $h=0.7$.}
  \label{fig:low_z_sfr_density}
\end{figure}

Figure~\ref{fig:low_z_sfr_density} places our estimate together with
other recent estimates from the literature. In order to make this
comparison we have adjusted the SFR density reported using the
\citet{1955ApJ...121..161S} IMF with a lower mass limit of 0.1\Msun\ 
and an upper mass limit of 100\Msun\ to our IMF by dividing all values
by 1.515. All values have been corrected to $h=0.7$, but no attempt
has been made to adjust the values for small differences in the
assumed cosmological model.

We have indicated different methods of estimating the SFR  using
different colours in the figure. It is clear that radio
determinations
\citep{2000ApJ...544..641H,2002AJ....124..675C,2002MNRAS.330..621S}
typically give  the highest estimates, although alternative 
calibrations might bring these down somewhat
\citep{Hopkins-et-al-2003}. 

Our value seems to be in good agreement with other \ha-based estimates
\citep[][]{1995ApJ...455L...1G,1998ApJ...495..691T,Glazebrook-cosmspec}.
In particular our value is in very good agreement with that derived
for the SDSS by \citet{Glazebrook-cosmspec} in their analysis of the
cosmic spectrum. We show the full range of their \ha-based estimate
for the local SFR in Figure~\ref{fig:low_z_sfr_density} and it is
clear we are in very good agreement despite the different approaches
taken.  It might seem surprising that our error-bar is no smaller than
many other determinations, which have used samples that are two orders
of magnitude smaller than ours. This is mostly due to the adoption of
a rather conservative systematic error for the aperture corrections.

Finally, to close off this section, we comment on
the contributions to the overall $\rhoSFR$ from the different classes
in Table~\ref{tab:SFR_density}. It is noteworthy that 
despite only contributing about 23\% of
the total stellar mass in the local universe, the SF class is the
dominant class in terms of SFR density. Over 50\% of
$\rhoSFR$ comes from this class. Likewise it is interesting that 
up to 15\% of the total SFR takes place in galaxies that show signs of
AGN activity \citep[cf.][]{2003astro.ph..4239K}.

\section{The overall properties of local star forming galaxies}
\label{sec:overall_prop}
\begin{figure}
  \centering
  \includegraphics[width=84mm]{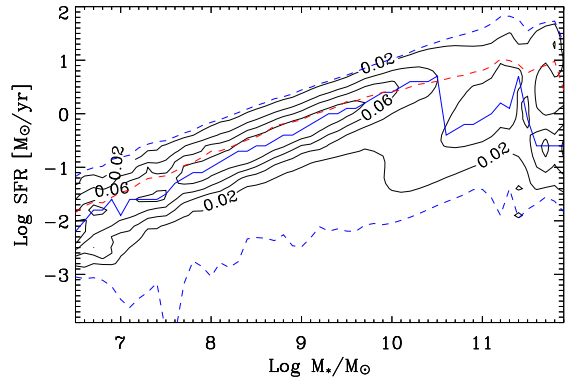}
  \caption{The relationship between the stellar mass and the SFR (both
    inside the fibre) for all galaxies with no AGN contribution. The
    figure has been volume weighted and normalised in bins of stellar
    mass. The contours are therefore showing the conditional likelihood
    of SFR given a stellar mass. The bin size is $0.1\times 0.1$ in
    the units given in the plot. The red line shows the average at a
    given stellar mass, whereas the blue line shows the mode of the
    distribution. The dashed lines show the limits containing 95\% of
    the galaxies at a given stellar mass.}
  \label{fig:mstar_vs_sfr}
\end{figure}

\newcommand{\rhoncolor}{black}

We now study the dependence of the SFRs on other physical parameters
of the galaxies. Most of the results we show below will be
familiar, as the general trends have been known for a long time. The
main importance of what follows is that for the first time it is
possible to derive the full distribution functions for these
quantities. This adds considerable quantitative information to what
was known before. One important point is that the results
shown in this  section are only very weakly dependent on the 
systematic uncertainties discussed previously.          
We will therefore ignore these in what follows.

We start with Figure~\ref{fig:mstar_vs_sfr}, which shows the SFR
distribution as a function of stellar mass for the SF, low S/N SF and
unclassifiable classes. The figure has been volume weigthed and
normalised in bins of stellar mass so it shows the conditional
probability of SFR given a stellar mass. The clear correlation between
SFR and stellar mass over a significant range in $\log \Mstar$ is
noticeable and will be a recurring theme in what follows. Note that
that at $\log \Mstar/M_{\odot} \ga 10$, the distribution of SFRs
broadens significantly and the correlation between stellar mass and
SFR breaks down.

\begin{figure*}
  \centering
  \includegraphics[angle=90,width=184mm]{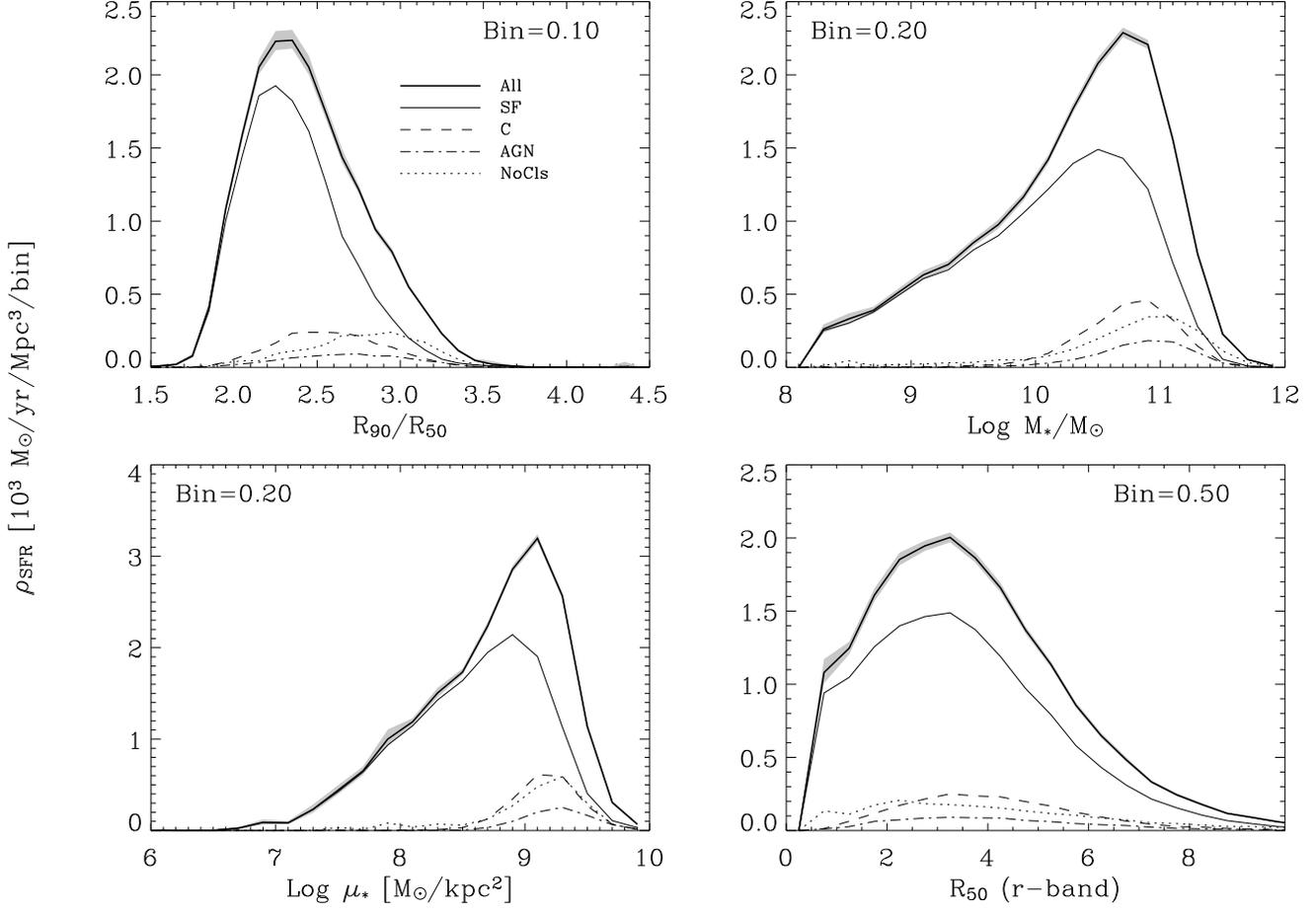}
  \caption{The total star formation density, $\rhoSFR$, as a function of
    \emph{Top left:} The central concentration of the galaxies. As
    discussed in section~\ref{sec:overall_prop} this is a proxy for a
    proper morphological classification. \emph{Top right: } The
    stellar mass and \emph{Bottom left:} The stellar surface mass
    density. The final panel shows $\rhoSFR$ as a function of the
    galaxy half-light radius. The shaded region in each panel shows
    the 68\% confidence region. The units are given on the left, and
    the bin size is given in each panel.}
  \label{fig:sfr_dens_fig1}
\end{figure*}

\begin{figure*}
  \centering
  \includegraphics[angle=90,width=184mm]{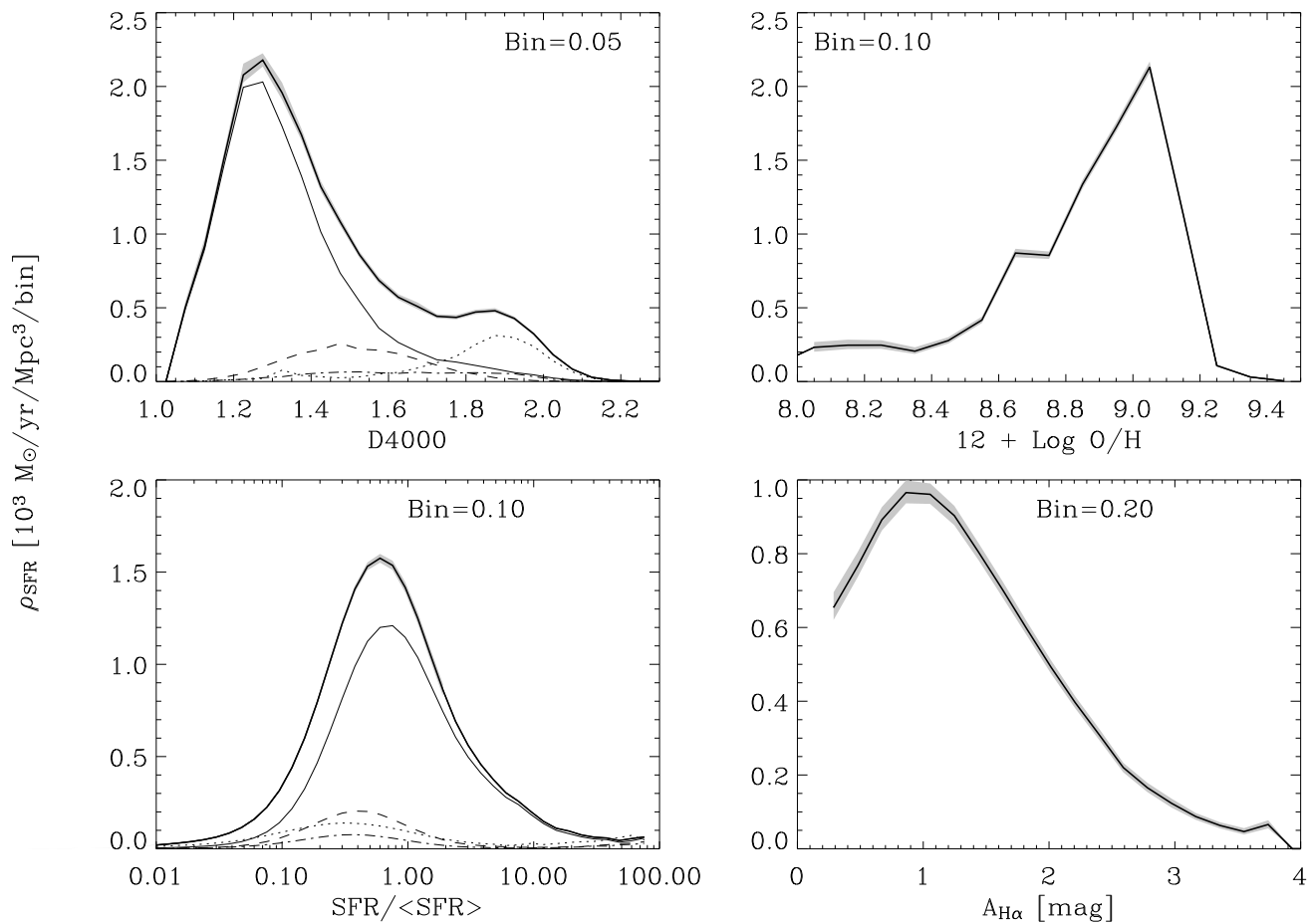}
  \caption{Like the previous figure but this time we show parameters
    determined inside the fibre along the $x$-axis and the
    distribution functions are therefore relative to the SFR density
    \emph{inside} the fibre. The top left shows the distribution with
    respect to the 4000\AA\ break, the gas-phase oxygen abundance is
    at the top right, the present-to-past average star formation rate (bottom
    left) and finally the dust attenuation at \ha\ in the bottom right
    panel.}
  \label{fig:sfr_dens_fig2}
\end{figure*}

In addition to the correlations between different galaxy parameters,
it is instructive to study 1-dimensional projections of our
multi-dimensional distribution functions. This enables us to address
the question of {\em how much} of the total SFR density takes place in
different kinds of galaxies.  To do this we carry out the same Monte
Carlo summation that we used to derive the total SFR density, but this
time in bins of mass, concentration, size, surface density etc.  The
results of this exercise are shown in Figures~\ref{fig:sfr_dens_fig1}
and~\ref{fig:sfr_dens_fig2}. These show the star formation density per
bin for a set of 8 different physical quantities. The bin size is
indicated in each panel. The shaded gray area shows the 68\%
(1$\sigma$) confidence interval determined from the Monte Carlo and
bootstrap summations.

\begin{figure*}
  \centering
  \includegraphics[angle=90,width=184mm]{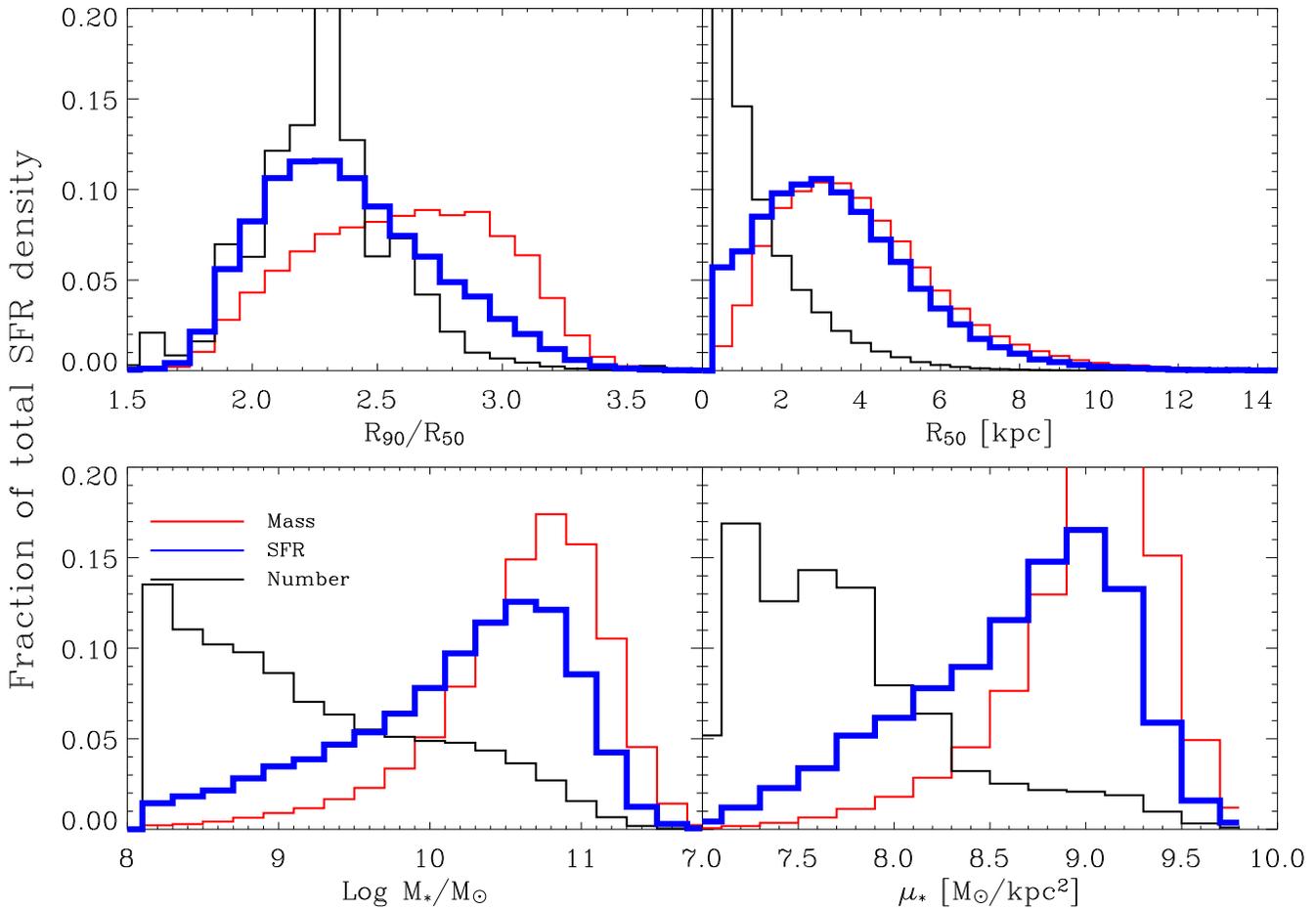}
  \caption{The contribution to the total number, mass and star formation
    density as a function of various galaxy parameters. The SF density
    is shown in blue, the mass density in red and the number density
    in \rhoncolor. \emph{Top left:} The contribution to the different
    densities as a function of the concentration of the galaxies.
    \emph{Top right:} The same, but as a function of  the half-light radii
    of the galaxies.
    \emph{Lower left:} The density contributions as a function of
    log stellar mass and  \emph{Lower
      right:} The contributions as a function of log
    of the stellar surface density in \Msun/kpc$^{2}$}
  \label{fig:sfr_dens_fig3}
\end{figure*}
\begin{figure*}
  \centering
  \includegraphics[angle=90,width=184mm]{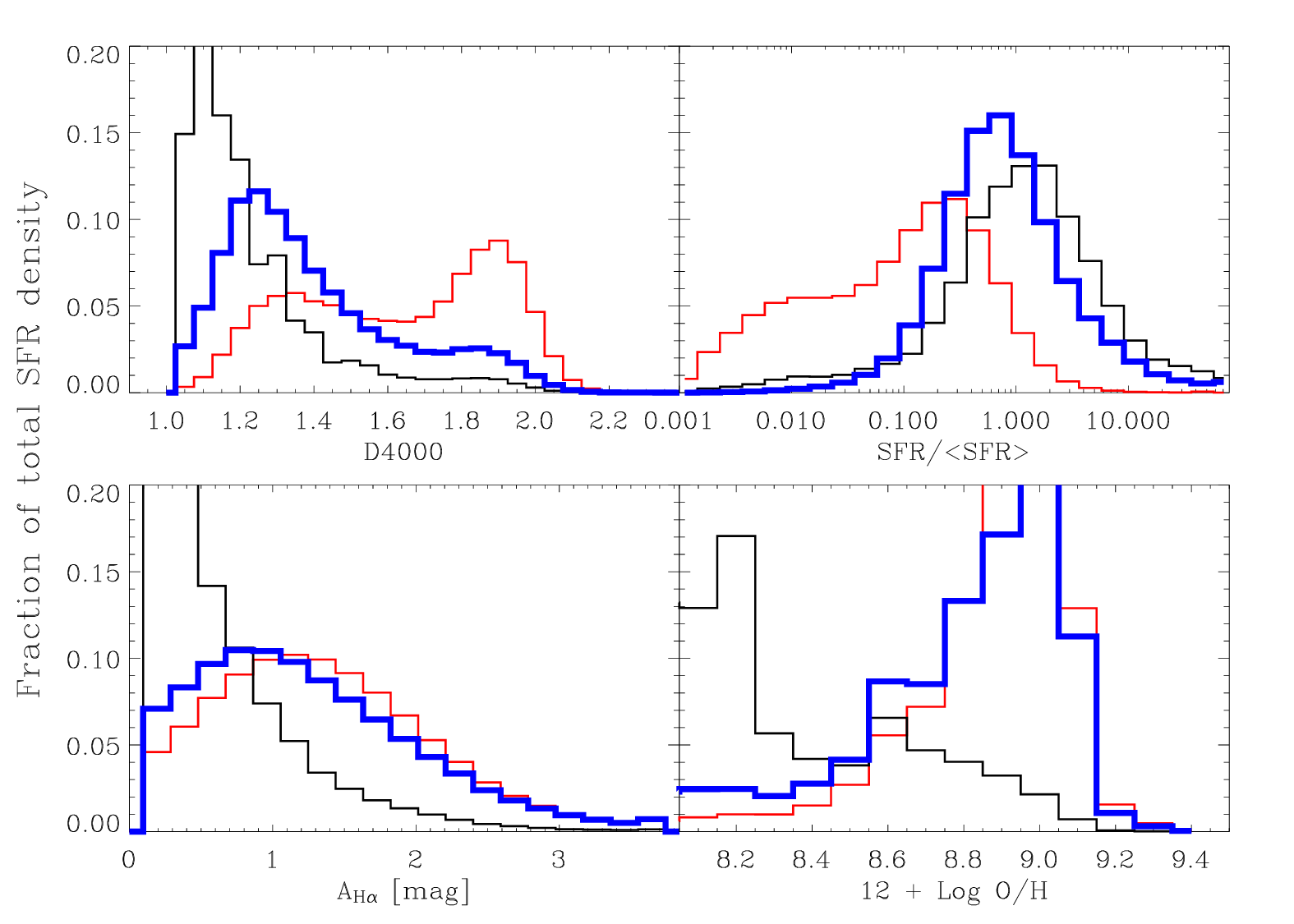}
  \caption{Like the previous figure, but this time showing the star
    formation density versus the 4000\AA\ break (upper left panel) the
    present to past-average SFR (upper right), the dust attenuation at
    \ha\ in magnitude (lower left) and the gas-phase oxygen abundance
    (lower right). The latter two use only the SF class.}
  \label{fig:sfr_dens_fig4}
\end{figure*}

The top left panel in Figure~\ref{fig:sfr_dens_fig1} shows $\rhoSFR$
as a function of the central concentration of the galaxies, defined as
the ratio of the radius containing 90\% of the \sdssr-band light to
that containing 50\%. As discussed by \citet{2001AJ....122.1238S} and
\citet{2001AJ....122.1861S}, this quantity correlates quite well with
morphological type for the galaxies in the SDSS. It related to other
similar quantitative measures of galaxy concentration discussed
previously
\citep[e.g.][]{Morgan-1958,1993MNRAS.264..832D,Abraham-et.al-94}.  The
plot shows that the $\rhoSFR$ distribution is broad and that it peaks
around the value for a pure disk $R_{90}/R_{50}=2.33$.  The SFR
density contributions split into individual galaxy sub-classes are
shown by thin lines as described in the legend. We see that the SF
class contributes primarily to the disk-like part of the distribution.
The star-formation in AGN occurs in galaxies that are more bulge-like.

The top right panel shows $\rhoSFR$ as a function of galaxy stellar
mass. We can observe several interesting features here:
\begin{itemize}
\item A considerable amount of star formation takes
place in low mass galaxies. The best-fit Schechter function has a
faint-end slope of $-0.45$ and a characteristic mass of $\log
M/M_\odot = 10.95$ with a possible transition to a steeper slope of
$\alpha_\mathrm{FS}\approx -0.55$ at $\log M/M_\odot < 9.8$. Approximately 50\% of
the total SFR takes place at $M_* > 2 \times 10^{10} M_\odot$ and
about 90\% at $M_* > 10^{9}M_\odot$.
\item The SF class almost completely dominates the SFR budget at masses less than
about $10^{10}\Msun$.
\item At masses $>10^{10}\Msun$ the majority of SFR takes
place in galaxies low S/N SF class with important contributions
estimated from galaxies that either cannot be classified or which show
signs of AGN in their fibre spectra.
\end{itemize}

Many of the same trends can also be seen in the bottom left panel which
shows $\rhoSFR$ as a function of the stellar surface mass density
\citep{2003MNRAS.341...54K}.  Taken together with the top left hand
panel, it is clear that the majority of the SFR comes from
high-surface brightness spirals.  These typically have half-light
radii around 3.0 kpc (slightly smaller than the Milky Way).

Turning to Figure~\ref{fig:sfr_dens_fig2} we now plot quantities that
are determined inside the fiber aperture on the $x$-axis.  We caution
that since we only know these quantities within the fibre aperture,
they may not be representative of the galaxy as a whole.  The top
right hand panel shows that most of the star formation takes place in
galaxies with low D4000.  12\% of the total SFR density comes from
galaxies with D4000$>1.8$, 2\% with D4000$>2.0$.  Note that the star
formation in these galaxies comes entirely from their outer parts and
they are probably spiral systems with significant bulges.  The bottom
left panel shows that the majority of star formation is taking place
in galaxies that have a present-to-past average star formation rate
between 0.3 and 0.5.

The top right panel shows the distribution as a function of
gas-phase oxygen abundance. Since the metallicity is derived from the
fits to the CL01 models, we only show results for the SF class.  It is
interesting that most stars are forming in galaxies with solar or
slightly super-solar metallicities, depending on the overall
calibration of the metallicity scale
\citep[e.g.][]{2003ApJ...591..801K}.  Finally the bottom right panel
shows the star formation density as a function of the attenuation at
\ha\ (again, only for the SF class). This shows a peak around 1.0
magnitudes of extinction in \ha, in good agreement with other
determinations \citep[e.g.][
C02]{1983ApJ...272...54K,2000MNRAS.312..442S,Hopkins-et-al-2003,Nakamura-Fukugita-2003}.
Notice that we calculate the SFR weighted distribution function of
$A_{H\alpha}$, whereas most earlier studies have focused on a sample
averaged value. It is clear from Figure~\ref{fig:dust_likelihoods}
that any such averaging will depend strongly on the sample used. This
has also been pointed out by \citet{Hopkins-et-al-2003}, who show that
a radio-selected subsample of the SDSS has considerably higher
attenuation than the average SDSS galaxy.

We now compare these SFR distributions with number and mass density
distributions.  Number distributions as function of many photometric
properties and as a function of local density have been discussed in
detail by \citet{2003ApJ...594..186B}  and by
\citet{2003ApJ...585L...5H}.  Likewise the mass distributions have
been discussed by \citet{2003MNRAS.341...33K,2003MNRAS.341...54K}.

Figure~\ref{fig:sfr_dens_fig3} is analogous to 
Figure~\ref{fig:sfr_dens_fig1} and shows the distribution of $\rhoSFR$
together with $\rho_{M_*}$, the mass density distribution, and
$\rho_N$, the number density distribution. In these plots
we have normalised the distributions to the total \emph{within the
 plot limits}. 
We remark that the
spikes seen in several of the number density distributions are due to
a handful of galaxies that are extremely close to our selection limits
with very high volume corrections. The
bootstrap errors on these spikes are large.

Figure~\ref{fig:sfr_dens_fig3} shows that although high concentration
galaxies are few in number, they contain the majority of the mass
\citep[see the discussion by ][]{2003MNRAS.341...33K}.  As discussed
previously, most of the star formation occurs in disk-like systems.
The top right hand panel shows that the majority of galaxies are small
\citep[cf.][]{2003ApJ...594..186B} and that the galaxies dominating
the star formation budget and those dominating the mass budget are
fairly similar in size.  The bottom left hand panel in the figure
shows the dependence on mass. We see that star formation is
typically occurring in galaxies that are somewhat less massive than
those that contain the bulk of the stellar mass in the local Universe.
As we noted previously, most of the star formation is taking place in
high surface brightness galaxies. However, the the bottom right panel
shows that the star formation is distributed over a wider range in
stellar surface density than the stellar mass.  This is presumably
because a significant fraction of the stellar mass is located in
ellipticals, which are not currently forming stars.

Figure~\ref{fig:sfr_dens_fig4} is analogous to
Figure~\ref{fig:sfr_dens_fig2} and shows properties measured within
the fiber aperture.  As discussed previously, the majority of the star
formation is occurring in galaxies with low D4000.  It is nevertheless
intriguing that the majority of galaxies in the universe have even
lower D4000 than those that dominate the SFR density. A very similar
result is illustrated in the top right-hand panel, which contrasts the
contributions to $\rho_N$, $\rho_{\mathrm{SFR}}$ and $\rho_{\Mstar}$
from galaxies with different present-to-past average star formation
rates.

The bottom panels show the density distributions as a function of the
attenuation at \ha\ (left) and as a function of gas phase oxygen
abundance (right). These quantities are only calculated for the SF
class.  We see that the majority of SF galaxies have low metallicity
and dust attenuation, two facts that are likely to be closely
connected. Most of the mass and most of the star formation are found
in galaxies with a wide distribution of $A_{H\alpha}$, centred around
1.0 magnitudes with a spread of $\sim 0.7$ magnitudes for the SFR
weighted distribution \citep[see also the related study
by][]{Hopkins-et-al-2003}. For the $A_{H\alpha}$ distribution weighted
by stellar mass, we find an average value of $\sim 1.3$ magnitudes and
a similar spread. The difference reflects the fact that the most
massive galaxies are more dusty than the galaxies dominating the SFR
budget.

\section{The specific star formation rate}
\label{sec:spec_sfr}

The total SFR of a galaxy is interesting and for certain
physical questions it is the relevant quantity. But given the strong
correlation between SFR and mass (cf.\ Figure~\ref{fig:mstar_vs_sfr}),
it is clear that by normalising the SFR by the stellar mass, one can
more easily study the relationship between star formation activity and
the physical parameters of the galaxies. In this section we
will turn our attention towards the star formation rate per unit mass,
the specific star formation rate. 
For compactness we will refer to the specific SFR as \sSFR\ below.

The specific star formation rate is closely related to
several other important physical quantities. It defines a
characteristic time-scale of star formation,
\begin{equation}
  \label{eq:tau_sfr}
  \sSFR = \tau_{\rm SFR}^{-1} = \frac{\dot{M}_*}{\Mstar}.
\end{equation}
Since the supernova rate is proportional to the star formation rate,
the specific star formation rate represents the current input of
supernova energy per star in the galaxy. Likewise, if the stellar mass
is proportional to the mass of the ISM, it is related to the
porosity of the ISM in the galaxy \citep{2002MNRAS.337.1299C}. Finally
$\sSFR$ does not depend on cosmology and is insensitive to the IMF as
long as the high mass slopes are comparable.

\begin{table*}
  \caption{The best estimate of the present-to-past average star
    formation rate, $b^V$. We have averaged the values form $\beta=3$
    and $\beta=0$ as appropriate for our systematic error estimates.}
  \begin{tabular}{|r|rrr|rrr|}\hline
    \multicolumn{1}{|c|}{Sample} & \multicolumn{3}{c}{Fibre} & 
    \multicolumn{3}{c|}{Total} \\ 
    &16\% & Mode & 84\%  & 16\% & Mode & 84\% \\ \hline 
         Total & 0.313 & 0.314 & 0.325 & 0.406 & 0.408 & 0.413 \\
            SF & 0.994 & 0.997 & 1.011 & 0.881 & 0.888 & 0.898 \\
    Low S/N SF & 0.198 & 0.201 & 0.202 & 0.421 & 0.432 & 0.437 \\
    Composites & 0.396 & 0.406 & 0.464 & 0.362 & 0.365 & 0.373 \\
           AGN & 0.210 & 0.216 & 0.232 & 0.229 & 0.233 & 0.239 \\
Unclassifiable & 0.030 & 0.031 & 0.032 & 0.131 & 0.133 & 0.138 \\
    \hline
  \end{tabular}
  \label{tab:b_values}
\end{table*}

The star formation per unit mass has either explicitly or implicitly
been used in numerous studies of field galaxies at $z<1$ \citep[e.g.][
and references
therein]{2000ApJ...536L..77B,1996AJ....112..839C,1997ApJ...489..559G,1997ARA&A..35..389E}
by using equivalent widths of \ha\ or \oii.  In the local Universe, it
has however often been rephrased in terms of the present to
past-average star formation rate
\citep[e.g.][]{1983ApJ...272...54K,1986FCPh...11....1S,1994ApJ...435...22K,2001AJ....121..753B,2002A&A...396..449G}.
This is a suggestive quantity, which immediately gives an indication of
the past star formation history of the galaxy and its relation to 
present-day activity. The main disadvantage of the present-to-past
average SFR is that it involves more assumptions than the specific
star formation rate as we will see below.

The conversion between the two can be succinctly summarised as
\begin{equation}
  \label{eq:b_vs_tau_SFR}
  b \equiv \frac{\mathrm{SFR}}{\langle\mathrm{SFR}\rangle}= \sSFR T (1-R),
\end{equation}
where $R$ is the fraction of the total stellar mass initially formed
that is return to the ISM over the lifetime of the galaxy, and $T$ is
the time over which the galaxy has formed stars ($T=t_H(z)- t_{\rm
  form}$, where $t_H(z)$ is the Hubble time at redshift $z$ and
$t_{\rm form}$ is the time of formation).

Given a choice of IMF, it is possible to reliably estimate $R$ using
stellar evolution theory.  We use the median $R$ estimated for our
galaxies which is $R\sim 0.5$.  For $T$ it is customary to take
$T=t_H$, ie.\ assume that all galaxies started to form stars at $t=0$.
We will adopt this convention below (with $T=t_H(0.1)$). This
assumption is sensible when comparing average $b$ values for large
samples, but might be more questionable for individual galaxies. In
general, this means that the $b$ values we calculate will be upper
limits.

Throughout this discussion, we will use $b$ and $\sSFR$ interchangeably,
depending on which is more appropriate. As
highlighted by \citet{2000A&A...358..869R}, the $b$ value is a
convenient parameterisation of ``burstiness'' and $b=1$ provides a
convenient divide between galaxies consistent with a constant or
declining SFR over the age of the universe and those that have a
higher SFR at present than they have typically had in the past.

\begin{figure*}
  \centering
  \includegraphics[angle=90,width=184mm]{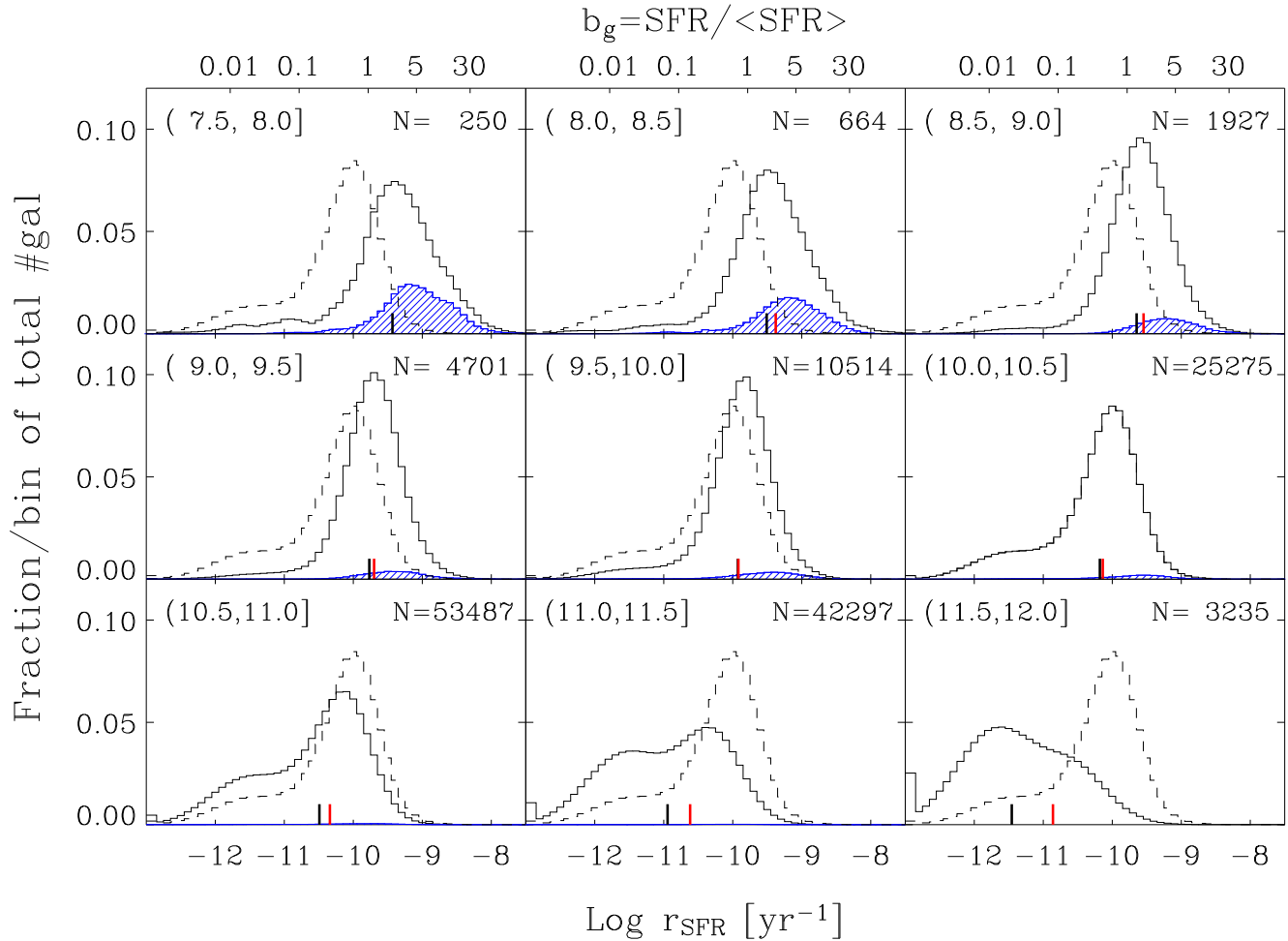}
  \caption{The observed specific star formation as a function of total
    stellar mass in the mass ranges indicated in the top left hand
    corner of each panel. Shaded areas show recent starbursts, see
    text for details. Each histogram is normalised to the total number
    of galaxies in the mass range, given as $N$ in each panel. The
    dashed histogram is the distribution in the $\log M_* \in (10.0,
    10.5]$ bin and is included for reference. The thick black bar
    shows the average $b$ value, $b^g$, in each mass range and the
  thick red bar shows $b^V$. The latter is not shown in the first 
  panel since we might not be statistically complete there. All
  objects with $\log \mathrm{SFR}/\Mstar \leq -12$ have been placed as a
  Gaussian around $\log \mathrm{SFR}/\Mstar = -12$ as discussed in the text.}
  \label{fig:spec_sfr_vs_mass}
\end{figure*}

Before turning to the results for our sample, we note that        
there are at least two reasonable definitions for the
\emph{average} $\sSFR$ and one must take care to specify which one
is discussing.

The average $\sSFR$ value of a sample of $N$ galaxies is simply
\begin{equation}
  \label{eq:average_b_galaxy}
  \sSFR^g = \left\langle\frac{\mathrm{SFR}}{M_*}\right\rangle =
  \frac{1}{\sum_i w_i} \sum_i w_i \frac{\mathrm{SFR}_i}{M_{*,i}},
\end{equation}
where the weights $w_i$ can be unity or e.g.\ $1/V_{\mathrm{max}}$
depending on the question asked. This is the appropriate quantity to
quote when talking about the typical galaxy.  In practice we calculate
this using the full likelihood distributions of $\sSFR$ for each
galaxy.  We also usually choose to quote the median rather than the
average as this is binning insenstive (cf.\ discussion in
Appendix~\ref{sec:manipulate_likelihoods}).

$\sSFR^g$ can be very different from the $\sSFR$ value for a
volume limited sample:
\begin{equation}
  \label{eq:b_vol_limited}
  \sSFR^V = \frac{\rho_{\mathrm{SFR}}}{\rho_{\Mstar}} = \frac{\sum_i w_i
    \mathrm{SFR}_i}{\sum_i w_i \Mstar},
\end{equation}
which is the appropriate quantity to quote for the                          
the star formation history of the universe as a whole.  
To calculate $\sSFR^V$ using our likelihood
distributions,  we  employ the same Monte Carlo summation
technique used in the calculation of $\rho_{\mathrm{SFR}}$ in
section~\ref{sec:sfr_density}. It is important to include the stellar
mass in this Monte Carlo code, because, as mentioned in
Appendix~\ref{sec:manipulate_likelihoods}, neglecting to do so will
bias our results. (To investigate the size of this bias we carried out
the calculation of $\sSFR^V$ with and without the full likelihood
distribution of $\log \Mstar$ and found that ignoring the uncertainty in
$\log \Mstar$ resulted in an overestimate of $\sSFR$ of about 15\%.) 

We begin by calculating $b^V$ for the sample as a whole.  This is a
straightforward application of the results in the previous section and
we show this quantity in Table~\ref{tab:b_values}. We see that for
$h=0.7$, the present day universe is forming stars at slightly higher
than 1/3 of its past average rate, depending on the time-span one
averages over in equation~(\ref{eq:b_vs_tau_SFR}). Once again we have
calculated the likelihood distribution using the Monte Carlo method
discussed in Appendix~\ref{sec:manipulate_likelihoods} with 100
bootstrap repetitions and 30 Monte Carlo sums for each repetition.
The resulting errors are also indicated in Table~\ref{tab:b_values},
but as with the total SFR density in section~\ref{sec:sfr_density} the
random errors are substantially smaller than the systematic ones.
Following the discussion in section~\ref{sec:sfr_density}, we adopt
the same estimates of systematic errors on the total SFRs and find a
present-to-past average SFR for the local universe of
\begin{equation}
  \label{eq:present_to_past_total}
  b^V(z=0.1) = 0.408^{+0.005}_{-0.002} \mathrm{(rand.)}^{+0.029}_{-0.090} \mathrm{(sys.)},
\end{equation}
with random errors corresponding  to the 68\% confidence interval.

It is interesting to compare $b^V$ for the different classes we have
defined. If we look at quantities determined inside the fibre,
it follows almost by definition that the SF class has the highest $b^V$
value and  the unclassifiable category  the lowest.
If we  look at the values derived from the total SFR, the SF
class still dominates, but it is interesting to note that the global
$b$ value for this class is the same as the central one. For the other
classes one typically finds that the global $b$ is higher than that
inside the fibre. This is, of course, partly a 
sample selection issue: galaxies classed as SF are those that
have relatively more star formation inside the fibre. It is also
interesting to note that  galaxies showing indications of AGN
activity form an intermediate category between the Composites and the
unclassifiable. This is again logical because AGN would have been classified
as composites if they had substantially higher SFR.

We now study how the specific star formation rate varies between
galaxies of different stellar mass.  We split the galaxy sample into
nine bins in \emph{total} stellar mass. We sum up the likelihood
distributions of $\sSFR$, ie.\ we calculate $\sSFR^g$ without the
normalisation in Equation~(\ref{eq:average_b_galaxy}). We show the
resulting distributions of $\sSFR^g$ in
Figure~\ref{fig:spec_sfr_vs_mass}.  The histograms are normalised to
the total number of galaxies within that bin.  No volume correction
has been applied, but the difference to that shown is very small
except in the lowest mass bins. (Note that we might not be complete in
mass in the lowest bin.) The contribution to the distributions that is
located outside the plotted range has been distributed on the two
lowest and two highest bins respectively.

The histogram for galaxies with $10.0 < \log \Mstar < 10.5$ is
repeated in each panel for reference. The shaded histograms show the
distributions of those galaxies classified as recent starbursts based
on their D4000 and H$\delta_A$ values by~\citet{2003MNRAS.341...33K}.
The thick black line shows the median volume weighted $\sSFR$ value
($\sSFR^g$), and the red line shows $\sSFR^V$. The latter is higher as
the total SFR is dominated by galaxies with extreme values of $\sSFR$.

Figure~\ref{fig:spec_sfr_vs_mass} shows that there is a reasonably
good agreement between the recent burst estimates of
\citet{2003MNRAS.341...33K}, which are sensitive to activity in the
last Gyr, and the activity indicated by emission lines. Thus galaxies
that are currently undergoing elevated star formation activity
(relative to the Hubble time) have typically done so for at least a
Gyr. It is also clear that the \emph{fraction} of galaxies that show
clear signatures of recent starbursts declines rapidly with mass. This
can either be due to a lower gas fraction in more massive galaxies
and/or the fact that periods of elevated star formation may be shorter
in duration in more massive galaxies.

Figure~\ref{fig:spec_sfr_vs_mass} also shows that the peak of the
distribution shifts rather little over a large range in mass ($\log
\Mstar \la 10.5$).  The peak value is very close to $b=1$. In other
words, most galaxies with $\log \Mstar \la 10.5$ are forming stars at a
rate that is consistent with their past average (ie.  at a constant
rate). Above $\mathrm{few}\times 10^{10}\Msun$, we see a clear change
in the shape of the distribution of $b^g$ with a transition towards a
low specific SFRs. This coincides well with the transition
mass defined by~\citet{2003MNRAS.341...33K}. Also note that the second
peak showing up at $\sSFR \sim 10^{-12}$ at high mass is broader than
the one at high $\sSFR$ because the SFRs are less well constrained
there. 

Despite this near-constancy of the peak, there is a marked change in
the shape of the tails. At low masses there is a substantial tail
towards higher $b$-values, ie.\ a trend towards more bursty star
formation activity.  This changes to a substantial tail towards low
$b$-values at high masses. Many massive galaxies are now forming stars
at a depressed rate compared to their past average.

\section{The birthrate parameter as a function of physical parameters}
\label{sec:b_vs_phys_param}

Section~\ref{sec:overall_prop} focused on the distribution of the SFR
density as a function of different physical parameters. We will now
study the relation between the specific SFR and these same quantities.
This will help us understand which factors control/are controlled by
the star formation activity in individual galaxies.

Most previous studies have focused on the trend with mass and Hubble
type. A key question has been which of these parameters correlate
better with $b$. The studies of local spirals by
\citet{1996A&A...312L..29G} and \citet{2001AJ....121..753B} indicated
that the driving variable was the stellar mass content.  Other studies
have made more of the correlation with Hubble type
\citep[e.g.][]{1994ApJ...435...22K}. Here we will improve on these
studies with a sample that is several orders of magnitude larger than
the previous studies and with more careful and internally consistent
modelling. We will not discuss the dependence on local density here.
Recent studies of the correlation between star formation and density
can be found in \citet{2003ApJ...584..210G} and
\citet{1998ApJ...504L..75B,2001ApJ...557..117B}.

\begin{figure*}
  \centering
  \includegraphics[width=184mm]{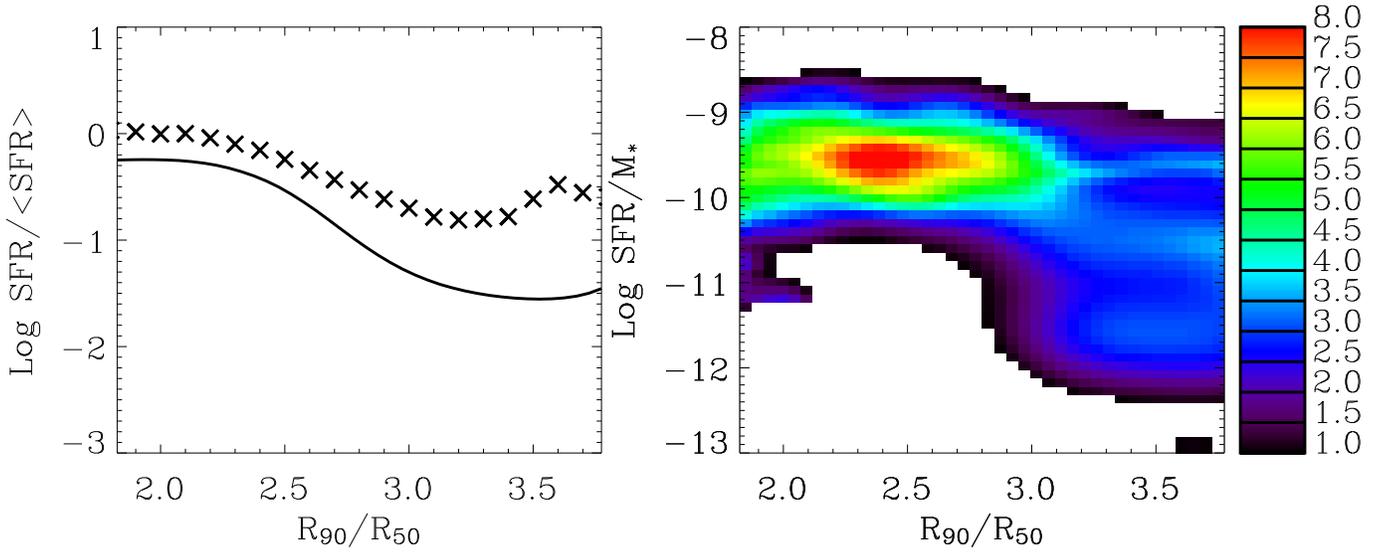}
  \caption{The specific SFR as a function of concentration.  The left panel
    contrasts $b^g$ with $b^V$. The continuous line in this plot shows
    the median of the unweighted $b^g$ (right panel) and the crosses
    show $b^V$ calculated from the data in
    Figure~\ref{fig:sfr_dens_fig3}. The right hand panel shows the
    (log of the) observed likelihood distribution of $\sSFR$ with
    respect to the concentration parameter, $R_{90}/R_{50}$,
    calculated as described in the text. The shading shows the
    conditional likelihood distribution (volume corrected) given a
    value for $R_{90}/R_{50}$. The contributions to the likelihood
    distributions below the plotted range have been put
    in the two lowest bins in \sSFR.}
  \label{fig:conc_vs_b}
\end{figure*}

\begin{figure*}
  \centering
  \includegraphics[width=184mm]{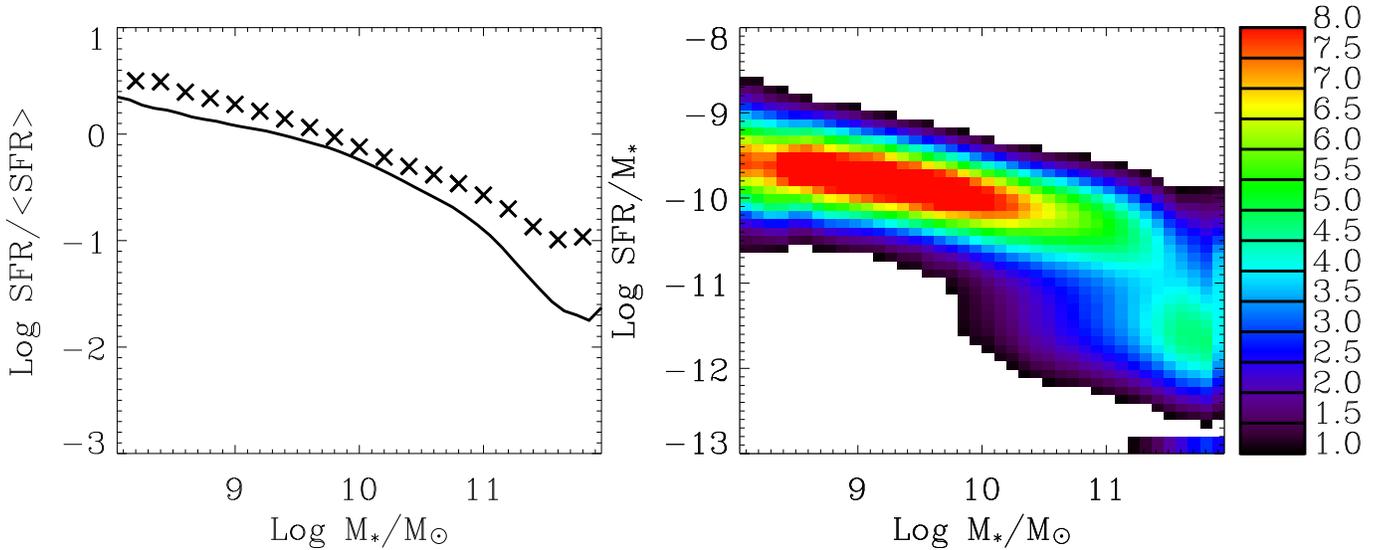}
  \caption{Similar to Figure~\ref{fig:conc_vs_b} but this time showing
    $b$ as a function of  the stellar mass.}
  \label{fig:mstar_vs_b}
\end{figure*}

\begin{figure*}
  \centering
  \includegraphics[width=184mm]{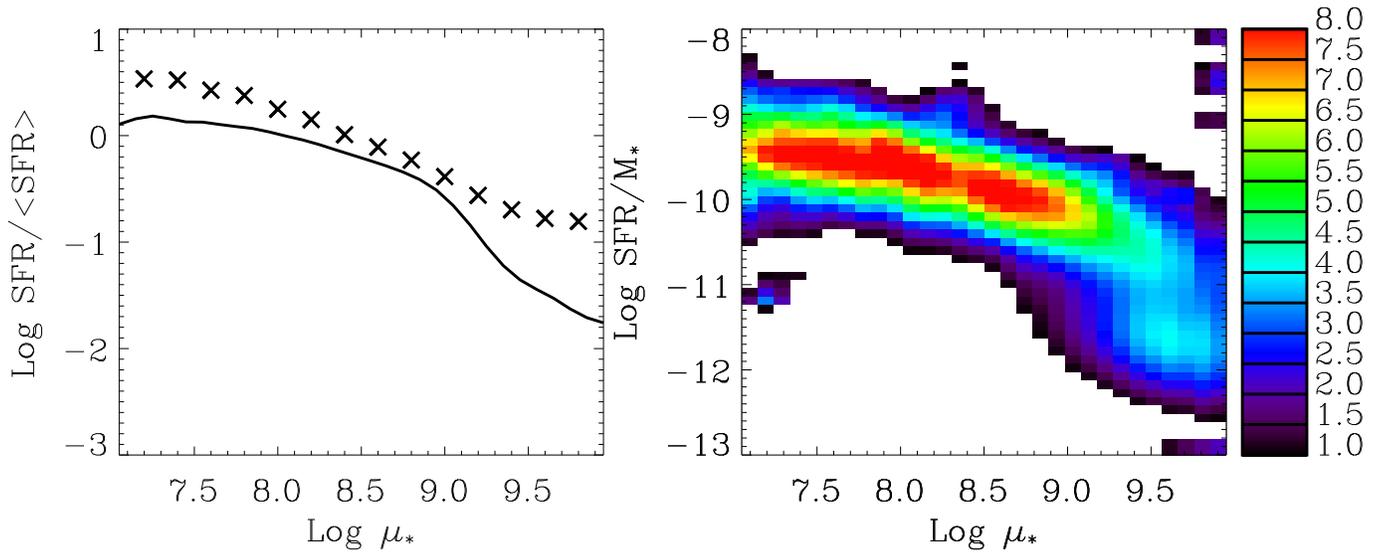}
  \caption{Similar to Figure~\ref{fig:conc_vs_b} but this time showing
    $b$ as a function of  stellar mass surface density.}
  \label{fig:logmu_vs_b}
\end{figure*}

\begin{figure*}
  \centering
  \includegraphics[width=184mm]{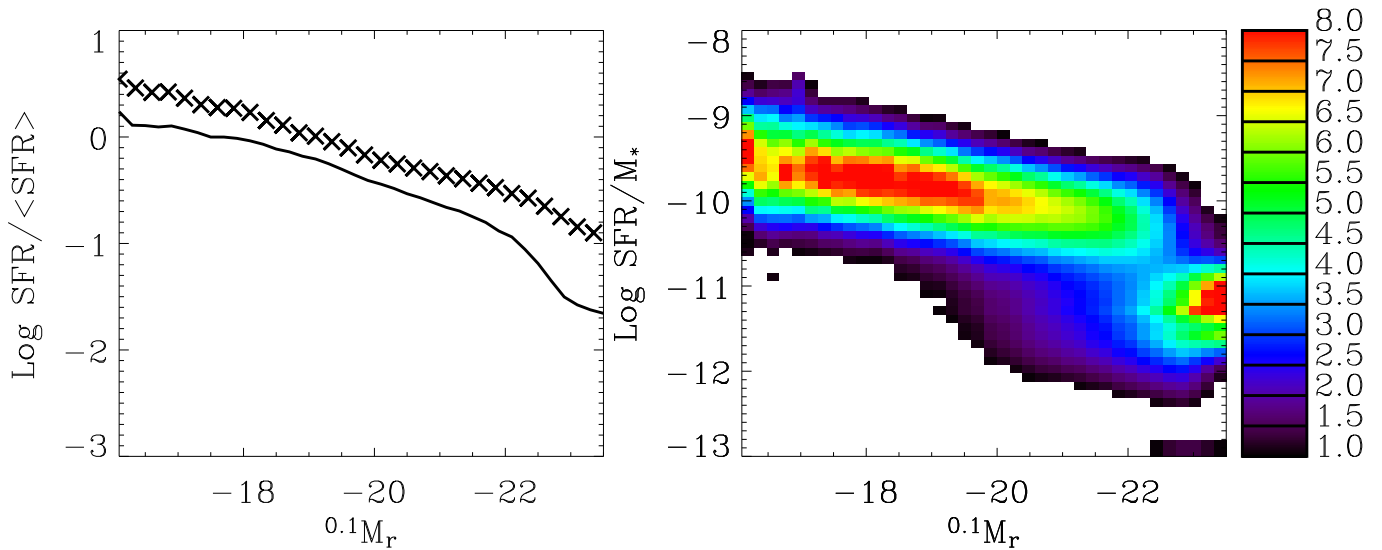}
  \caption{Similar to Figure~\ref{fig:conc_vs_b} but this time showing
    $b$ as a function of   $^{0.1}M_r$ magnitude.}
  \label{fig:MR_vs_b}
\end{figure*}

\begin{figure*}
  \centering
  \includegraphics[width=184mm]{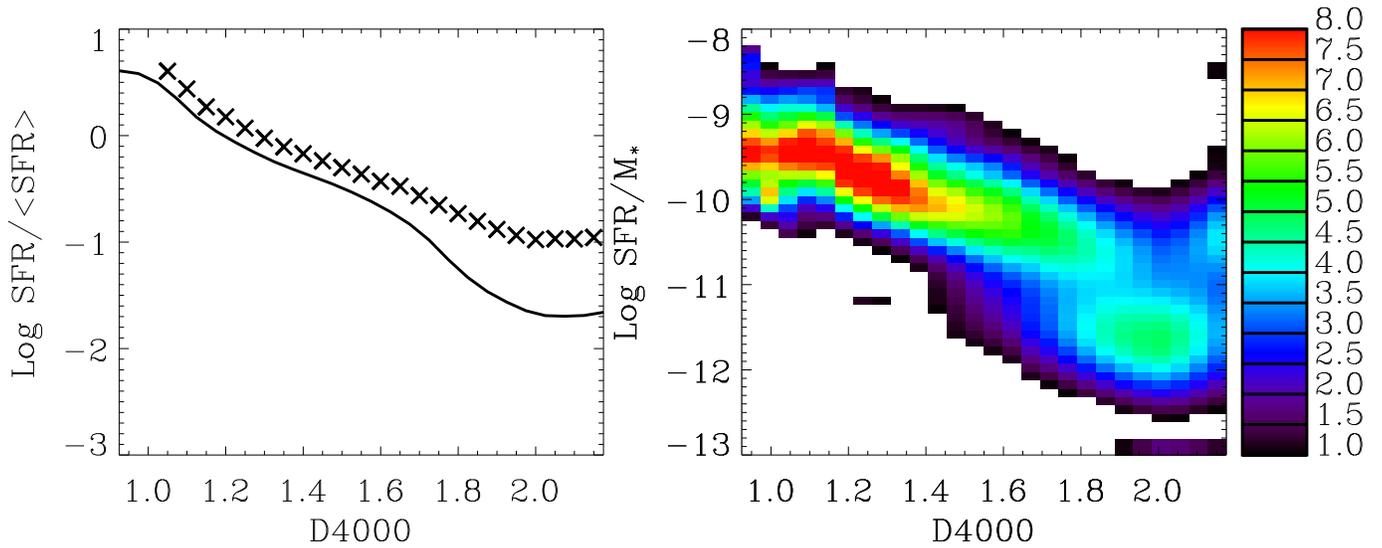}
  \caption{Similar to Figure~\ref{fig:conc_vs_b} but this time showing
    $b$ as a function of  D4000. Notice the clear correlation, despite a
    substantial scatter. In this figure the solid line shows $b^g$
    determined in the fibre whereas the crosses are for the total.}
  \label{fig:d4000_vs_b}
\end{figure*}

\begin{figure*}
  \centering
  \includegraphics[width=184mm]{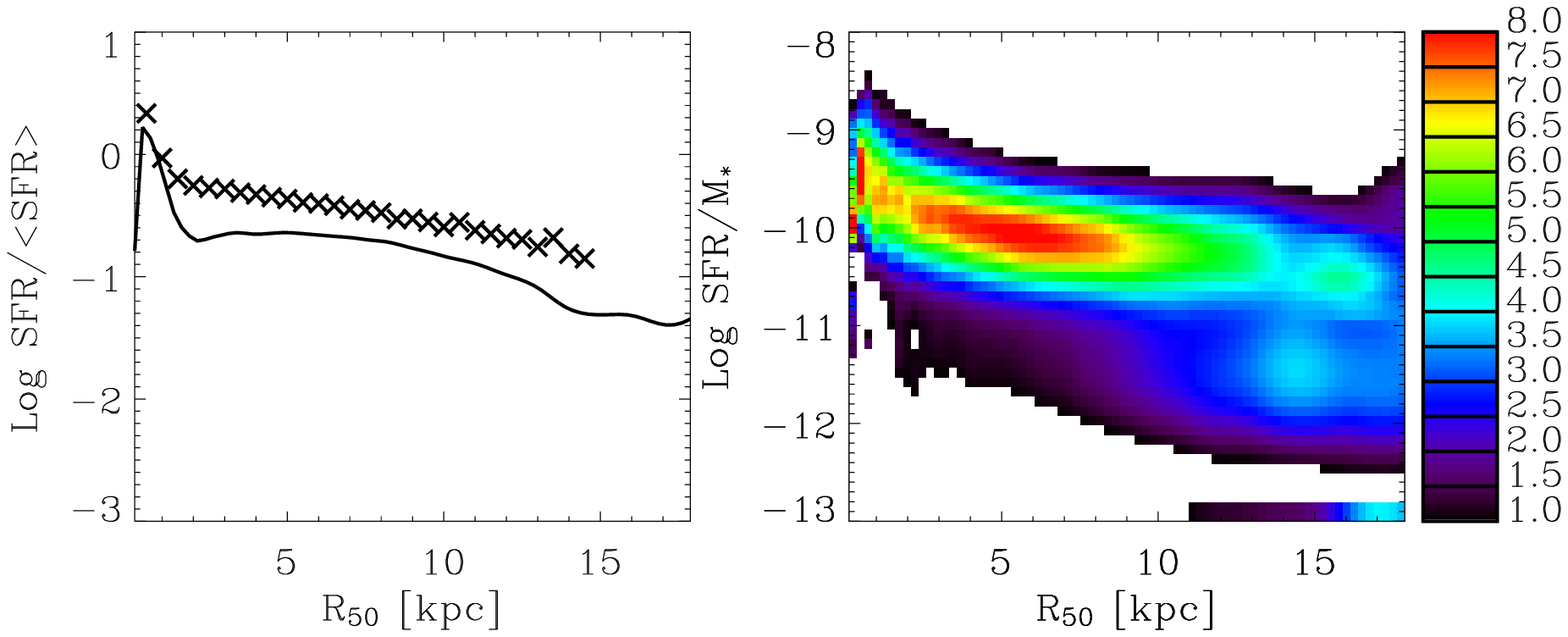}
  \caption{Similar to Figure~\ref{fig:conc_vs_b} but this time showing
    $b$ as a function of the half-light radius of the galaxies.}
  \label{fig:r50_vs_b}
\end{figure*}

Throughout this analysis, we will show results using $b^g$ and $b^V$
derived from our \emph{total} SFR estimates.  In
Figure~\ref{fig:conc_vs_b}, we show the trend of $b$ with central
concentration/Hubble type. The left hand panel in
Figure~\ref{fig:conc_vs_b} contrasts $b^g$ (solid line) with $b^V$
(crosses). The solid line shows the median of the unweighted $b^g$
calculated from the right panel using the formalism discussed in
Appendix~\ref{sec:manipulate_likelihoods}. We do not show the
confidence intervals, but as is evident from the full likelihood
distributions, they are wide. The crosses show $b^V$ in each bin in
$R_{90}/R_{50}$. This is calculated using Monte Carlo summation and is
basically the ratio of the black and red curves in
Figure~\ref{fig:sfr_dens_fig3}. In general $b^V$ is larger than $b^g$
because $\sSFR$ has substantial tails towards large values. $b^V$ is
therefore dominated by galaxies away from the median.

The right panel of Figure 21 shows the conditional likelihood
distribution of $\sSFR$ given the concentration parameter,
$C=R_{90}/R_{50}$, for the local universe. This has been constructed
by assuming that $P(C)$ is independent of $P(\sSFR)$ and assuming a
5\% error on $C$ (this latter assumption is not important for the
results). Each distribution has been weighted by 1/V$_{\mathrm{max}}$.
The distribution in the right panel then follows from a straight
addition of all the individual $P(C, \sSFR)$ and normalisation of the
distribution in each bin of $C$ to produce the conditional likelihood
distribution. As in Figure~\ref{fig:spec_sfr_vs_mass} contributions to
the likelihood distributions outside the plotted range in \sSFR\ has
been placed in the two lowest bins in \sSFR.

These two panels illustrate the bimodal nature of the distribution of
galaxies.  This has recently been described and quantified in a number
of papers
\citep[e.g.][]{2001AJ....122.1861S,2003ApJ...594..186B,2003MNRAS.341...54K,2003astro.ph..9710B}.
As can be seen, galaxies split into two basic populations: concentrated
galaxies with low specific star formation rates, and low mass, less
concentrated galaxies with high specific star formation rates. The
distribution at high concentration is broad however because there is
also a fairly important contribution from concentrated, actively star
forming galaxies at very low redshift. When the figure is plotted for
the sample directly, ie.\ without volume weighting, the two regimes
are much more clearly separated.

Figure~\ref{fig:mstar_vs_b} shows, with the same layout, the
distribution of $b$ as a function of stellar mass.  We see that
although there is a clear decline in $b$ with increasing stellar mass,
the trend in $b^g$ is weak over a large range in stellar mass until it
suddenly drops at around $10^{11} \mathrm{h}_{70}^{-2}\Msun$. The
$b^V$ value shows a much smoother trend. This is caused by the large
spread in $\sSFR$ at given mass. If we compare Figures
~\ref{fig:mstar_vs_b} and \ref{fig:conc_vs_b}, we see that the Hubble
type correlates somewhat less well with the $b$ parameter than
the stellar mass does.

Figure~\ref{fig:logmu_vs_b} shows $b$ as a function of stellar mass
surface density, $\mu_*$. The connection between this and \Mstar\ has
been discussed in detail by \citet{2003MNRAS.341...54K}. Those authors
show that $\mu_*$ correlates well with $\Mstar$, except at the very
highest masses where $\mu_*$ reaches a saturation value. It is
therefore not surprising that we see a very similar trend with $\mu_*$
as we saw with $\Mstar$ in Figure~\ref{fig:mstar_vs_b}.  When we plot
the figure without volume weighting it is very similar but the low SFR
peak becomes more prominent at high $\mu_*$ than at high $\Mstar$,
suggesting that the surface density of stars is more important than
the stellar mass in determining when star formation is turned off.

Figure~\ref{fig:MR_vs_b} again presents a similar view. This time the
$b$ parameter is plotted as a function of absolute \sdssr-band
magnitude at $z=0.1$. The correlation here is fairly good. For
galaxies fainter than about $M_r=-20$, the $b_g$ parameter increases
by about 40\% per magnitude.

Figure~\ref{fig:d4000_vs_b} shows the relationship between $b$ and the
4000\AA-break. The left panel shows this to be a fairly tight relation
--- one reason why \citet{2003MNRAS.341...33K} were so successful in
constraining the $M/L$ ratio for galaxies using the 4000\AA-break and
the $H\delta_A$ index. We note that this relation does  have the
potential to constrain  star formation in high redshift galaxies, 
as the Balmer break can be measured out to considerably higher
redshift than \ha. From Figure~\ref{fig:d4000_vs_b} we see that a
measurement of D4000 allows one to constrain $b_g$ to within a factor
of three (1$\sigma$) for D4000$<1.6$. Although this is not very precise,
it should be weighed  against the fact that it is
very difficult to estimate star formation rates to better than within
a factor of two with any technique. It is however important to caution
that the D4000 index intrinsically depend on the SFR in the last
giga-year rather than the current SFR, as well as the metallicity. It
is therefore necessary to assume that the present SFR is similar to
the average activity in the last Gyr to make recourse to the relation
in Figure~\ref{fig:d4000_vs_b}.

We caution that D4000 is measured within the fibre, whereas our SFR
estimates are for the entire galaxy. If we make the same plot using
SFR estimated within the fibre, we still recover a tight correlation
between $b^g$ and D4000. Moreover, we find that $b^g$ and $b^V$ now
coincide.  It is galaxies that have high D4000 within the fibre but
significant star formation outside which make $b^V$ considerably
higher than $b^g$ at high D4000 in Figure~\ref{fig:d4000_vs_b}.

We finish by showing $b$ versus $R_{50}$ in
Figure~\ref{fig:r50_vs_b}. It is clear that the low and high $b$
galaxies co-exist over the entire range in $R_{50}$. But there is
a general trend for higher $b^V$ at the smallest sizes and for large
galaxies to have typically lower values of $b^V$.

\subsection{Comparison with previous studies}
\label{sec:previous_comparisons}

\begin{figure}
  \centering
  \includegraphics[width=84mm]{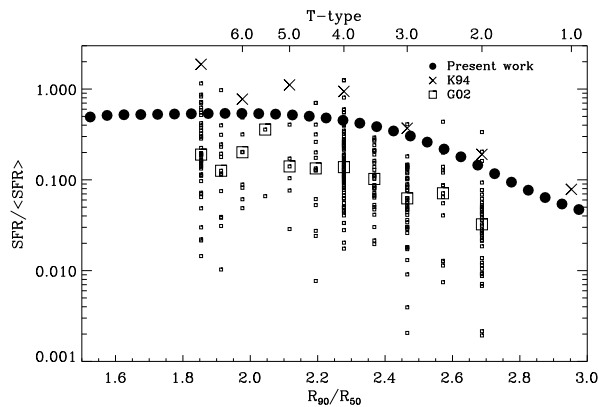}
  \caption{The present to past-average star formation rate for the
    sample plotted against central concentration in the
    \sdssr-band. The top $x$-axis shows the mean $T$-type implied by
    the concentration according to the recipe described in the
    text. The crosses show the sample
    of~\citet{1994ApJ...435...22K} and the boxes that of
    \citet{2002A&A...396..449G}.}  
  \label{fig:conc_vs_b_against_others}
\end{figure}

Before we turn to a discussion of these results, it is
enlightening to compare our results with previous work in more
detail. The main comparison samples are those presented by
\citet{1994ApJ...435...22K} and \citet{2002A&A...396..449G} in their
studies of star formation histories in spiral galaxies. They calculate
$b$-parameters for a range in morphological types and in
Figure~\ref{fig:conc_vs_b_against_others} we plot this distribution of
b versus approximate morphological type for our sample. We have
taken the relation between RC3 T-type and concentration from
\citet{2001AJ....122.1238S} to convert concentration measures into
approximate T-types. We fitted a linear relation between $T$ and $C$
from their Table 3 which is indicated on the upper axis. In the same
plot we show the results of \citet[][ their Table
4]{1994ApJ...435...22K} as crosses and those of
\citet{2002A&A...396..449G} as squares, with big squares corresponding
to the median value (BCGs are not shown). The data we show for our
study are the unweighted $b^g$ values which should be appropriate
for comparison to local studies.

There is clearly a considerable uncertainty in the $x$-axis, but the
general agreement in trends and turn-over between the different
samples indicates that the conversion between $C$ and
$T$-type derived by \citet{2001AJ....122.1238S} seems to be fairly
reliable. It is also clear
that the values from \citet{2002A&A...396..449G} are considerably
lower than ours. This may be due in part to differences in galaxy density in the
two samples, but is more likely due to differences in the method used
to determine the stellar mass. Gavazzi et al use an average
dark matter mass to stellar mass ratio to derive stellar masses. 
We have chosen not to attempt to adjust for 
differences in methodology, as our main interest is in comparing the
overall trend.

\section{Discussion}
\label{sec:discussion}

The preceding sections have provided a detailed inventory of the star
formation activity in the local universe. One possibly
surprising result is that the dominant
contribution to the SFR density comes from galaxies with on-going star
formation in their inner few kiloparsecs (the SF and low S/N SF classes).
These galaxies contribute $\sim 72$\% of the total SFR density, 
whereas galaxies with signs of AGN
(Composites and AGN) contribute about 15\% of the total.
The remaining 12\% of $\rhoSFR$ is located in galaxies we are unable to
classify based on their fibre spectra. These are almost certainly
galaxies with substantial bulges, in which stars
are forming in the outer disk. A thorough examination of this question
has to be postponed to a different article, but it is clear that the
relative importance of star formation in the outer disk is a function
of the mass or luminosity of the galaxy. When we look at narrow
redshift ranges we find that the median aperture corrections increase by a
factor of 2 from stellar masses around $10^9\Msun$ to $\Mstar \sim
10^{11}\Msun$. The natural question is how this relates to star
formation in bulges as compared to disks, but this will have to await
a study of the resolved properties of our galaxies.

We found that galaxies in the SF class have SFR profiles that follow
the \sdssr-band light. For these galaxies, it is possible, and
accurate, to remove aperture bias by simple scaling with the
\sdssr-band flux \citep[cf.][]{Hopkins-et-al-2003}.

We have, for the first time, been able to calibrate the
conversion factor from \ha\ to SFR as a function of mass and we show that
the standard \citet{1998ARA&A..36..189K} conversion factor is a very
good average correction for most galaxies in the local Universe.    
By  lucky coincidence,  trends in the conversion
factor and in the intrinsic Case B ratio of \ha\ to \hb\ 
tend to cancel each other out. As a result,  it will
be sufficient to take a fixed Case B ratio and a fixed
$\eta^0_{H\alpha}$ for most studies, even though neither of the two assumptions is
correct. 

One of the main questions we want to address is what excites 
large-scale star formation in galaxies.
Seen from this angle the different parameters we have studied fall
into two categories. 
First there is the incidental set -- those parameters
which are controlled by star formation activity rather than the other
way around. Among these are D4000 and  colour. Of more interest are
those parameters that may have some direct effect on star
formation such as \Mstar, $\mu_*$, Hubble type and size.

We have found that the stellar mass correlates well with the SFR, a
result found previously for a much smaller sample by
\citet{1996A&A...312L..29G}. This is quite likely because even though
the gas {\em fractions} of galaxies decline with increasing mass, the
total gas masses still increase as a function of mass
\citep{2001AJ....121..753B}.

Because of this overall correlation, the SFR density distribution
functions are rather similar to the mass density distribution
functions discussed by \citet{2003MNRAS.341...33K}. We find that the
majority of star formation takes place in high surface brightness
disk-dominated galaxies spanning a range of concentrations. These span
a broad range of masses, peaking for galaxies slightly less massive
than $10^{11}\Msun$ and with half-light radii around 3 kpc. The star
forming gas has typically solar or slightly super-solar metallicity.

Our most striking result is contained in
Figure~\ref{fig:sfr_dens_fig1}, which shows that the majority of star
formation in the local universe takes place in massive galaxies. This
is a robust prediction of hierarchical models and the shape of our
distribution matches that of
\citet{2003MNRAS.339..312S} (although these authors show $\rhoSFR$
versus the halo mass). This clearly warrants a more in-depth comparison
and we will follow this up in a forthcoming paper.

Another interesting question is whether star formation in the local
universe is predominantly bursty or relatively quiescent. As we saw in
Figure~\ref{fig:mstar_vs_b}, the answer depends strongly on the mass
of the galaxy.  It is safe to say, however, that the vast majority of
star formation in the local universe takes place in galaxies that are
not undergoing extreme starbursts.  Quantitatively we find that 28\%
of $\rhoSFR$ comes from galaxies with
SFR$/\langle\mathrm{SFR}\rangle>1$ and 50\% at
SFR$/\langle\mathrm{SFR}\rangle>0.54$, but only 3\% of $\rhoSFR$ comes
from galaxies with SFR$/\langle\mathrm{SFR}\rangle>10$.  The
traditional definition of star bursts is however $b\ge 2$--$3$ and
with that definition around 20\% of the local SFR density takes place
in star bursts. We should caution that our method of inferring the SFR
from optical emission lines is likely to miss part of the SFR
occurring in extremely dusty FIR-dominated galaxies. Given that such
systems are very rare at low redshifts
\citep[e.g.][]{1996ARA&A..34..749S}, this does not significantly
affect our results.

Although IR-luminous galaxies are rare in the local universe, most
local star formation occurs in galaxies with substantial optical
attenuation in their central parts. It is important to realise
that this quantity might suffer from aperture biases so the galaxy
averaged attenuation could be lower. The relationship between dust
attenuation and SFR activity has been discussed by many                     
authors. The most recent discussion is given by
\citet{Hopkins-et-al-2003} for a subsample of our SDSS data. Our
results are in good agreement with theirs. Our main result is that the
SFR-weighted dust attenuation at \ha\ is $A_{H\alpha} \sim 1.0\pm 0.5$
magnitudes, with a good correlation between dust attenuation and SFR.
The exact relationship between dust and star formation remains unclear.
We see evidence for a strong correlation between stellar mass and
dust attenuation, presumably reflecting a correlation between
metallicity and dust content. The correlation between SFR and
dust may simply reflect a more fundamental correlation between dust and
metallicity. This question clearly merits further investigation.
Another consequence of the moderately strong dust attenuation present
in most galaxies is that
the colours of the galaxies that dominate the SFR density have
intermediate values --- this is why we needed a second colour to get a
good constraint on star formation activity in
section~\ref{sec:aperture_effects}.

To isolate the influence of other factors that may regulate
the star formation, it is clearly necessary to
take out the dominant trend with mass. One should probably normalise
by the mass of the gas reservoir, but as this                         
information is not  available we have normalised by stellar mass. 
Previous studies by \citet{2002A&A...396..449G} and
\citet{2001AJ....121..753B} have shown that SFR$/\Mstar$ correlates with
mass. We confirm this trend. The median value of $\sSFR$ is
fairly constant over a wide range at low masses and then decreases smoothly
at masses greater than $\Mstar\sim10^{10}\Msun$. The
simplest interpretation is that low mass galaxies form
stars at an approximately constant rate, whereas the star formation
history is increasingly peaked towards earlier times for more massive
galaxies.

These trends have arguably been seen before. They agree with the
``downsizing'' scenario of \citet{1996AJ....112..839C}. What has not
been apparent up to now is the fact that the extremes of the distribution
change noticeably. As we go up in mass, there are fewer and fewer
galaxies with very high $\sSFR$ and a corresponding tail towards low
$\sSFR$ starts to appear.

It is clear from the previous section that no single factor stands out
as controlling $\sSFR$. This is not entirely surprising
since the properties we have used are simple and somewhat limited.
It might turn out  that more detailed
properties correlate better with $b$, for example
lopsidedness \citep{2000ApJ...538..569R},  local density
\citep{2003ApJ...584..210G} or  merging state
\citep[e.g.][]{2000MNRAS.311..565L}. We will study
these relations further in future work.

We mentioned previously that the $\sSFR^V$ value for the sample as a
whole provides a strong constraint on the star formation history of
the universe.  This has been discussed in some detail by
\citet{2001AJ....121..753B}, but that study did not use a SFH that is
compatible with look-back studies -- it assumed an exponential star
formation history for all galaxies. Our $b^V$ value with our cosmology
corresponds to an effective $\tau = 7_{-1.5}^{+0.7}$Gyr (including
systematic errors) if one considers the whole universe to have an
exponential SFH.

We will however adopt the parameterisation of the SFH of the universe
used by \citet{2002ApJ...569..582B}. In general form this reads
\begin{equation}
  \label{eq:SFH_formula}
  \mathrm{SFR} \propto \left\{ 
    \begin{array}{ll}
      (1+z)^\alpha & \mbox{for $z\in [z_d, z_c]$} \\
      (1+z)^\beta & \mbox{for $z<z_d$} \\
      0 & \mbox{for $z>z_c$}\\
    \end{array}
  \right.,
\end{equation}
where we will adopt $z_d=1$, $z_c=5$ to be consistent with
\citet{2002ApJ...569..582B} and \citet{Glazebrook-cosmspec}. Those
authors showed that the local ``cosmic spectrum'' is well fit by a
model with $\beta=3$ and $\beta=0$. 

To calculate $b^V$ for the SFH defined by
Equation~(\ref{eq:SFH_formula}) we simply integrate over time and
define
\begin{equation}
  \label{eq:mstar_from_model}
  \langle \mathrm{SFR} \rangle = \frac{1}{t_0} \int_{0}^{t_0}
  \mathrm{SFR}(t) dt, 
\end{equation}
with $t_0$ being the age of the universe at $z=0.1$. This, together with
equation~(\ref{eq:SFH_formula}), then gives the present to past-average
SFR for the model
\begin{equation}
  \label{eq:b_model}
  b^V_{\mathrm{model}} = t_0 \frac{\mathrm{SFR}(t_0)}{\int_{0}^{t_0}
  \mathrm{SFR}(t) dt}.
\end{equation}

It is interesting to note that because of the way we measure $\Mstar$
and the SFR, there is a dependence on the Hubble parameter
in our $b$ value calculations. That dependence cancels out in
equation~(\ref{eq:SFH_formula}). To constrain $\alpha$ and $\beta$
we thus need a prior on $H_0$. We will take the simple approach and choose
$h=0.7$. We use the constraint on $b^V$ given in
equation~(\ref{eq:present_to_past_total}). 

\begin{figure}
  \centering
  \includegraphics[width=84mm]{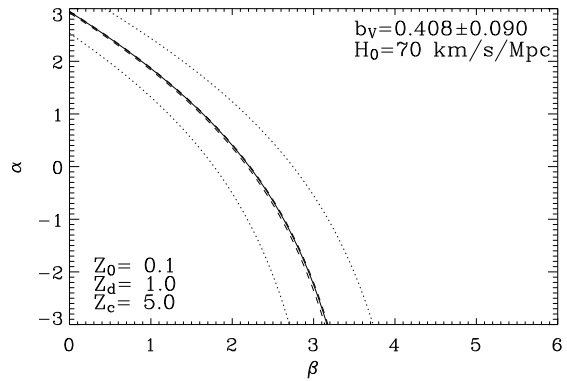}
  \caption{The constraint on $\alpha$ and $\beta$ (see
    Equation~\ref{eq:SFH_formula}) from the $b^V$ value calculated for
  our sample. The solid line shows the locus of $b^V = 0.400$ whereas
  the dashed line shows the 68\% confidence limits from the likelihood
  distributions and the dotted the locus of our conservative
  systematic error estimates.}
  \label{fig:alpha_beta_constrains}
\end{figure}

Figure~\ref{fig:alpha_beta_constrains} shows the resulting constraints
on $\alpha$ and $\beta$. Since the $b^V$ value only constrains the
$z=0.1$ value of the SFR and the integral under the curve, there is 
a strong degeneracy even with a fixed functional form. This
degeneracy cannot be lifted without independent constraints, but it
is still of considerable interest. Most noticeably it is clear that
any really strong evolution, $\beta > 4$, at $z<1$ is excluded by the
current data. 

The dotted lines in Figures~\ref{fig:alpha_beta_constrains} show the
effect of systematic uncertainties on these constraints (for
simplicity we have symmetrised these). The dashed
lines show the purely statistical errors (1$\sigma$). It is clear that
if we could remove all systematic uncertainties, the method would give
a very powerful constraint on the star formation history of the
universe. It is therefore worth asking what is required to further
reduce the systematic uncertainties.

A thorough test of our aperture corrections 
using resolved spectroscopy would substantially reduce the
systematic uncertainty in our SFR estimates. To reduce 
uncertainties due to evolution, one could limit oneself to a
narrow range in redshift, but that would mean a corresponding increase
in the uncertainty in extrapolating to faint galaxies. Alternatively,
it might be possible to directly constrain the evolution inside the
sample, but that is a rather delicate operation due to the complex
aperture effects.  We have postponed this to a later paper.

Systematic uncertainties notwithstanding, the results  presented
here provide remarkably good constraints on the cosmic star formation
history using only data at $z<0.22$. It is clear that a similar study
carried out at higher redshift will be of great interest and provide a
very interesting complement to recent studies of the mass-assembly of
the universe
\citep{2000ApJ...536L..77B,2001ApJ...562L.111D,2003ApJ...587...25D,2003ApJ...594L...9F,2003astro.ph..7149R}.
 
In general it is our firm belief that future studies of high redshift
data should try as far as possible to study the evolution of the full
distribution functions in question rather than summarising the
distributions to a single value. As we have seen, important changes
might occur in the tails of the distributions with little influence on
the median value. These physically interesting details can only be
extracted through the use of distribution functions. It has been
standard to study luminosity functions. With the techniques introduced
in the present paper it is now possible to extend the analysis to a
wide variety of distribution functions. This has been possible due to
the unprecedented dataset provided by the SDSS survey.

\section{Acknowledgements}
\label{sec:acks}

We thank Ivan Baldry, Andrew Hopkins, Karl Glazebrook and Greg Rudnick
for stimulating discussions and valuable input on parts of this work.
J.B. and S.C. thank the Alexander von Humboldt Foundation, the Federal
Ministry of Education and Research, and the Programme for Investment
in the Future (ZIP) of the German Government for their
support. J.B. would also like to acknowledge the receipt of an ESA
post-doctoral fellowship.

The research presented here has benefitted greatly from the NYU Large
Scale Structure database and the efforts of Mike Blanton to maintain
this. 

Funding for the creation and distribution of the SDSS Archive has been
provided by the Alfred P. Sloan Foundation, the Participating
Institutions, the National Aeronautics and Space Administration, the
National Science Foundation, the U.S. Department of Energy, the
Japanese Monbukagakusho, and the Max Planck Society. The SDSS Web site
is http://www.sdss.org/.

The SDSS is managed by the Astrophysical Research Consortium (ARC) for
the Participating Institutions. The Participating Institutions are The
University of Chicago, Fermilab, the Institute for Advanced Study, the
Japan Participation Group, The Johns Hopkins University, Los Alamos
National Laboratory, the Max-Planck-Institute for Astronomy (MPIA),
the Max-Planck-Institute for Astrophysics (MPA), New Mexico State
University, University of Pittsburgh, Princeton University, the United
States Naval Observatory, and the University of Washington.


\appendix

\section{Manipulation of likelihood distributions}
\label{sec:manipulate_likelihoods}

Throughout this paper we use the full likelihoods of all parameters.
This means that in many situations it is necessary to combine
likelihood distributions. For the benefit of the
reader we summarise some of the most important results here. We
will use standard notation and write e.g. $f_Z(z)$ for the likelihood
of a value $z$ in the likelihood distribution $Z$. Thus we represent
the addition of two likelihood distributions as $Z=X+Y$. This sum is
discussed in most introductory statistics textbooks and the well know
result is that
\begin{equation}
  \label{eq:z_eq_x_plus_y}
  f_Z(z) = \sum_i  f_{XY} (x_i, z-x_i) \stackrel{\mathrm{indep.}}{=}
  \sum_i f_X (x_i) f_Y(z-x_i),
\end{equation}
where the latter equality is valid in the case of independent
variables.

In the present work we do not often use the straight addition of
distributions. The general case is not normally given in
statistics textbooks. It is however easy to derive. We will focus our
attention on the cumulative distribution function (CDF)
$F(z\leq\hat{z})$. In the continuous case this is given as 
\begin{equation}
  \label{eq:f_cumul}
  F(z\leq \hat{z}) = \int_{\cal R}\!\!\int f_{XY} (x, y) dx\,dy,
\end{equation}
where $\cal R$ is the region where the constraint $z\leq \hat{z}$ is
fulfilled. If we write the general composition of two likelihood
distributions as $Z=g(X,Y)$ then the constraint $z\leq\hat{z}$
translates into $z\leq g(x,y)$. For well-behaved distributions we can
invert this to write $y \leq h(x,z)$. In general we can then write
\begin{equation}
  \label{eq:f_cumul_two}
  F(z\leq \hat{z}) = \int_{-\infty}^{\infty} \int_{-\infty}^{h(x,z)}
  f_{XY}(x, y)\, dx\, dy.
\end{equation}

A simple change of variable presents itself and by setting $y =
h(x,u)$ we get 
\begin{equation}
  \label{eq:f_cumul_three}
  F(z\leq \hat{z}) = \int_{-\infty}^{\infty} \int_{-\infty}^{z}
  f_{XY}\left(x, h(x,u)\right) J(h)\, dx\, du,
\end{equation}
with $J=\partial h/\partial u$ being the Jacobian of the variable transformation. By taking
the derivative of this we can get the likelihood distribution
$f_Z(z)$: 
\begin{equation}
  \label{eq:f_of_z}
  f_Z(z) = \frac{\partial F}{\partial z} = \int_{-\infty}^{\infty}
  f_{XY}\left(x, h(x,z)\right)\, \left|J(h)\right|\, dx, 
\end{equation}
which can be evaluated for any given $g(X,Y)$. The extension to
discrete distribution functions is straightforward (see below). 

As an example we can take the addition of two likelihood distributions
for the SFR which we need for the aperture corrections. Here we have
\begin{equation}
  \label{eq:adding_SFR}
  Z = \log_{10} (10^X + 10^Y),
\end{equation}
since the likelihood distributions are in log space. Thus calculating
the Jacobian and writing it as a discrete distribution
\begin{eqnarray}
  \label{eq:adding_log_sfr}
  \lefteqn{f_Z (z_j) = } & & \nonumber \\
  & & \sum_i f_X(x_i) f_Y\left(\log_{10}(10^{z_j}-10^{x_i})\right)
  \frac{10^{z_j}}{10^{z_j}-10^{x_i}}. 
\end{eqnarray}

This formalism extends fairly straightforwardly to the iterated
composition of many (independent) likelihood distributions
\begin{eqnarray}
  \label{eq:comp_n}
  f_Z(z) & = & \int_{-\infty}^{\infty} \cdots \int_{-\infty}^{\infty}
  f_{X_1}(x_1) f_{X_2}\left[h_{2} (x_1, x_2)\right] \ldots \nonumber \\
  & & f_{X_N}\left[h_N(x_{N-1}, x_N)\right] \left[\prod_{i=2}^{N}
  \left|\frac{\partial h_i}{\partial x_i}\right|\right] d^N x_i,
\end{eqnarray}
but this is not a practical approach for arbitrary likelihood
distributions due to combinatorial explosion (the sum/integral has to
be done over all combinations of variables that combine to a
particular value of $z$). Instead it is easier in this situation to
calculate the combined likelihood distribution in an approximate
manner by carrying out a Monte Carlo sampling of the distributions. In
this case we first form the CDF, $F(x)$, for each galaxy. We
then draw,  for each galaxy, a random number, $p$, between 0 and 1 and
solve $F(x)=p$ numerically using linear interpolation. 

We use the latter approach for instance when evaluating the star
formation density of the local universe. The main disadvantage of a
Monte Carlo approach is that one needs to carry out a large number of
repetitions to ensure a sufficient sampling of the final likelihood
distribution and to estimate the confidence intervals with good
accuracy.

Before continuing we would like to emphasise a technical issue which
is of considerable importance for some combinations of likelihood
distributions: Ignoring the likelihood distribution of a particular
variable (in effect setting its error to zero) will in general
\emph{bias} the estimate of any combination of this variable with any
other unless they are both Gaussians.  A specific example relevant for
our work is the ratio of two variables whose likelihood distributions
are kept in log space.

In particular the calculation of $\sSFR$ in
section~\ref{sec:spec_sfr} falls into this category. Here we have
$Z=10^{Y}/10^{X}$ where $X$ is the likelihood distribution of the
stellar mass and $Y$ that of the SFR. If we now for illustrative
purposes assume that these are Gaussian in \emph{log} space: $\log
\mathrm{SFR} \sim N\left(\log \mathrm{SFR}_0,
  \sigma_{\mathrm{SFR}}\right)$ and likewise for $\log M_*$ then a
straightforward calculation gives us that the mean of $Z$
\begin{equation}
  \label{eq:mean_z}
  \langle z \rangle = \int_0^{\infty} z f_Z(z) dz,
\end{equation}
can be written
\begin{equation}
  \label{eq:mean_z_answer}
  \langle z \rangle = e^{\log \mathrm{SFR/M_*}_0 +
    \left(\sigma_{M_*}^2 + \sigma_{\mathrm{SFR}}^2\right)/2\ln 10}.
\end{equation}

Clearly setting $\sigma_{M_*}=0$ would bias the average value for the
specific SFR in this situation. The exact amount depends on the
details of the likelihood distribution and the bias could vanish
entirely. In practice (as mentioned in section~\ref{sec:spec_sfr}) we
find that for the calculation of the specific SFR, ignoring the
uncertainty in $\Mstar$ would bias our results about 15\% high.

Finally we would like to mention the various ways one can summarise a
distribution and what we typically use \citep[see
also][]{2003MNRAS.341...33K}. The \textbf{average} is the simplest as
it is calculated as
\begin{equation}
  \label{eq:average_likelihoods}
  \langle x \rangle = \sum_i x_i f_X(x_i),
\end{equation}
since $f_X$ is normalised to a total of one. The mode is equally
simple
\begin{equation}
  \label{eq:mode_definition}
  x_{\mathrm{mode}} = x_i \mbox{ such that } \max_j f_X(x_j) = f_X(x_i). 
\end{equation}
In the event of a tie we use the smallest value satisfying the criterion.
These two, but in particular the mode are affected by the binning
chosen for the likelihood distributions.

More commonly used in the present paper are the percentiles of the
distributions, defined in the standard way as 
\begin{equation}
  \label{eq:def_percentiles}
  F_X(x_p) = p,
\end{equation}
where $F_X$ is the CDF, $p$ is the
percentile --- $p=0.5$ gives the median, $p=0.16$ and $p=0.84$ give
the upper and lower ``$1\sigma$'' limits chosen and $x_p$ is the value
to solve for.

To avoid severe binning effects (in particular for the extremes of the
distributions) we calculate these percentiles by solving
equation~(\ref{eq:def_percentiles}) using linear interpolation between
grid points. In this situation we interpolate the CDF and this is a
robust procedure. For the average a straight application of
equation~(\ref{eq:average_likelihoods}) would require interpolation of
the probability density function (PDF) which might depend sensitively
on the interpolation scheme. Instead we calculate the average by
interpolating the CDF and drawing $10^4$ realisations of this for each
galaxy and calculating the average from this. We have experimented
with different numbers of realisations and interpolation schemes and
found that the estimate of the average so calculated is robust to
these changes within the accuracy needed for our project.

\section{Minor contributions to the emission line spectra}
\label{sec:PNe_contribution}

We will here discuss, briefly, the contribution of supernova remnants
(SNRs) and planetary nebulae (PNe) to the overall emission line flux
of galaxies. The SNRs are interesting in their own right, but due to
their short lifetime and the fact that they co-exist with star forming
regions they are unlikely to affect the line-radiation significantly.
This is borne out in studies of resolved \hii-regions in the LMC
\citep{2000ApJS..128..511O} which show that SNRs only affect the
emission line spectra at a fairly low level.

As discussed by CL01, the contribution of PNe to the line radiation
can be ignored in galaxies with active star formation, however in the
current work we have galaxies with very low \ha-luminosities and it is
therefore necessary to ask if these fluxes can come from PNe. It is
straightforward to assess this as the PNe \oiii\ luminosity function
seems to be universal \citep[e.g.][ and references
therein]{2002ApJ...577...31C}. We can therefore use this to determine
the \oiii-flux from the PNe population in a given galaxy. Using the
PNe LF in \citet{2002ApJ...577...31C} we find that the maximum
contribution to the total \oiii-flux of the sample by PNe is of the
order of 1.5\%. This is in general totally negligible in comparison
with other sources of uncertainty, but it is worth noting that we do
predict that among galaxies with very weak emission lines, $S/N<5$ in
\oiii, around 20\% of the galaxies could have PNe contributing more
than 20\% of the \oiii-flux thereby complicating studies of galaxies
with very weak emission lines.  Since the \oiii/\ha\ ratio in PNe is
in general considerably larger than for star forming galaxies, the
effect on \ha\ and hence the star formation estimates, is therefore
entirely negligible except at $S/N < 1$ in \ha\ so has no effect
whatsoever for our results here.

A significant amount, typically 30--50\%, of the emission line flux in
galaxies comes from the diffuse ionized gas
\citep[DIG][]{1994ApJ...426L..27L,1997ApJ...491..114W,2003ApJ...586..902H}.
This is implicitly taken into account by the CL01 models because as
CL01 show the combination of the (low-ionization) DIG and the (higher
ionization) \hii-regions is typically represented by a lower effective
ionization parameter \citep[see also][]{2000ApJ...544..347M}. Although
we do not explicitly incorporate leaky \hii-regions as done by e.g.\ 
\citet{2003ApJ...586..902H}, we can produce similar line-ratios by
changing $\xi$. Thus the models do not need any extra component to
take into account DIG and as long as the DIG is heated by OB stars as
seems necessary, its presence should not bias our SFR estimates. The
interpretation of the fitted ionization parameter and dust-to-metal
ratio is more complex and we will discuss this issue in more detail in
an upcoming paper.

\end{document}